\documentclass[11pt]{article}
\usepackage{epsfig}
\usepackage{latexsym}
\usepackage{graphicx}
\usepackage{color}
\usepackage{amsmath,amssymb}
\usepackage{cite}

\def\simlt{\stackrel{<}{{}_\sim}}  
\def\simgt{\stackrel{>}{{}_\sim}}  
\makeatletter
\def\section{\@startsection {section}{1}{\z@}{-3.5ex plus -1ex minus
 -.2ex}{2.3ex plus .2ex}{\large\bf}}
\def\subsection{\@startsection{subsection}{2}{\z@}{-3.25ex plus -1ex
minus -.2ex}{1.5ex plus .2ex}{\normalsize\bf}}
\makeatother
\newcommand{\captionfonts}{\small}
\makeatletter  
\long\def\@makecaption#1#2{%
  \vskip\abovecaptionskip
  \sbox\@tempboxa{{\captionfonts #1: #2}}%
  \ifdim \wd\@tempboxa >\hsize
    {\captionfonts #1: #2\par}
  \else
    \hbox to\hsize{\hfil\box\@tempboxa\hfil}%
  \fi
  \vskip\belowcaptionskip}
\makeatother   

\catcode`@=11
\def\marginnote#1{}
\newcount\hour
\newcount\minute
\newtoks\amorpm
\hour=\time\divide\hour
by60
\minute=\time{\multiply\hour by60 \global\advance\minute
by-\hour}
\edef\standardtime{{\ifnum\hour<12 \global\amorpm={am}
\else\global\amorpm={pm}\advance\hour by-12 \fi
 \ifnum\hour=0
\hour=12 \fi
 \number\hour:\ifnum\minute<10
0\fi\number\minute\the\amorpm}}
\edef\militarytime{\number\hour:\ifnum\minute<10
0\fi\number\minute}
\def\draftlabel#1{{\@bsphack\if@filesw
{\let\thepage\relax
 \xdef\@gtempa{\write\@auxout{\string
\newlabel{#1}{{\@currentlabel}{\thepage}}}}}\@gtempa
 \if@nobreak
\ifvmode\nobreak\fi\fi\fi\@esphack}
\gdef\@eqnlabel{#1}}
\def\@eqnlabel{}
\def\@vacuum{}
\def\draftmarginnote#1{\marginpar{\raggedright\scriptsize\tt#1}}
\def\draft{\oddsidemargin
0.0truein
 \def\@oddfoot{\sl preliminary draft \hfil
\rm\thepage\hfil\sl\today\quad\militarytime}
 \let\@evenfoot\@oddfoot
\overfullrule 3pt
 \let\label=\draftlabel
\let\marginnote=\draftmarginnote
\def\@eqnnum{(\theequation)\rlap{\kern\marginparsep\tt\@eqnlabel}
\global\let\@eqnlabel\@vacuum}
}
\catcode`@=12
\def\dj{\hbox{d\kern-0.347em \vrule width 0.3em height 1.252ex depth
-1.21ex \kern 0.051em}}
\def\bma#1{\mbox{\boldmath{$#1$}}}

\def\ee{{\rm e}\,}

\def\Dirac{\,\raise.15ex\hbox{/}\mkern-13.5mu D}
\def\dirac{\,\raise.15ex\hbox{/}\kern-.57em \partial}
\def\aslash{\,\raise.15ex\hbox{/}\mkern-13.5mu A}
\def\shalf{{\ifinner {\textstyle \frac{1}{2}}\else \frac{1}{2} \fi}} 
\def\sthreehalfs{{\ifinner {\textstyle \frac{3}{2}}\else \frac{3}{2} \fi}} 
\def\sshalf{{\ifinner {\scriptstyle \frac{1}{2}}\else \frac{1}{2} \fi}} 
\def\sfourth{{\ifinner {\textstyle \frac{1}{4}}\else frac{1}{4} \fi}}
\def\sphifour{{\ifinner {\textstyle \frac{1}{4!}}\else \frac{1}{4!} \fi}}
\def\lsim{\stackrel{<}{_\sim}}

\def\XXint#1#2#3{{\setbox0=\hbox{$#1{#2#3}{\int}$}
     \vcenter{\hbox{$#2#3$}}\kern-.5\wd0}}

\def\bea{\begin{eqnarray}} \def\eea{\end{eqnarray}}
\def\be{\begin{eqnarray}} \def\ee{\end{eqnarray}} \def\nn{\nonumber}

\newcommand{\promille}{%
  \relax\ifmmode\promillezeichen
        \else\leavevmode\(\mathsurround=0pt\promillezeichen\)\fi}
\newcommand{\promillezeichen}{%
  \kern-.05em%
  \raise.5ex\hbox{\the\scriptfont0 0}%
  \kern-.15em/\kern-.15em%
  \lower.25ex\hbox{\the\scriptfont0 00}}
\newcommand{\beq}{\begin{eqnarray}}
\newcommand{\eeq}{\end{eqnarray}}
\newcommand{\sli}{\newline \vspace*{-3.5mm}}

\newcommand{\gsim}{\raisebox{-0.13cm}{~\shortstack{$>$ \\[-0.07cm] $\sim$}}~}

\begin{document}

\thispagestyle{empty}

\begin{center}
\hfill CERN-PH-TH/2010-020, KA-TP-08-2010

\begin{center}

\vspace{1.7cm}

{\LARGE\bf Composite Higgs Search at the LHC}
\end{center}

\vspace{1.4cm}

{\bf J.R. Espinosa$^{\,a,\,b}$}, {\bf C. Grojean$^{\,b, \, c}$} and  {\bf M. 
M\"uhlleitner$^{\,d}$}\\

\vspace{1.2cm}

${}^a\!\!$
{\em {ICREA, Instituci\`o Catalana de Recerca i Estudis Avan\c{c}ats,
Barcelona, Spain}}\\
at {\em { IFAE, Universitat Aut{\`o}noma de Barcelona,
08193 Bellaterra, Barcelona, Spain}}
\\
${}^b\!\!$
{\em {CERN, Physics Department, Theory Unit, CH--1211 Geneva 23, Switzerland}}
\\
${}^c\!\!$
{\em {Institut de Physique Th\'eorique, CEA Saclay, F--91191 
Gif-sur-Yvette, France}}
\\
${}^d\!\!$
{\em {Institut f\"ur Theoretische Physik, Karlsruhe Institut of Technology, 76128 Karlsruhe, Germany}
}\\

\end{center}

\vspace{0.8cm}

\centerline{\bf Abstract}
\vspace{2 mm}
\begin{quote}
\small
The Higgs boson production cross-sections and decay rates depend, within
the Standard Model (SM), on a single unknown parameter, the Higgs mass. In
composite Higgs models where the Higgs boson emerges as a pseudo-Goldstone
boson from a strongly-interacting sector, additional parameters control
the Higgs properties which then deviate from the SM ones. These deviations modify the LEP and Tevatron
exclusion bounds and significantly affect the searches for the Higgs boson
at the LHC.  In some cases, all the Higgs couplings are reduced, which
results in deterioration of the Higgs searches but the deviations of the
Higgs couplings can also allow for an enhancement of the gluon-fusion
production channel, leading to higher statistical significances. The search in the 
$H \to \gamma \gamma$ channel can also be substantially improved due to an
enhancement of the branching fraction for the decay of the Higgs boson into a pair of photons.
\end{quote}

\vfill

\newpage

\section{Introduction}
\label{sec:Intro}

The massive nature of the weak gauge bosons requires new degrees of
freedom and/or new dynamics around the TeV scale to act as an ultraviolet (UV) moderator
and ensure a proper decoupling at high energy of the longitudinal
polarizations $W^\pm_L, Z_L$. It is remarkable that a simple elementary
weak-doublet not only provides the three Nambu--Goldstone bosons that will
become the spin-1 longitudinal degrees of freedom but also contains an
extra physical scalar field, the notorious Higgs boson, that screens the
gauge-boson non-Abelian self-interaction contributions to scattering
amplitudes and hence offers a consistent description of massive spin-1
particles. The minimality of this ElectroWeak Symmetry Breaking (EWSB)
sector comes as a result of a highly constrained structure among the
couplings of the Higgs doublet to the other Standard Model (SM) particles:
a single parameter, the mass of the physical Higgs boson, dictates all the
physical properties of the Higgs sector. Despite intensive searches over
the last 20 years, no experimental results have been able to establish the
reality of this theoretical paradigm. However, a harvest of electroweak
precision data accumulated during these experimental searches, together
with the absence of large flavor-changing neutral currents, suggests that
violent departures from this minimal Higgs mechanism are unlikely, and
rather call for smooth deformations, at least at low energy.

This provides a plausible motivation for considering a light Higgs boson
emerging as a pseudo-Goldstone boson from a strongly-coupled sector, the
so-called Strongly Interacting Light Higgs (SILH)
scenario~\cite{Giudice:2007fh,Contino:2010mh}\footnote{SILH models have
some similarities with models where the role of the Higgs is played by a
composite dilaton resulting from the spontaneous breaking of scale
invariance. See Refs.~\cite{dilaton} for a recent discussion.}. At low
energy, the particle content is identical to the SM one: there exists a
light and narrow Higgs-like scalar but this particle is a bound state from
some strong dynamics~\cite{Kaplan:1983fs, othercompositeHiggs} and a mass
gap separates the Higgs boson from the other usual resonances of the
strong sector as a result of the Goldstone nature of the Higgs.  
Nevertheless, the rates for Higgs production and decay differ
significantly from those in the minimal Higgs incarnation.  The aim of the
present work is to look at how the searches for a Higgs boson are affected
by the modifications of its couplings. Reference~\cite{Falkowski:2007hz}
already studied the modification induced by the strong dynamics to the
gluon-fusion Higgs production and it was argued that it could have an
impact on the Higgs searches\footnote{See also Refs.~\cite{otherhgg} for
an analysis of the gluon-fusion Higgs production in similar but different
contexts.}. We extend this analysis and estimate the experimental
sensitivities in the main LHC search channels studied by ATLAS and CMS.

In the attempt of providing a simple theoretical picture to parametrize
the Higgs couplings in composite models, Ref.~\cite{Giudice:2007fh}
constructed an effective Lagrangian involving higher dimensional operators
for the low energy degrees of freedom and concluded that, as far as the
LHC studies are concerned, the Higgs properties are essentially governed
by its mass plus two new parameters. The effective SILH Lagrangian should
be seen as an expansion in $\xi=(v/f)^2$ where $v=1/\sqrt{\sqrt{2} G_F}\approx 246$~GeV and $f$ is the
typical scale of the Goldstone bosons of the strong sector. Therefore, it
can be used to describe composite Higgs models in the vicinity of the SM
limit, $\xi \to 0$. To reach the technicolor limit~\cite{reviewtechnicolor}, $\xi \to 1$, a
resummation of the full series in $\xi$ is needed. Explicit models, built
in five-dimensional (5D) warped space, provide concrete examples of such a resummation. In
our analysis, we will rely on two 5D models that exhibit different
behaviors of the Higgs couplings that, we hope, will be representative of
the various composite Higgs models. In these explicit models, the two
extra parameters that generically control the couplings\footnote{We will qualify these couplings as {\it anomalous} couplings since they differ from the SM ones.} of a composite
Higgs boson are related to each other and the deviations from the SM Higgs
couplings are only controlled by the parameter $\xi=(v/f)^2$ which varies
from 0 to 1\footnote{Similar deviations of the Higgs couplings are also present in extra-dimensional models where the Higgs mixes with the {\it radion} field~\cite{Giudice:2000av}, however, in that case, the deviations do not originate from strong interactions.}. In that sense, our analysis is an exploration of the
parameter space of composite models along some special directions only. A
complementary, but more general, analysis relying on the two parameters of
the SILH Lagrangian is also possible, but it would be restricted to the
range of validity of the $(v/f)^2$ expansion and would not allow to
approach the technicolor limit. For these reasons, we did not pursue it
further.

Composite Higgs models offer a continuous deformation of the SM Higgs
paradigm. Another possible deformation consists in playing with the
anomalous dimension of the Higgs field like in Higgsless
models~\cite{higgsless}, gaugephobic models~\cite{Cacciapaglia:2006mz},
unHiggs models~\cite{Stancato:2008mp} and conformal technicolor
models~\cite{Luty:2004ye}, whose effective 4D descriptions might involve
some non-local operators to take into account the non-canonical dimension
of the Higgs boson (see Refs.~\cite{reviewsNewPhysics} for reviews of
models of new physics at the TeV scale).

It should be stressed that the couplings of the Higgs boson in the SILH
scenario are not the most general ones that would be allowed by the
general principles of quantum field theory and the local and global
symmetries of the models considered: for instance, the important anomalous
couplings will have the same Lorentz structure as the SM ones. In
principle, some couplings with a different Lorentz structure could also be
expected, but these ones would be generated only via the exchange of heavy
resonances of the strong sector and not directly by the strong dynamics of
the Goldstone bosons, therefore they would be parametrically suppressed,
at least by a factor $(f/m_\rho)^2$ ($m_\rho>2.5$~TeV is the typical mass
scale of these resonances), and are irrelevant for our analysis. For
similar reasons and due to the Goldstone nature of the Higgs, a direct
coupling of the Higgs boson to two gluons or two photons will always
induce sub-leading effects compared with the ones we are
considering\footnote{\label{footnote:hgg} This statement will change when
the SM fermions, and in particular the top quark, have a direct coupling
to the strong sector. Then some top-partners are expected to give sizeable
corrections to the $H\gamma\gamma$ and $Hgg$ vertices (we will explore 
this
possibility in a future work). On the contrary, when the fermions are
elementary, all these corrections can be recast into a correction to the
Yukawa couplings only and the $H\gamma\gamma$ and $Hgg$ loop-induced
vertices do not depend on the details of the resonance spectrum, hence, as
announced, the Higgs physical properties do depend only on two extra
parameters in addition to its mass. This structure is specific to SILH
models and does not hold in general:  indeed totally model-independent
operator analyses~\cite{generalhgg,hgg} lead to the conclusion that the
dominant effects should appear in the vertices $H\gamma\gamma$ and
$Hgg$.}.

Higgs anomalous couplings are not by themselves a direct probe of the
strong sector at the origin of EWSB. For that, one would need to wait for
the direct production of the heavy resonances of the strong sector or to
rely on the processes with two Goldstones in the final state, as in the
$WW$ scattering or in the double Higgs production by boson
fusion~\cite{Contino:2010mh}, where the composite nature of the Higgs
boson would manifest itself by a residual growth of the amplitudes above
the Higgs mass. Nevertheless, the relative importance of the various Higgs
production and decay channels can bring first insights on the dynamics
that controls the Higgs sector.

The paper is organized as follows: in Section~\ref{sec:couplings}, we give
the general parametrization of the couplings of a composite Higgs as
derived from the SILH Lagrangian of Ref.~\cite{Giudice:2007fh} and, for
the two explicit 5D composite Higgs models we will consider, we give the
exact form of these couplings valid for values of $\xi$ interpolating
between the SM and the technicolor limits. The deviations in the Higgs
decay rates are presented and the bounds on the Higgs mass at LEP and
Tevatron are studied (Section~\ref{sec:constraints}).
Section~\ref{sec:searches} contains our main results: we first discuss the
modifications, due to the composite nature of the Higgs boson, of the
Higgs production cross-sections including the next-to-leading order QCD
corrections and then we re-examine the various search channels for a Higgs
boson at the LHC computing the changes in their statistical
significance. At low values of $\xi$, the searches are
made more difficult due to a general reduction of all the Higgs couplings,
but for larger values of $\xi$, it is possible to increase the
significance thanks, in particular, to an enhanced Higgs production by
gluon fusion, though this enhancement is model-dependent. Finally, in
Section~\ref{sec:Conclusions}, we combine the various search channels
and we present our conclusions. In the appendix, we collect the various
estimators of the statistical significance we use in our analysis.

\section{General parametrization of the Higgs couplings}
\label{sec:couplings}

\subsection{SILH couplings}

The effective Lagrangian describing a SILH involves higher dimensional
operators. There are two classes of higher dimensional operators:
(i)~those that are genuinely sensitive to the new strong force and will
affect qualitatively the physics of the Higgs boson and (ii)~those that
are sensitive to the spectrum of the resonances only and will simply act
as form factors. Simple rules control the size of these different
operators (see Ref.~\cite{Giudice:2007fh})  and the effective Lagrangian
generically takes the form
\bea
\mathcal{L}_{\rm SILH} & = & \frac{c_H}{2f^2} \left( \partial_\mu |H|^2 
\right)^2
+ \frac{c_T}{2f^2}  \left(   H^\dagger{\overleftrightarrow D}_\mu H\right)^2 
- \frac{c_6\lambda}{f^2} |H|^6
+ \left( \frac{c_yy_f}{f^2} |H|^2 {\bar f}_L Hf_R +{\rm h.c.}\right) 
\nonumber\\ 
&&+\frac{ic_Wg}{2m_\rho^2}\left( H^\dagger  \sigma^i \overleftrightarrow 
{D^\mu} H \right )( D^\nu  W_{\mu \nu})^i
+\frac{ic_Bg'}{2m_\rho^2}\left( H^\dagger  \overleftrightarrow {D^\mu} H 
\right )( \partial^\nu  B_{\mu \nu})  +\ldots 
\label{eq:silh}
\eea
where $g, g'$ are the SM EW gauge couplings, $\lambda$ is the SM Higgs
quartic coupling and $y_f$ is the SM Yukawa coupling to the fermions
$f_{L,R}$. All the coefficients, $c_H, c_T \ldots$, appearing in
Eq.~(\ref{eq:silh}) are expected to be of order one unless protected by
some symmetry. For instance, in every model in which the strong sector
preserves custodial symmetry, the coefficient $c_T$ vanishes and only
three coefficients, $c_H, c_y$ and $c_6$, give sizable contributions to
the Higgs (self-)couplings.  The operator $c_H$ gives a correction to the
Higgs kinetic term which can be brought back to its canonical form at the
price of a proper rescaling of the Higgs field, inducing a universal shift
of the Higgs couplings by a factor $1-c_H\, \xi/2$. For the fermions, this
universal shift adds up to the modification of the Yukawa interactions
\begin{eqnarray}
&g_{Hf\bar{f}}^{\xi} = g_{Hf\bar{f}}^\textrm{\tiny SM}\times [1-(c_y 
+ c_H/2) 
\xi],\\
&g_{HVV}^{\xi} = g_{HVV}^\textrm{\tiny SM} \times (1-c_H\, \xi/2),\ \
g_{HHVV}^{\xi} = g_{HHVV}^\textrm{\tiny SM} \times (1-2 c_H\, \xi)
\end{eqnarray}
where $V=W,Z$, $g_{Hf\bar{f}}^\textrm{\tiny SM}=m_f/v$ ($m_f$ denotes the fermion mass), $g_{HW^+W^-}^\textrm{\tiny SM}=gM_W$, $g_{HZZ}^\textrm{\tiny SM}=\sqrt{g^2+g'^2} M_Z$, $g_{HHW^+W^-}^\textrm{\tiny SM}=g^2/2$ and $g_{HHZZ}^\textrm{\tiny SM}=(g^2+g'^2)/2$. As announced in the Introduction, all the dominant corrections, i.e., the ones controlled by the strong operators, preserve the Lorentz structure of
the SM interactions, while the form factor operators will also introduce
couplings with a different Lorentz structure.

\subsection{Higgs anomalous couplings in two concrete models}

The Holographic Higgs models of Refs.~\cite{Contino:2003ve, Agashe:2004rs,
Contino:2006qr} are based on a five-dimensional theory in Anti de-Sitter
(AdS) space-time.  The bulk gauge symmetry $SO(5)\times U(1)_X \times
SU(3)$ is broken down to the SM gauge group on the UV boundary and to
$SO(4) \times U (1)_X \times SU(3)$ on the IR. Since the symmetry-breaking
pattern of the bulk and IR boundary is given by $SO(5)\rightarrow SO(4)$,
we expect four Goldstone bosons parametrized by the $SO(5)/SO(4)$
coset~\cite{Agashe:2004rs}:
\begin{equation}
\Sigma = \langle \Sigma\rangle e^{\Pi/f} \ ,
 \qquad \langle \Sigma\rangle =  (0,0,0,0,1) \ ,
 \qquad 
 \Pi = 
\left(
\begin{array}{cc}
0_{4} & \mathcal{H} \\
- \mathcal{H}^T & 0\\
\end{array}
\right)
\, ,
\end{equation}
where $\mathcal{H}$ is a real 4-component vector, which transforms as a doublet
under the weak $SU(2)$ group and can be associated with the Higgs. The
couplings between the Higgs boson and the gauge fields are obtained from
the pion kinetic term
\begin{equation}
{\cal L}_{\rm kin}=\frac{f^2}{2}(D_\mu\Sigma)(D^\mu\Sigma)^T\, .
\label{eq:kin}
\end{equation}
In  the unitary gauge where
$
\Sigma = \left(\sin H/f ,0,0,0,\cos  H/f \right)
$, Eq.~(\ref{eq:kin}) gives
\begin{equation}
{\cal L}_{\rm Kin}=\frac{1}{2}\partial_\mu H\partial^\mu H+m^2_W(H) 
\left[ W_\mu 
W^\mu+\frac{1}{2\cos^2\theta_W}
Z_\mu Z^\mu\right] \hspace{.2cm} \textrm{with } \hspace{.2cm} 
m_W(H)=\frac{gf}{2}\sin\frac{H}{f}\, .
\label{eq:MCHMgauge}
\end{equation}
Expanding Eq.~(\ref{eq:MCHMgauge}) in powers of the Higgs field, we obtain the
Higgs couplings to the gauge fields
\begin{equation}
\label{rescalehVV}
g_{HVV}=g_{HVV}^\textrm{\tiny SM}\ \sqrt{1-\xi}\ , \hspace{1cm}
g_{HHVV}=g_{HHVV}^\textrm{\tiny SM}\ (1-2\xi)\ ,
\end{equation}
with the compositeness parameter $\xi$ defined as
\begin{equation}
\xi=\left(\frac{v}{f}\right)^2=\sin^2 \frac{\langle H \rangle}{f}\ .
\end{equation}
The couplings of the Higgs boson to the fermions can be obtained in the
same way, but they will depend on the way the SM fermions are embedded
into representations of the bulk symmetry. In the MCHM4
model~\cite{Agashe:2004rs} with SM fermions transforming as spinorial
representations of $SO(5)$, the interactions of the Higgs to the fermions
take the form
\begin{equation}
{\cal L}_{\rm Yuk}=-m_f(H){\bar f}f \hspace{.2cm} \textrm{ with } 
\hspace{.2cm}  
m_f(H)= M \sin\frac{H}{f}\ .
\label{eq:MHCM4}
\end{equation}
We then obtain
\begin{equation}
\label{rescalehff1}
\textrm{MCHM4:} \hspace{1cm} g_{Hff}=g_{Hff}^\textrm{\tiny SM}\ \sqrt{1-\xi}\ .
\end{equation}

In the MCHM5 model~\cite{Contino:2006qr} with SM fermions transforming as
fundamental representations of $SO(5)$, the interactions of the Higgs to the
fermions take the following form ($M$ is a constant of mass-dimension one)
\begin{equation}
{\cal L}_{\rm Yuk}= -m_f(H){\bar f}f \hspace{.2cm} \textrm{ with } 
\hspace{.2cm}  
m_f(H)= M \sin\frac{2H}{f}\ .
\label{eq:MHCM5}
\end{equation}
We then obtain
\begin{equation}
\label{rescalehff2}
\textrm{MCHM5:} \hspace{1cm} g_{Hff}=g_{Hff}^\textrm{\tiny SM}\ 
\frac{1-2\xi}{\sqrt{1-\xi}}\ .
\end{equation}

In both models, the Higgs couplings to gauge boson are always reduced
compared to the SM ones, as expected from the positivity
theorem~\cite{Low:2009di} on the $c_H$ coefficient of the SILH Lagrangian.
On the contrary, the two models exhibit different characteristic behaviors
in the Higgs couplings to fermions: in the vicinity of the SM, i.e., for
low values of $\xi$, the couplings are reduced, and the reduction is more
important for MCHM5 than for MCHM4, but, for larger values of $\xi$, the
couplings in MCHM5 are raising back and can even get much larger than the
SM ones. This latter effect is at the origin of an enhancement of the
Higgs production cross-section by gluon fusion, enhancement that will
significantly affect the Higgs searches.

In the previous expressions for the anomalous Higgs couplings we keep the full $\xi$-dependence, without expanding in small $\xi$. In general, higher-order derivative operators for $\Sigma$ would induce momentum dependent corrections to these couplings but,  as discussed in Ref.~\cite{Giudice:2007fh}, such contributions will be suppressed by powers of $p^2/m_\rho^2$, and we neglect such effects.

\subsection{Branching ratios and total widths}
The partial widths in the composite Higgs models can be easily obtained
from the SM partial widths by rescaling the couplings involved in the
Higgs decays. Since in MCHM4 all Higgs couplings are modified by the same
universal factor $\sqrt{1-\xi}$, the branching ratios are the same as in
the SM model. The total width will be different though by an overall
factor $1-\xi$.\sli

In MCHM5, all partial widths for decays into fermions are obtained from
the SM widths by multiplication with the modification factor of the Higgs
Yukawa coupling squared,
\beq
\Gamma(H\to f\bar{f}) = \frac{(1-2\xi)^2}{(1-\xi)} \;
\Gamma^{SM} (H\to f\bar{f}) \; .
\eeq
The Higgs decay into gluons is mediated by heavy quark loops, so that
the multiplication factor is the same as for the fermion decays:
\beq
\Gamma(H\to gg) = \frac{(1-2\xi)^2}{(1-\xi)} \;
\Gamma^{SM} (H\to gg) \; .
\eeq
For the Higgs decays to massive gauge bosons $V$ we obtain
\beq
\Gamma(H\to VV) = (1-\xi) \; \Gamma^{SM}
(H\to VV) \; .
\eeq
The Higgs decay into photons proceeds dominantly via $W$-boson and
top and bottom loops. Since the couplings to gauge bosons and
fermions scale differently in MCHM5, the various
loop contributions have to be multiplied with the corresponding Higgs
coupling modification factor. The leading order width is given by
\beq
\label{Hgaga}
\Gamma(H\to \gamma\gamma) = 
\frac{\Gamma^{SM}(H\to \gamma\gamma)}{
[I_\gamma(M_H)+J_\gamma(M_H)]^2}\left[
\frac{1-2\xi}{\sqrt{1-\xi}}I_\gamma(M_H)
+\sqrt{1-\xi}J_\gamma(M_H)\right]^2\ ,
\eeq
where
\begin{equation}
\begin{array}{c}
I_\gamma(M_H)=\frac{4}{3} F_{1/2}(4M_t^2/M_H^2), \hspace{.4cm} 
J_\gamma(M_H)=F_{1}(4M_W^2/M_H^2),\\
F_{1/2}(x)\equiv -2x\left[1+(1-x)f(x)\right], \hspace{.4cm} 
F_{1}(x)\equiv 2+3x\left[1+(2-x)f(x)\right], \\
f(x)\equiv \arcsin[1/\sqrt{x}]^2\  \mathrm{for}\ x\ge 1 \hspace{.2cm} \mathrm{and} \hspace{.2cm}
f(x)\equiv -\frac{1}{4} \left[ \log \frac{1+\sqrt{1-x}}{1-\sqrt{1-x}} - i \pi \right]^2\  \mathrm{for}\ x<1.
\end{array}
\end{equation}

Both decays into gluons and photons are loop-induced and might in
principle be affected by possible new particles running in the loops. The
set-ups we are considering, however, assume that the only chiral degrees
of freedom the Higgs couples to are the SM ones (see
footnote~\ref{footnote:hgg}). This will certainly be modified if the top
quark, for instance, is a composite particle since additional top-partners
would then also be expected to have a significant coupling to the Higgs (see for instance Ref.~\cite{Spira:1997ce}).
Under our original assumption, the corrections to the $H\gamma\gamma$ and
$Hgg$ vertices originate from the modified Yukawa interactions only and
the loop-decays can be safely computed in the framework of our effective
theory. The higher order corrections to the decays are unaffected as long as QCD
corrections are concerned, since they do not involve the Higgs couplings.
\sli

We have calculated the Higgs branching ratios with the program HDECAY
\cite{hdecay}  where we have implemented the modifications due to the
composite model described above. The program HDECAY includes the most
important higher order corrections to the various Higgs decays and
includes the off-shell effects in the Higgs decays into massive gauge
bosons and a top quark pair. \sli

\begin{figure}[th]
\begin{center}
\epsfig{figure=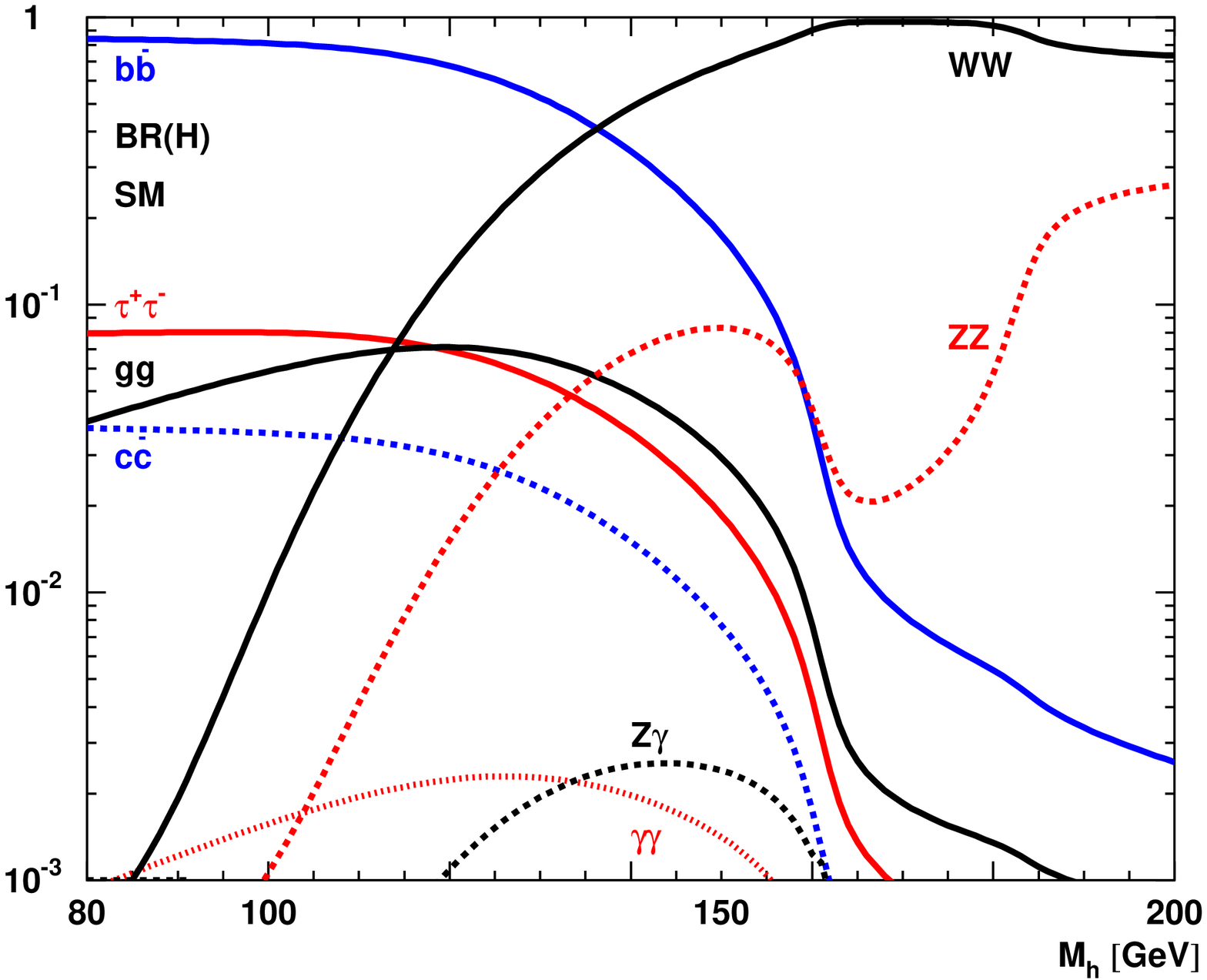,width=7.cm}
\hspace*{0.5cm}
\epsfig{figure=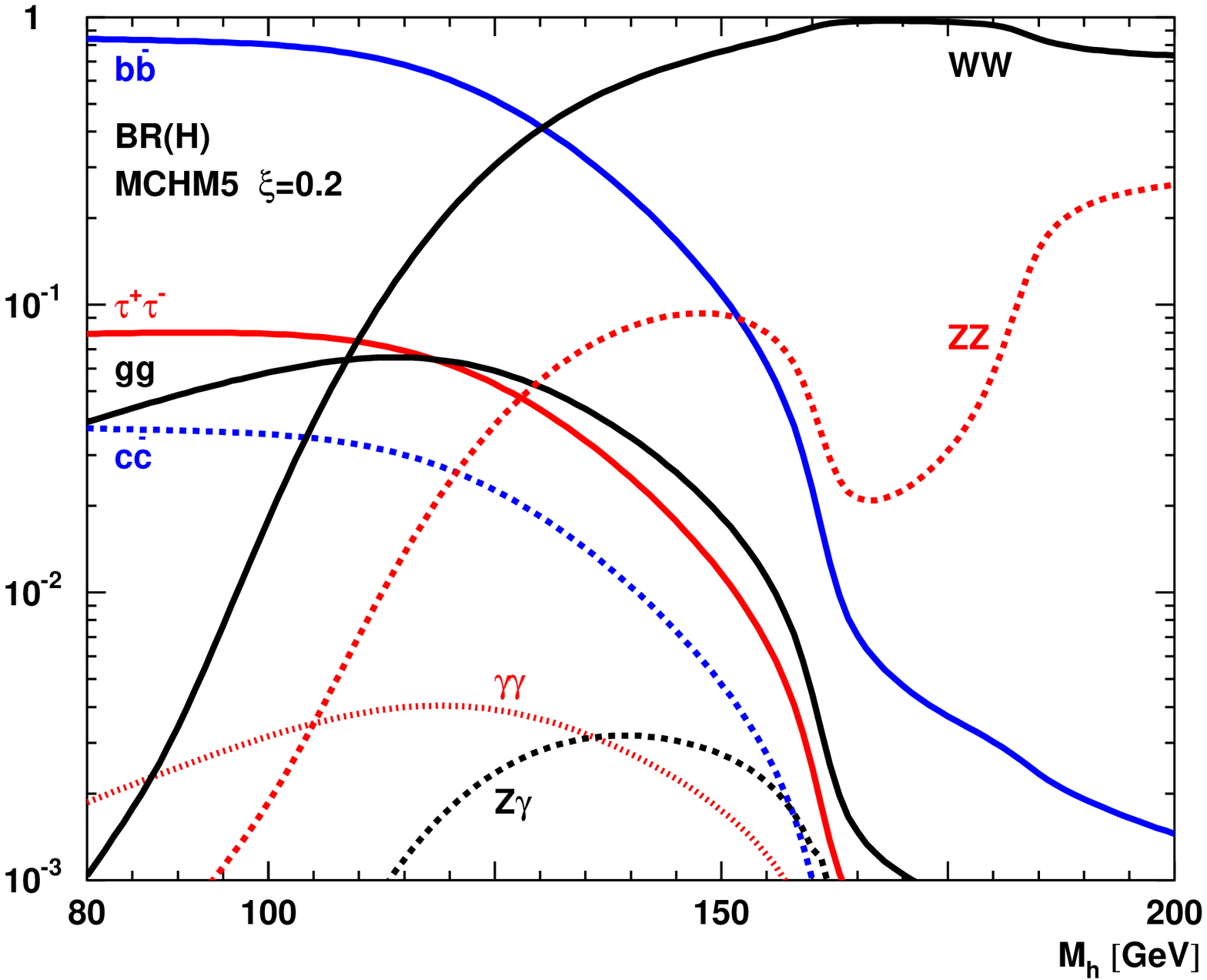,width=7.cm}
\vskip 0.5cm
\epsfig{figure=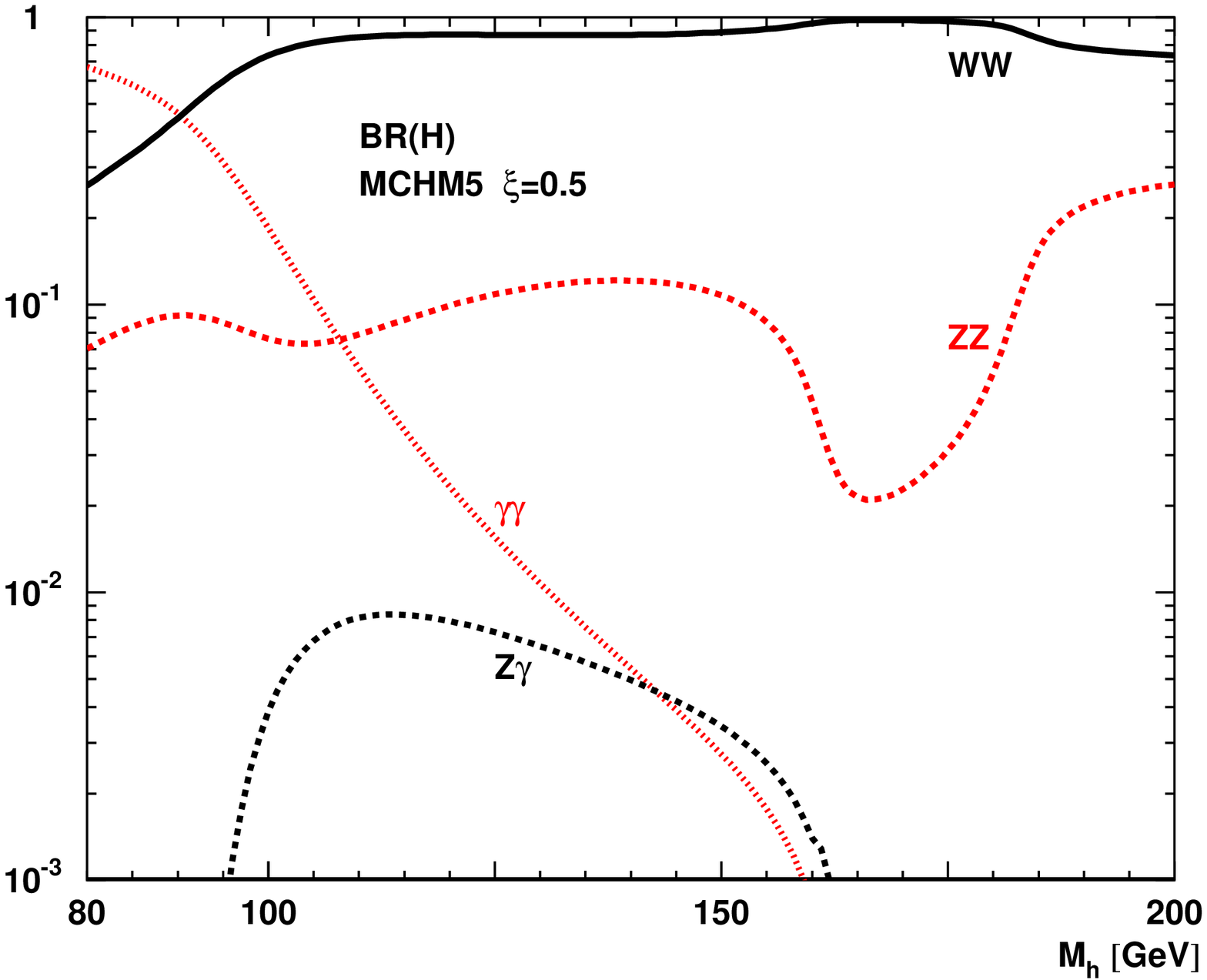,width=7.cm}
\hspace*{0.5cm}
\epsfig{figure=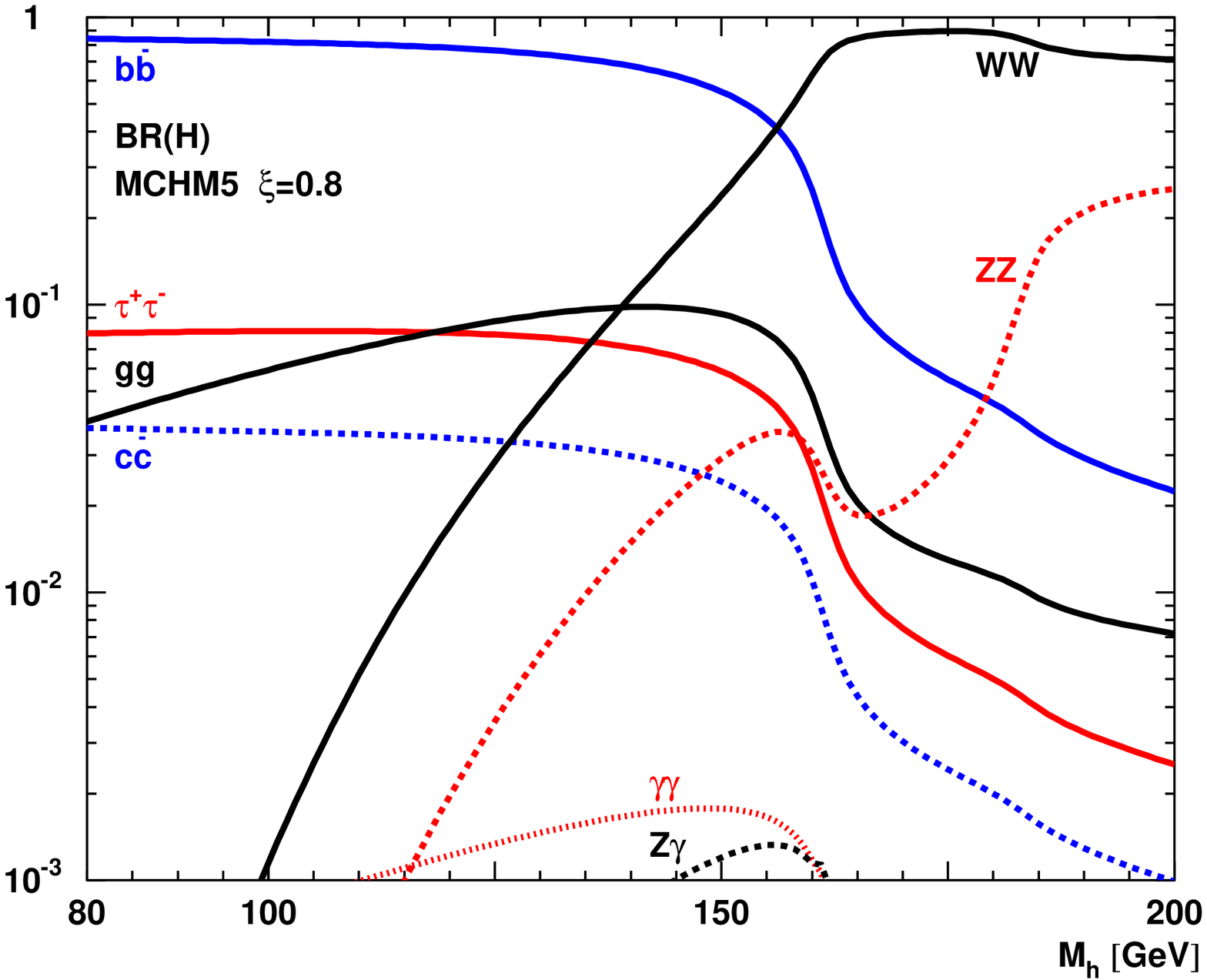,width=7.cm}
\caption{\label{fig:BRs}
Higgs branching ratios as a function of the Higgs boson mass in
the SM ($\xi=0$, upper left) and MCHM5 with $\xi=0.2$ (upper right), 0.5 
(bottom left) and 0.8 (bottom right).}
\end{center}
\end{figure}

Figure~\ref{fig:BRs} shows the branching ratios in the SM and those of
MCHM5 for three representative values of $\xi=0.2,0.5,0.8$. The Higgs mass
range has been chosen between 80 and 200 GeV, which is the mass range
favoured by composite Higgs models. Notice that the lower mass range has
not been excluded yet completely by the LEP bounds (see
Section~\ref{sec:constraints}).

The SM branching ratios show
the typical behaviour dictated by the Higgs mechanism, which predicts
the Higgs couplings to the matter particles to be proportional to the
mass of these particles. 
A Higgs boson in the intermediate mass range, ${\cal O}(M_Z) \le M_H
\le {\cal O}(2 M_Z)$, dominantly decays into a $b\bar{b}$ pair and a
pair of massive gauge bosons, one or two of them being virtual. Above
the gauge boson threshold, it almost exclusively decays into $WW,ZZ$,
with a small admixture of top decays near the $t\bar{t}$ threshold
(not visible in this Figure, being outside the mass range plotted here).
Below $\sim 140$ GeV the decays into $\tau^+\tau^-$, $c\bar{c}$
and $gg$ are important besides the dominant $b\bar{b}$ decay. The
$\gamma\gamma$ decay, though very small, provides a clear 2-body
signature for the Higgs production in this mass range. The branching
ratios in MCHM4 are exactly the same as in the SM, since all
couplings scale with the same modification factor, which then drops 
out in the branching ratios.

\begin{figure}[ht]
\begin{center}
\epsfig{figure=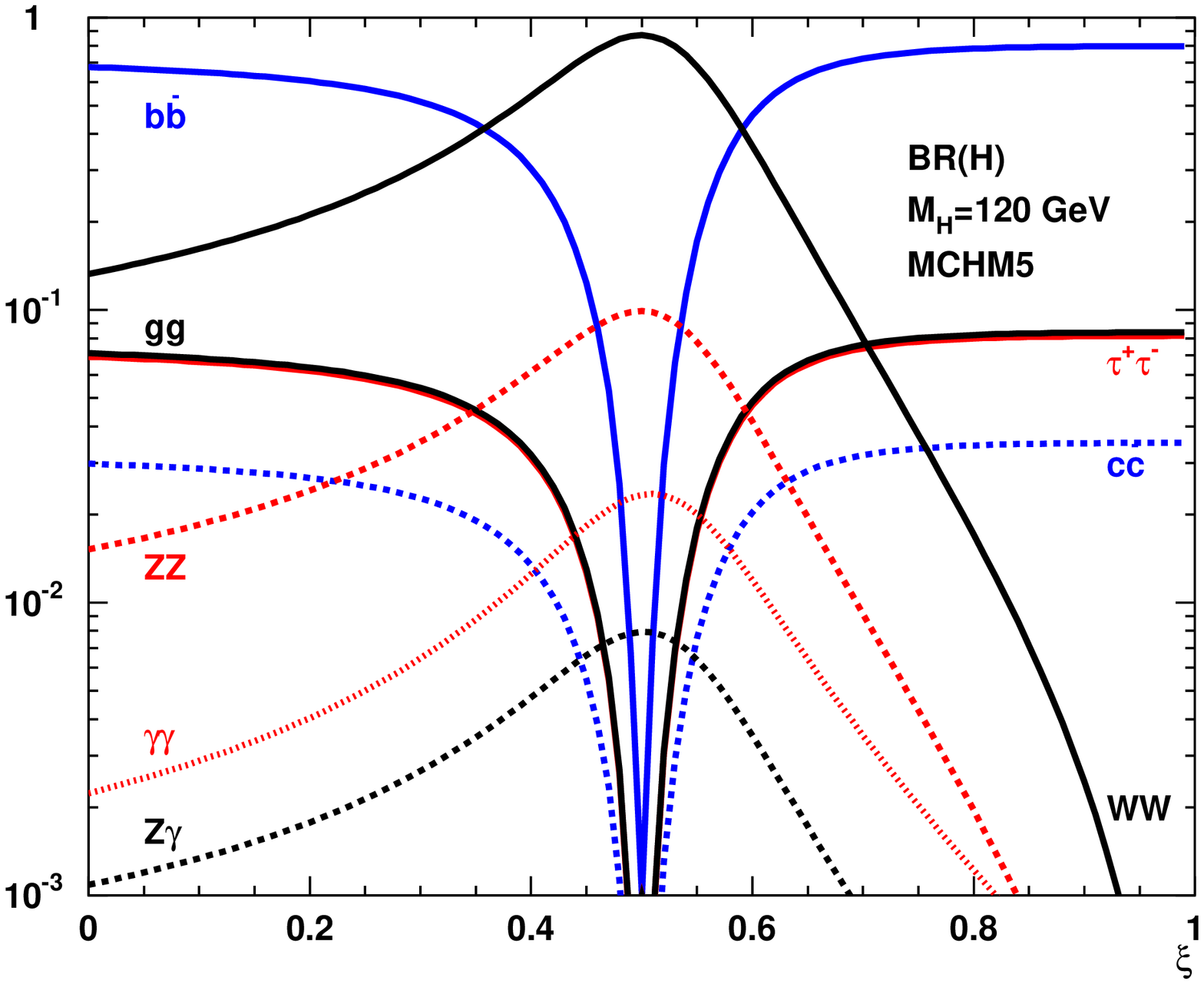,width=7cm}
\hspace*{0.5cm}
\epsfig{figure=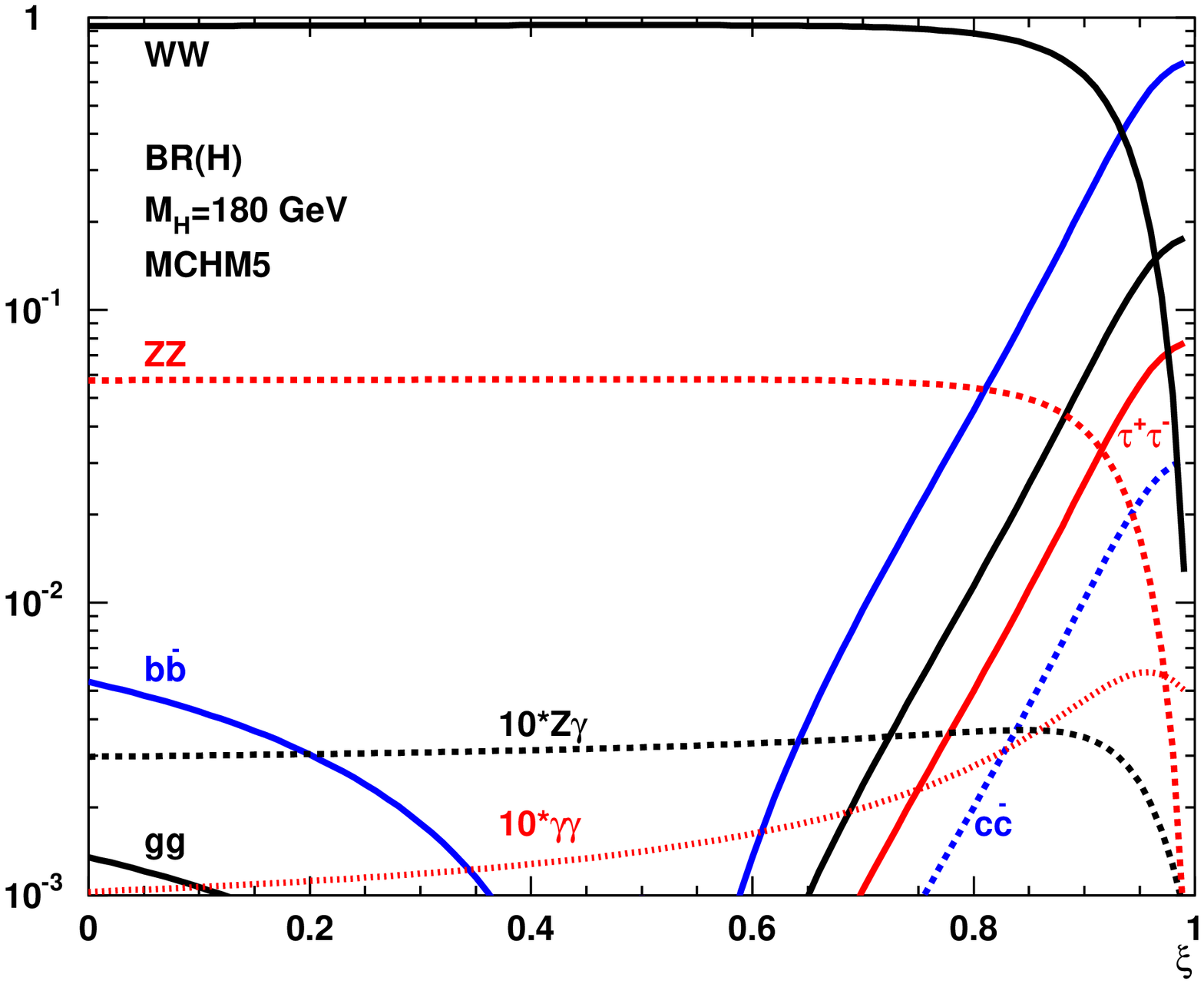,width=7cm}
\caption{\label{fig:branxi}
The branching ratios of MCHM5 as a function of $\xi$ for
 $M_H=120$~GeV (left) and $M_H=180$~GeV (right).}
\end{center}
\end{figure}

As can be inferred from Figs.~\ref{fig:BRs}, for non-vanishing $\xi$
values, the branching ratios (BRs) in MCHM5 can change considerably. The
behaviour can be easily understood by looking at Fig.~\ref{fig:branxi}
which shows the same branching ratios as a function of $\xi$ for two
representative values of the Higgs boson mass, $M_H=120$ GeV and 180 GeV.
The BRs into fermions are governed by the $(1-2\xi)^2/(1-\xi)$ prefactor
of the corresponding partial widths: as $\xi$ increases from 0, there is
first a decrease of the fermionic BRs, until they vanish at $\xi=0.5$ and
then grow again with larger~$\xi$. The same behaviour is observed in the
decay into gluons, which is loop-mediated by quarks. The decays into gauge
bosons show a complementary behaviour: for small $\xi$, due to the
decreasing decay widths into fermions, the importance of the vector boson
decays becomes more and more pronounced until a maximum value at $\xi=0.5$
is reached. Above this value the branching ratios into gauge boson
decrease with increasing Higgs decay widths into fermion final states: the
Higgs boson becomes gaugephobic in the technicolor limit ($\xi\to 1$).
\vspace*{0.5cm}

Coming back to Fig.~\ref{fig:BRs}, we see that for small $\xi\ =0.2$, the
decays into massive and massless gauge bosons set in at lower Higgs mass
values and are more important than in the SM. The BR into $\gamma\gamma$,
especially important for low Higgs mass searches at the LHC, is larger
now. The branching ratio into $b\bar{b}$ gets less important, an effect
that is more visible at higher Higgs masses.  This behaviour culminates at
$\xi=0.5$, where only decays into gauge bosons are present due to the
closure of the decays into fermions\footnote{ 
Instead the Higgs boson could decay into fermions through an electroweak 
particle-loop (note that the interference term with the tree-level decay is absent since the tree-level amplitude is vanishing) and the decay could in principle compete with the $\gamma\gamma$ 
decay, which is also loop mediated and plays an important role for 
small Higgs masses. However, in addition to the loop suppression, the decay 
into light fermions has an additional suppression factor of order $m_f^2/M^2$ 
where $m_f$ is the light fermion mass and $M$ is a mass of electroweak size that can 
be either the Higgs mass, the top mass or the $W$ mass depending on the diagram 
involved. We have checked numerically that this loop decay channel into $b\bar{b}$ is about 2 orders of magnitude subdominant compared to the $\gamma\gamma$ decay. Similarly, a fermiophobic Higgs boson can also decay radiatively into two gluons, but this 2-loop EW process will be totally negligible.}. In particular, the Higgs decay mode
into photons can reach large values up to $\sim 70$~\% at 80 GeV in this
case. Note also that,  for $\xi=0.5$,  the decay into a pair of gluons is also absent  since the Higgs does not couple to the top quark. In practice, however, such a decay can be mediated by the heavy vector resonances of the strong sector.
Also the branching ratios into massive gauge bosons are significant
at low Higgs masses, while above the gauge boson thresholds they approach
their SM values. For large $\xi\ =0.8$, the low Higgs mass region is
dominated by the decays into heavy fermions. The branching ratios extend
to somewhat higher Higgs mass values than in the SM. The onset of the
gauge boson decays is postponed to Higgs mass values larger than in the
SM.

\begin{figure}[ht]
\begin{center}
\epsfig{figure= 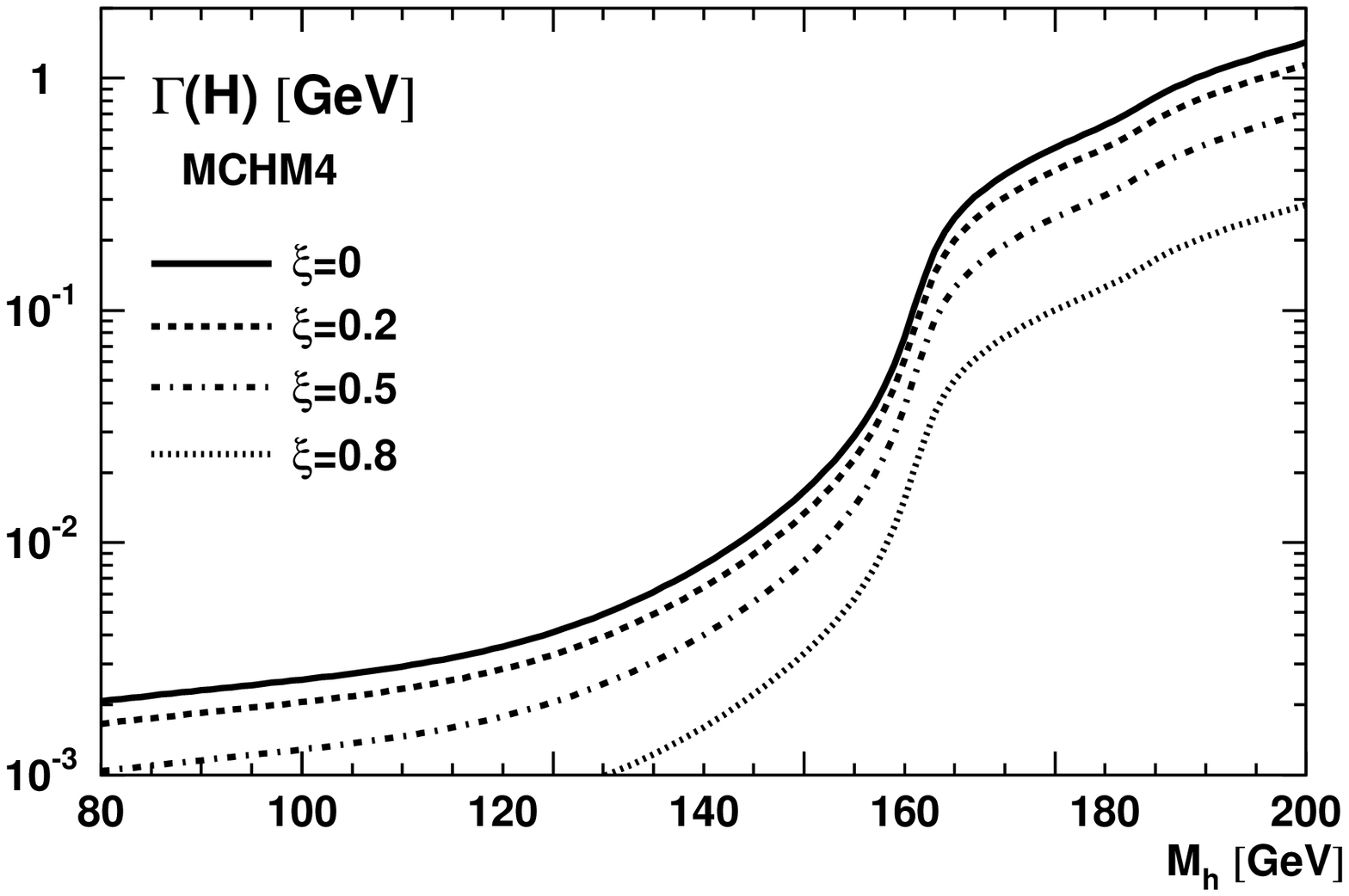,width=7cm}
\hspace*{0.5cm}
\epsfig{figure= 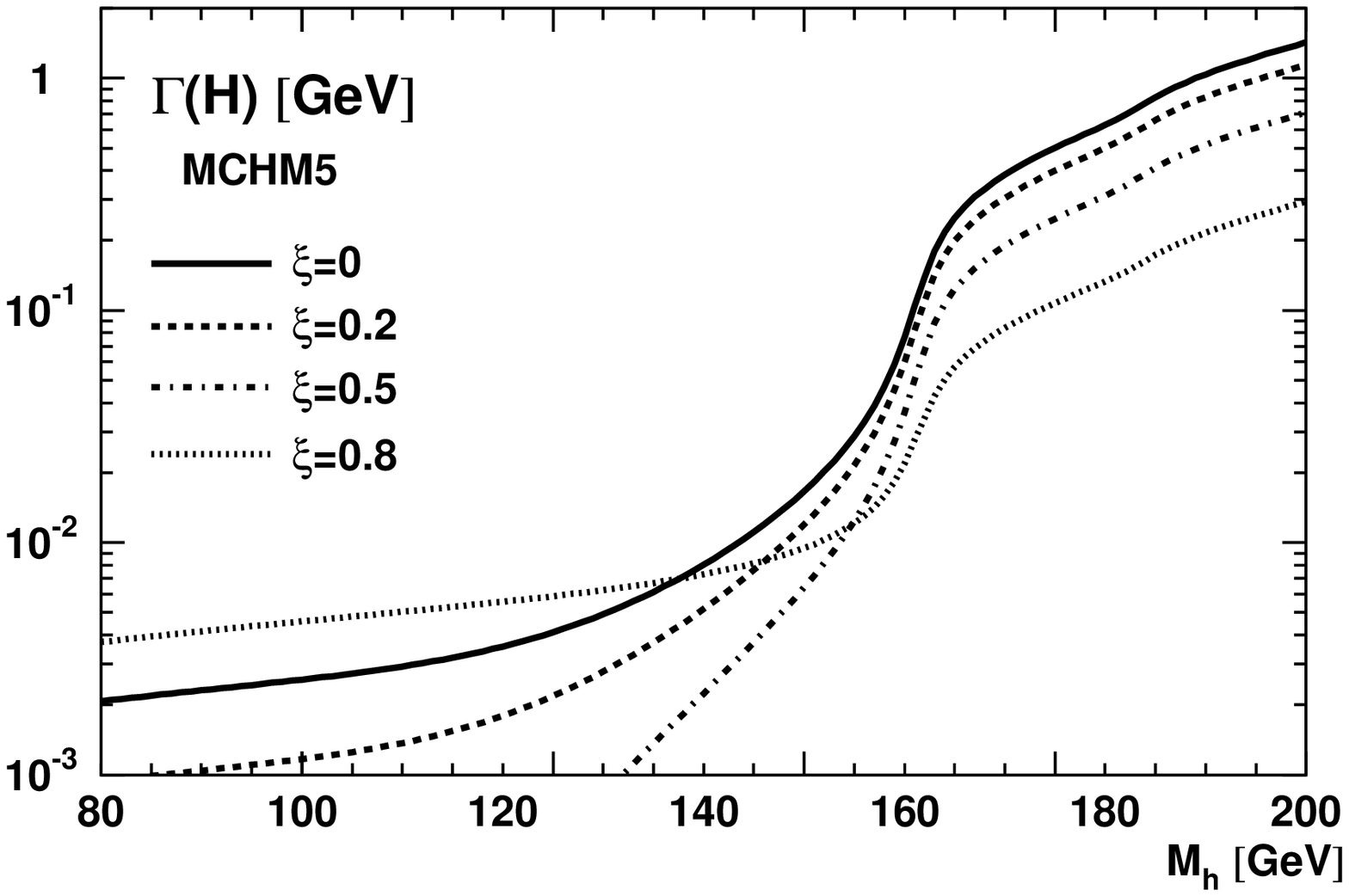,width=7cm}
\caption{\label{fig:SMHw}
The Higgs total width $\Gamma_H$ (in GeV) vs. $M_H$ (in GeV) in the SM 
(continuous line) and for $\xi=0.2$ (dashed), $\xi=0.5$ (dot-dashed) and 
$\xi=0.8$ (dotted) in MCHM4 (left) and MCHM5 (right).}
\end{center}
\end{figure}

\begin{figure}[ht]
\begin{center}
\epsfig{figure=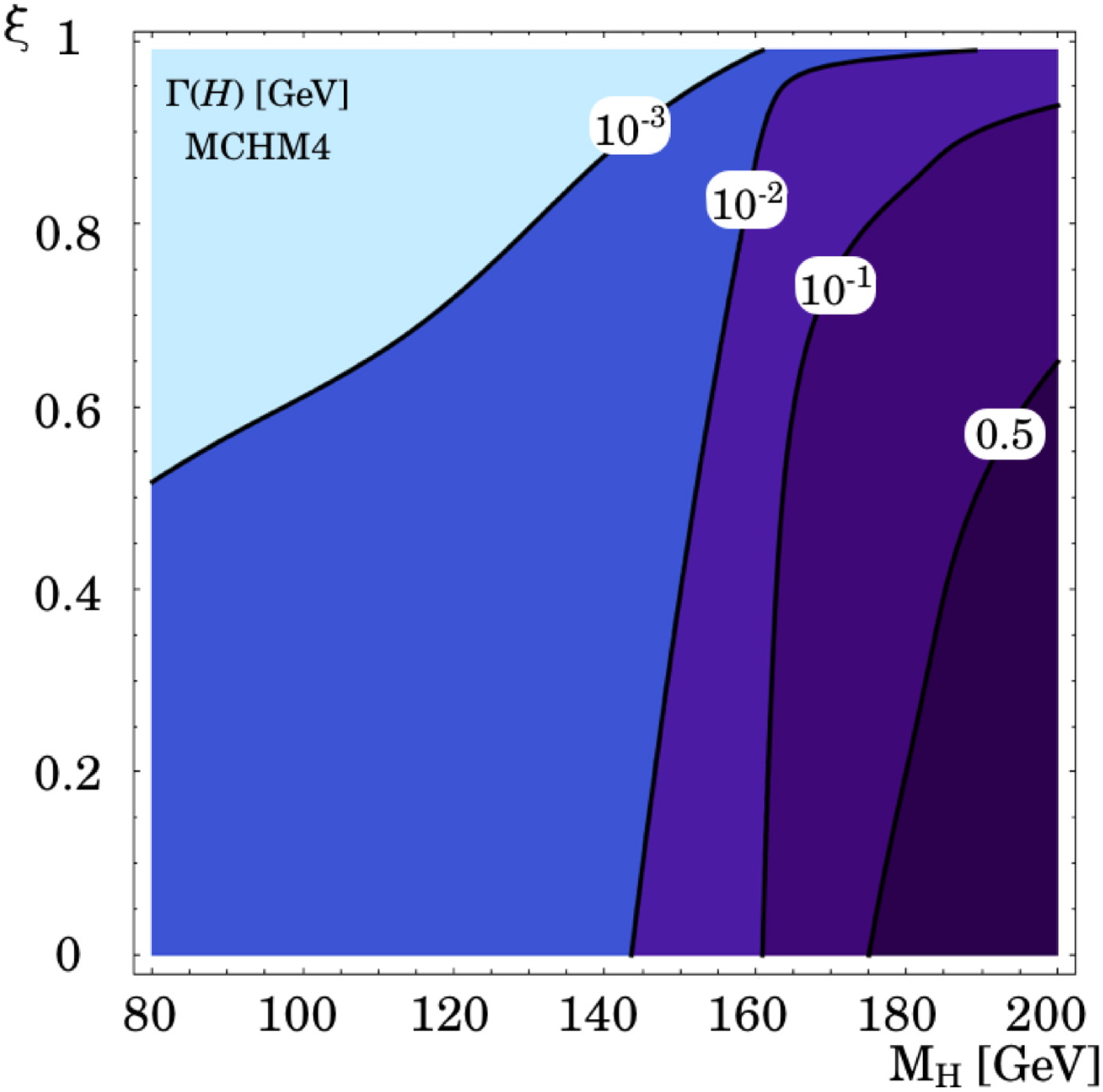,width=7cm}
\hspace*{0.5cm}
\epsfig{figure=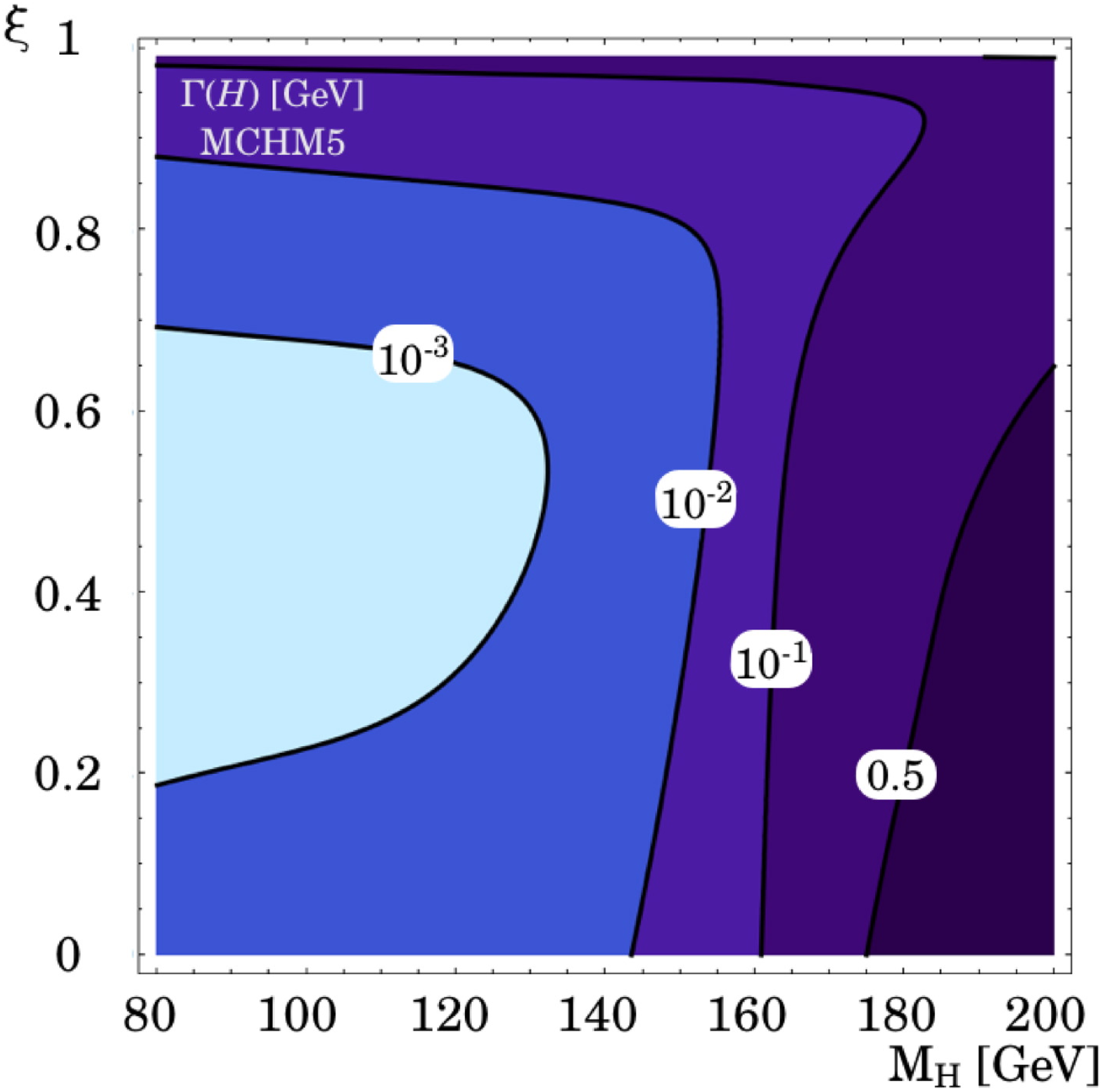,width=7cm}
\caption{\label{fig:Hw}
Contour plots of the Higgs total width,
$\Gamma_H$, in the plane
$(M_H,\xi)$ for MCHM4 (left) and MCHM5 (right). 
The contours correspond to $\Gamma_H=10^{-3}, 10^{-2}, 10^{-1}$ and 
$0.5$~GeV.  }
\end{center}
\end{figure}

Figure~\ref{fig:SMHw} shows the Higgs width as a function of $M_H$ in the
SM and for $\xi=0.2, 0.5$ and $0.8$ both in MCHM4 (left plot) and MCHM5
(right plot). Below $\sim 150$~GeV, the width is rather small and
increases rapidly as the vector boson decay channels open up.  The Higgs
width in MCHM4 and MCHM5 is also plotted in Fig.~\ref{fig:Hw} in the
$(M_H,\xi)$ plane. In MCHM4, the total width decreases monotonously with
rising $\xi$ due to the rescaling of the couplings with $\sqrt{1-\xi}$.  
In MCHM5, the total width develops a pronounced minimum at $\xi=0.5$ for
low Higgs mass values (the light region on the right plot of
Fig.~\ref{fig:Hw}).  The origin of this minimum is of course the reduced
couplings to fermions which even vanish identically at $\xi=0.5$. For
larger values of $\xi$, the fermionic channels reopen and the total width
rises with growing $\xi$. At large Higgs masses the total width is
dominated by gauge boson decays at low $\xi$ values, since we are above
the gauge boson threshold here. At large $\xi$ values the role is taken
over by the fermion decays, which do not become as large as the gauge
boson decays, however, so that also in the limit $\xi \to 1$, for large
Higgs mass values the total width remains below the SM value at $\xi=0$. A
small total width may be of advantage for Higgs boson searches since more
stringent mass cuts could be applied in that case. However, in our analysis, we will simply 
study how the Higgs searches rescale with $\xi$ and we will not try to optimize the cuts used in the SM searches to a different Higgs width.


\section{Constraints from LEP, the Tevatron and electroweak precision data}
\label{sec:constraints}

\begin{figure}[t]
\begin{center}
\epsfig{figure=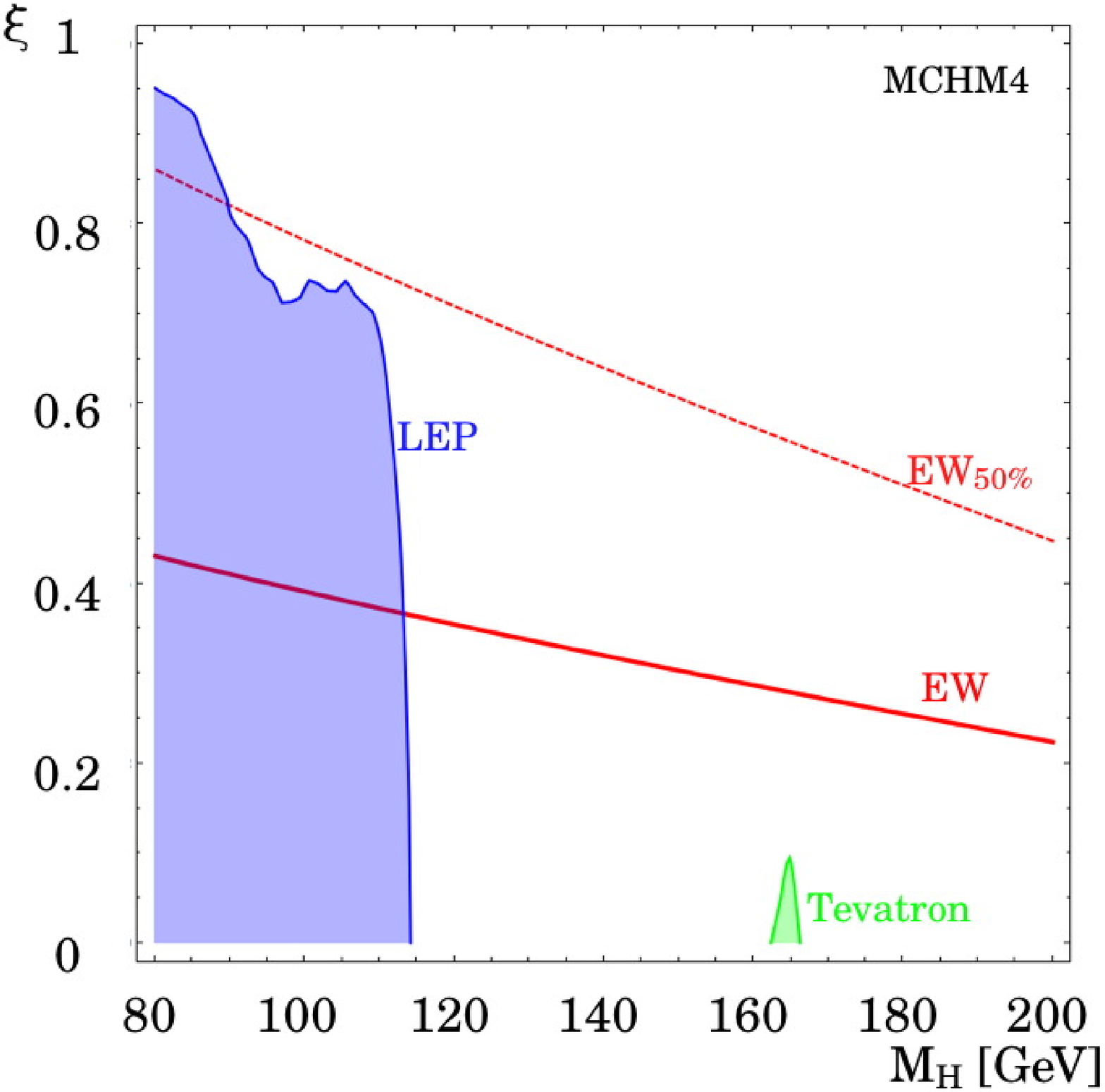,height=7cm,width=7cm}
\hspace*{0.5cm}
\epsfig{figure= 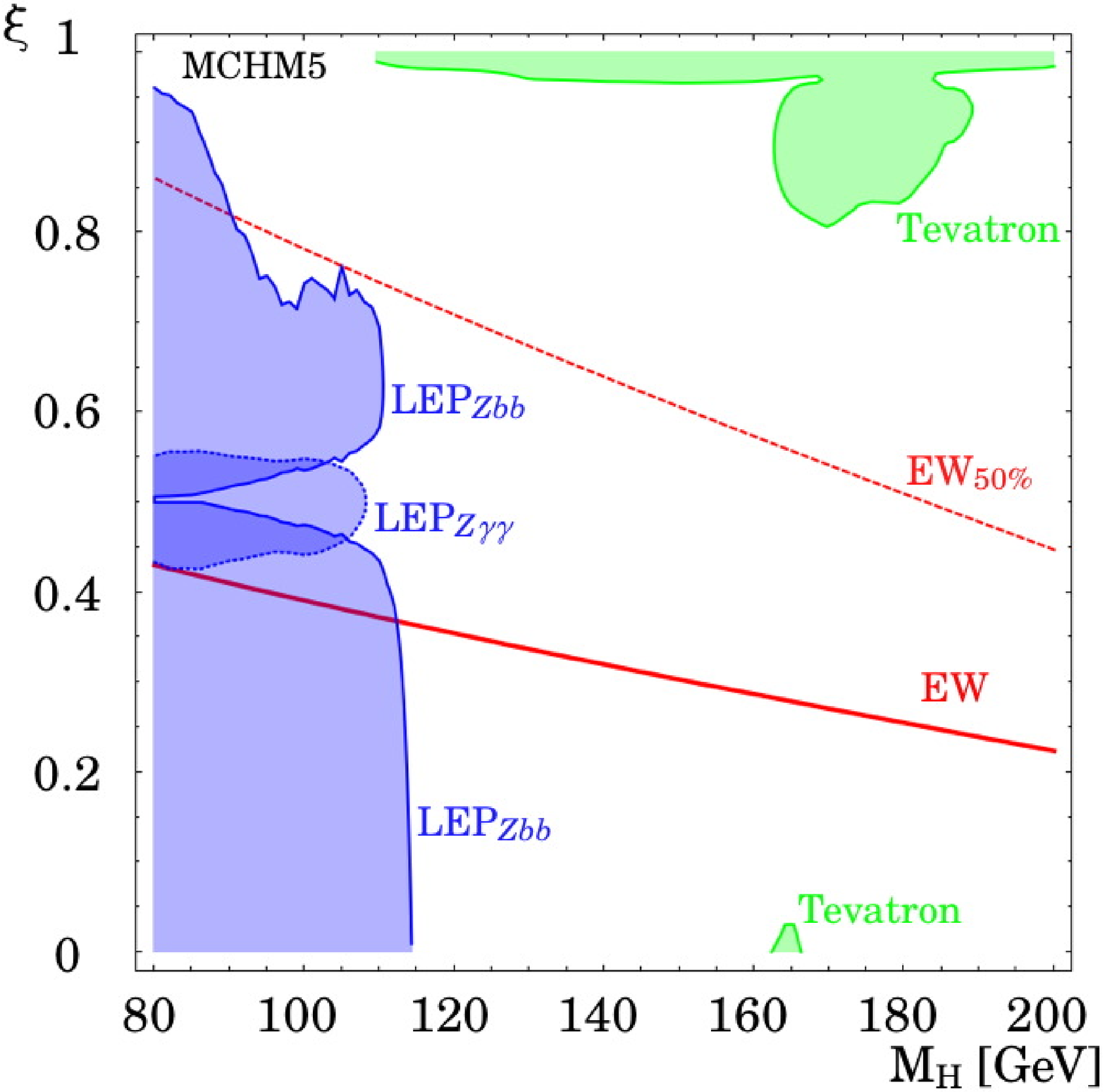,height=7cm,width=7cm}
\caption{\label{fig:bounds}
Experimental limits from Higgs searches at LEP (blue/dark gray) and the Tevatron 
(green/light gray) in the plane $(M_H,\xi)$ for MCHM4~(left) and MCHM5~(right). EW 
precision data prefer low value of $\xi$: the red continuous line 
delineates the region favored at 99\% CL (with a cutoff scale fixed at 
2.5~TeV) while the region below the red dashed line 
survives if there is an additional 50\% cancellation of the oblique 
parameters.}
\end{center}
\end{figure}

Higgs searches at LEP and the Tevatron set constraints on the parameter
space $(M_H,\xi)$ of the composite Higgs models we consider.  
Figure~\ref{fig:bounds} shows the excluded regions for MCHM4~(left) and
MCHM5~(right). To generate the plots we have used the Higgsbounds
program~\cite{higgsbounds}, cross-checking the results wherever possible
and modifying it suitably to take into account the latest changes in
Tevatron limits.

At LEP, the most relevant search channel is $e^+e^-\to ZH\to Z b\bar{b}$
\cite{LEPHbb}, which is sensitive both to the Higgs-gauge coupling (in
Higgs-strahlung production) and to the Higgs-fermion coupling (in the Higgs
decay). The former coupling is reduced in both models and explains why the
SM lower Higgs mass limit $M_H>114.4$ GeV is degraded in the composite
models, as shown in Figs.~\ref{fig:bounds}. In MCHM5, the Higgs-fermion
coupling vanishes at $\xi=0.5$ implying that the limit from the above
process is lost in the neighbourhood of this $\xi$ value. In this region
the process $e^+e^-\to ZH\to Z\gamma\gamma$ can be exploited
\cite{LEPHgg}. LEP sets a limit on $(\sigma_{ZH}/\sigma^{SM}_{ZH}) \times
BR(H\to \gamma\gamma)$ which does not translate into a limit on $M_H$ in
the SM but is useful in our composite model to cover the $\xi=0.5$ hole in
the $H\to b\bar{b}$ LEP limit (see Fig.~\ref{fig:bounds}, right).

At the Tevatron, the most relevant search is through the $H\to WW$ decay,
which in the SM excludes at 95\% C.L. the mass range 162 GeV $< M_H<$ 166
GeV~\cite{TEVH}. In our composite models this excluded band shrinks to
zero quickly once $\xi$ slightly exceeds zero, as then the production
cross-section (dominated by the gluon-fusion process) is reduced.  In MCHM5, however, Tevatron can
exclude a region with $M_H\sim 165-185$ GeV and large $\xi>0.8$ through
the channel $H\to WW$ with the $W$'s decaying leptonically \cite{TEVH}.
This occurs thanks to the enhancement of Yukawa couplings at large $\xi$,
which boosts the gluon-fusion production mechanism while the $WW$
branching ratio is still high (unless one is really close to $\xi=1$, when
fermionic decays take over). For such large values of $\xi$ ($\xi\simgt
0.97$) in MCHM5, the decay $H\to\tau\tau$ would lead to an observable
signature at the Tevatron \cite{TEVtautau} and the corresponding parameter
region is also excluded (see Fig.~\ref{fig:bounds}, right). Nevertheless,
this region is already at the border of the regime of nonperturbative Yukawa
couplings (see below) where the validity of our computations is not guaranteed.
In any case, these Tevatron  exclusion bounds should be regarded as rough estimates. Indeed, the Tevatron collaborations combine different search channels in a very sophisticated way. The relative importance of the search channels in our concrete models changes, however, with varying $\xi$. For each $\xi$ value, the search channels would have to be combined at the same level of sophistication as done by Tevatron analyses. This is clearly beyond the scope of our work, though. Nevertheless, the bounds presented in Fig~\ref{fig:bounds} serve to get an approximate picture of exclusion regions due to Tevatron searches.

In the SM, the Higgs mass is notoriously constrained not only by direct
searches but also by EW precision data. As is well known~\cite{Peskin:1991sw}, 
the oblique parameters are indeed logarithmically sensitive to the Higgs 
mass. In composite models, there are three main contributions to the oblique
parameters whose origin can be easily understood using the SILH effective
Lagrangian~(\ref{eq:silh}): (i)~The operator $c_T$ gives a contribution to
the $T$ parameter, $\hat{T}=c_Tv^2/f^2$, which would impose a very large
compositeness scale; however, assuming that the custodial symmetry is
preserved by the strong sector, the coefficient of this operator is
vanishing automatically. The explicit models we are considering fulfill
this requirement. (ii)~A contribution to the $S$ parameter is generated by
the form factor operators only, $\hat{S}=(c_W+c_B) M_W^2/M_\rho^2$, and
will simply impose a lower bound on the mass of the heavy resonances,
$m_\rho \geq 2.5$~TeV. Throughout this paper, we have assumed that the
mass gap between the Higgs boson and the other resonances of the strong
sector is large enough to satisfy this bound. (iii) Finally, there is a
third contribution to the oblique parameters that will constrain the
parameter space of our set-up: since the couplings of the Higgs to the SM
vectors receive some corrections of the order $\xi$, the nice cancelation
occurring in the SM between the Higgs and the gauge boson contributions to
$S$ and $T$ does not hold anymore and they are both logarithmically
divergent~\cite{Barbieri:2007bh} (the divergence in $T$ will eventually be
screened by resonance states if the strong sector is invariant under the
custodial symmetry). $S$ and $T$, or equivalently $\epsilon_{1,3}$~\cite{Altarelli:1990zd}, can be
easily estimated from the SM $\log(M_H)$ pieces \begin{multline}
\textrm{SM:} 
\left\{
\begin{array}{c}
\delta \epsilon_1 \approx 8.6\cdot 10^{-4} \times \log(M_H/M_Z)\\
\delta \epsilon_3 \approx 5.4 \cdot 10^{-4} \times \log(M_H/M_Z)
\end{array}
\right.\\
\Rightarrow
\textrm{SILH:}
\left\{
\begin{array}{c}
\delta \epsilon_1  \approx 8.6\cdot 10^{-4} \times \left[\log(M_H/M_Z) - 
\xi \log (M_H/\Lambda) \right]\\
\delta \epsilon_3  \approx 5.4 \cdot 10^{-4} \times \left[\log(M_H/M_Z) - 
\xi \log (M_H/\Lambda) \right]
\end{array}
\right.
\end{multline}
Therefore, EW precision data prefer low values of the compositeness
parameter $\xi$. In Fig.~\ref{fig:bounds}, we have plotted the upper bound
on $\xi$ as a function of the Higgs mass (continuous red line) obtained
from the 99\% CL limits on $\epsilon_{1,3}$. Allowing a partial
cancellation of the order of 50\% with contributions from other states, 
the upper bound on $\xi$ is relaxed by a factor of about 2 (dashed red 
line).

Finally, it should also be mentioned that the limit $\xi \to 1$ is
not fully consistent with basic perturbative requirements, in particular for MCHM5.
Indeed, in deriving the Yukawa coupling of the top, we fixed the top mass to its experimental value, which requires some 5D coupling to become very large in the limit $\xi \to 1$. The exact perturbative limit depends on the details of the models and the way the top mass is actually generated. A simple estimate can be inferred by writing Eq.~(\ref{eq:MHCM5}) with $M=\lambda f$ where $\lambda$ is a dimensionless coupling that should be bounded from above. We will simply require $\lambda< 4\pi$, which gives
\begin{equation}
\xi < 1- \left( m_t/(8\pi v)\right)^2 \approx 0.999\ .
\end{equation}
This limit, though certainly not very accurate, gives an idea on the maximal possible value for $\xi$.

\section{LHC Searches}
\label{sec:searches}

In the composite Higgs models, the Higgs boson search channels can be
significantly changed compared to the SM case, due to the modified
production cross-sections and branching ratios.  As an extreme example, in
MCHM5, the Higgs couplings to fermions will be absent for $\xi=0.5$. In
this case, the Higgs boson production through gluon fusion, which is
dominant in the SM, cannot be exploited\footnote{In principle, when the Higgs boson decouples from the fermions, the gluon-fusion production process can still receive a contribution from the heavy resonances of the strong sector, but this contribution is negligible when the masses of these resonances are above 2--3~TeV as required by EW precision data. The 2-loop EW gluon-fusion process is also totally negligible.}. On the other hand the branching
ratios into gauge bosons will be enhanced due to the absence of the decay
into $b\bar{b}$ final states.  In order to identify which search channels
become important and which search strategy should be applied, we produced
contour plots (in the $(M_H,\xi)$ parameter plane) of the expected
significances for different search channels in the two composite Higgs
models discussed above. Before we present our results, we will first
discuss how the production cross-sections of a composite Higgs boson
change.


\subsection{Higgs boson production cross-sections\label{XSCs}}

At the LHC, the relevant Higgs production processes (depicted in 
Fig.~\ref{XSdiag}) are (for reviews, see Refs.~\cite{fortsch,Djouadi:2005gi})
\begin{description}
\item[Gluon fusion] The gluon-fusion process $gg \to H$~\cite{georgi}
constitutes the most important Higgs production cross-section in the SM.
At leading order, it is mediated by top and bottom quark loops. The
next-to-leading order (NLO) QCD corrections have been obtained including
the full mass dependence of the loop particles~\cite{graudenz} as well as
in the heavy top quark limit \cite{graudenz,laenen}. The NLO corrections
increase the total cross-section by 50-100 \%. The next-to-next-to leading
order (NNLO) corrections have been determined in the heavy top quark limit
enhancing the total cross-section by another 20\% \cite{harlander}. These
results have been improved by soft-gluon resummation at
next-to-next-to-leading log (NNLL) accuracy adding another $\sim 10$\% to
the total cross-section~\cite{resum}. Recently, the top quark mass effects
on the NNLO loop corrections have been investigated~\cite{ozeren} and confirmed
the heavy top limit as a reliable approximation in the small and
intermediate Higgs mass range. Furthermore, the electroweak (EW)
corrections have been evaluated and turned out to be small \cite{ewcorr}.

\begin{figure}
\begin{center}
\hspace{-1cm}
\epsfig{figure=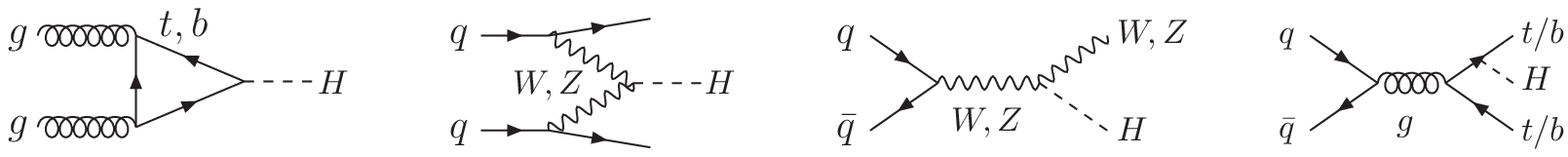,bbllx=110,bblly=625,bburx=600,bbury=670,width=16cm,clip=
}
\end{center}
\caption{\label{XSdiag}Generic diagrams contributing to Higgs production 
in gluon fusion, weak-boson fusion, Higgs-strahlung and associated production
 with heavy quarks.}
\end{figure}

We have calculated the gluon-fusion cross-section including the 
NLO QCD corrections with the full mass dependence of the loop
particles. This corresponds to the approximation used in the CMS
analyses. Since the gluon-fusion cross-section is mediated
by heavy quark loops and the NLO QCD corrections do not affect the
Higgs couplings, the gluon-fusion composite-Higgs production
cross-section is obtained from the NLO QCD SM cross-section by the
squared rescaling factor for the Higgs Yukawa coupling
[see Eqs.~(\ref{rescalehff1}) and (\ref{rescalehff2})],
hence\footnote{Equations~(\ref{eq:ggfus}) agree with the results obtained 
in Ref.~\cite{Falkowski:2007hz}, which confirms the claim of 
footnote~\ref{footnote:hgg} that all the corrections to  the 
$H\gamma\gamma$ and $Hgg$ vertices originate from the modified  Yukawa 
interactions.}
\beq
\begin{array}{llll}
\sigma_{NLO} (gg\to H) &=& (1-\xi) \,\, \sigma^{SM}_{NLO} (gg \to H) &
\qquad \mbox{MCHM4}, \\[0.1cm]
\sigma_{NLO} (gg\to H) &=& \frac{(1-2\xi)^2}{(1-\xi)} \, 
\sigma^{SM}_{NLO} (gg \to H) &
\qquad \mbox{MCHM5}.
\label{eq:ggfus}
\end{array}
\eeq
The NLO SM gluon-fusion cross-section has been obtained with the
program HIGLU~\cite{higlu}.

\item[{\boldmath $W/Z$} boson fusion] The next-important SM Higgs boson
production cross-sections are the $W$ and $Z$ boson-fusion processes $qq
\to qq + W^*W^*/Z^* Z^* \to qqH$~\cite{wzfusion}. They also play a
role for Higgs boson searches in the intermediate mass range, since the
additional forward jets allow for a powerful reduction of the
background processes. The NLO QCD corrections are of order 10\% of the
total cross-section~\cite{wzfusqcd,fortsch}.  The full NLO QCD and EW
corrections to the differential cross-sections result in modifications
of the relevant distributions by up to ~20\%~\cite{wzfusew}.\sli

We have calculated the Higgs boson production in gauge boson fusion at
NLO QCD, which is the approximation used in the ATLAS and CMS analyses. 
Since
the QCD corrections do not involve Higgs interactions, the NLO QCD
production cross-section for the composite Higgs model can be obtained
from the SM NLO QCD process by multiplication with the same 
rescaling factor as for the Higgs gauge coupling squared 
[see Eq.~(\ref{rescalehVV})], i.e.,
\beq
\sigma_{NLO} (qqH) = (1-\xi) \,\, \sigma^{SM}_{NLO} (qqH)
\qquad \mbox{for MCHM4 and MCHM5} \;. \label{eq:vbf}
\eeq
We have obtained the SM production cross-section at NLO with the
program VV2H~\cite{programs}. \sli

\item[Higgs-strahlung] The Higgs-strahlung off $W,Z$ bosons
$q\bar{q} \to Z^*/W^* \to H + Z/W$ provides alternative production
modes in the intermediate mass range $M_H \lsim 2 M_Z$
\cite{higgsrad}. The NLO QCD corrections are positive and of ${\cal
  O}(30\%)$~\cite{qcdhrad,fortsch} while the NNLO corrections are small
\cite{nnlohrad}. The full EW corrections are known and decrease the total 
cross-section by ${\cal O}$(5-10\%)~\cite{ewhrad}. \sli

The NLO QCD corrections do not involve the Higgs couplings, so that
the composite Higgs-strahlung cross-section at NLO QCD is obtained
from the corresponding SM cross-section by the same rescaling factor
as for the Higgs gauge boson coupling squared:
\beq
\sigma_{NLO} (VH) = (1-\xi) \,\, \sigma^{SM}_{NLO} (VH)
\qquad \mbox{for MCHM4 and MCHM5} \;,
\label{eq:hrad}
\eeq
where $V$ denotes $W,Z$.
The NLO QCD SM Higgs-strahlung cross-section has been obtained with
the program V2HV~\cite{programs}. \sli

\item[Higgs radiation off top quarks] plays a role only for the
production of a light SM Higgs boson with masses below $\sim
150$~GeV. The LO cross-section~\cite{lotth} is moderately increased 
($\sim 20$\%) at the LHC by the NLO QCD corrections~\cite{nlotth}. 
The production of a composite Higgs boson in association with a top
quark pair at NLO QCD is obtained from the SM cross-section via
\beq
\begin{array}{llll}
\sigma_{NLO} (Ht\bar{t}) &=& (1-\xi) \,\, \sigma^{SM}_{NLO}
(Ht\bar{t}) & \qquad \mbox{MCHM4}, \\[0.1cm]
\sigma_{NLO} (Ht\bar{t}) &=& \frac{(1-2\xi)^2}{(1-\xi)} \,
\sigma^{SM}_{NLO} (Ht\bar{t}) &  \qquad \mbox{MCHM5}.
\end{array}
\label{eq:tth}
\eeq
The LO SM cross-section has been obtained by means of the program HQQ
\cite{programs}. Subsequently it has been dressed with the $K$-factor 
quantifying the increase of the SM cross-section due to NLO corrections. 
In MCHM5, this cross-section may provide an interesting search channel for
large values of $\xi$ near one, where the enhancement factor compared
to the SM cross-section becomes significant. \sli

\begin{figure}[ht]
\begin{center}
\epsfig{figure=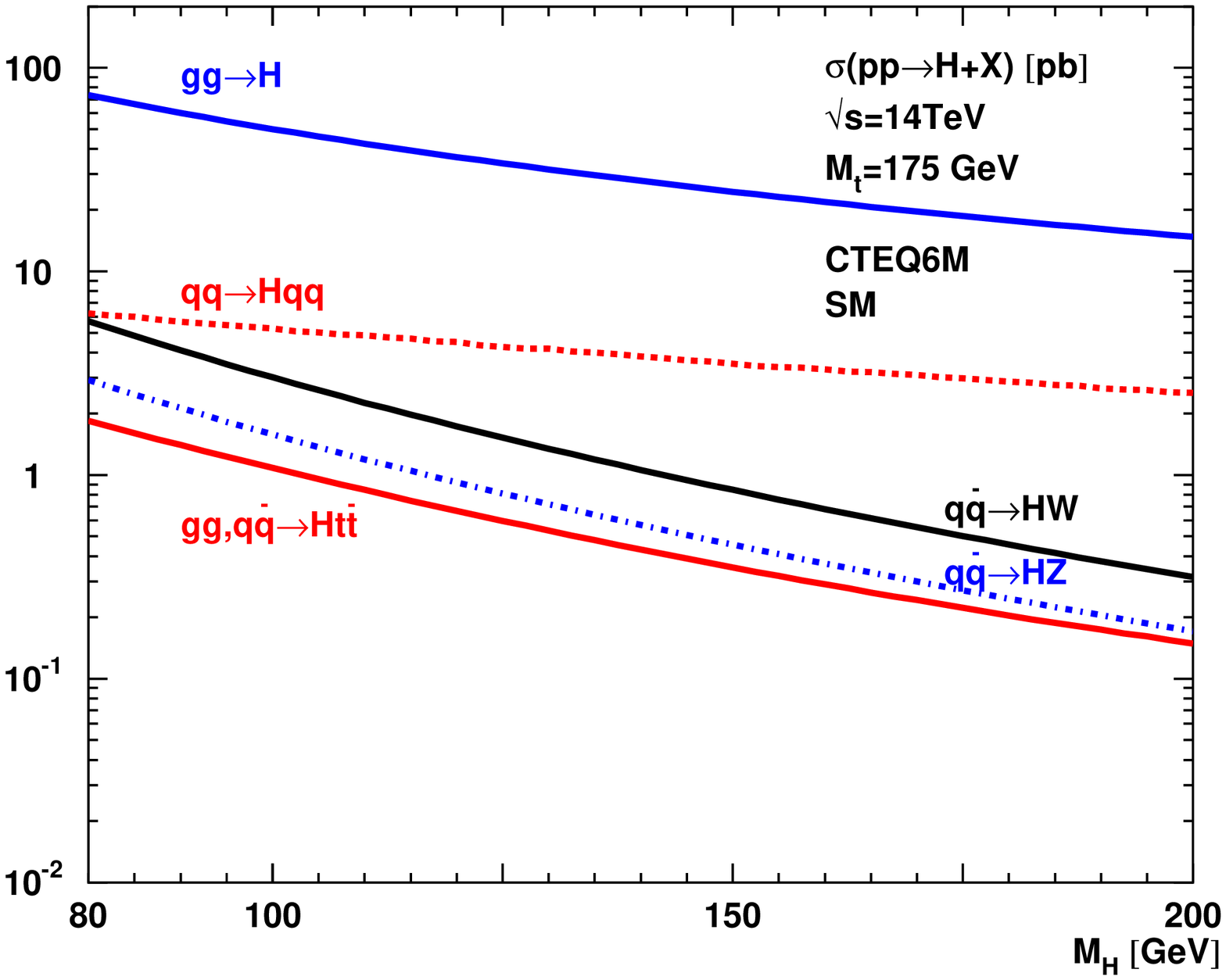,width=7cm}
\hspace*{0.5cm}
\epsfig{figure=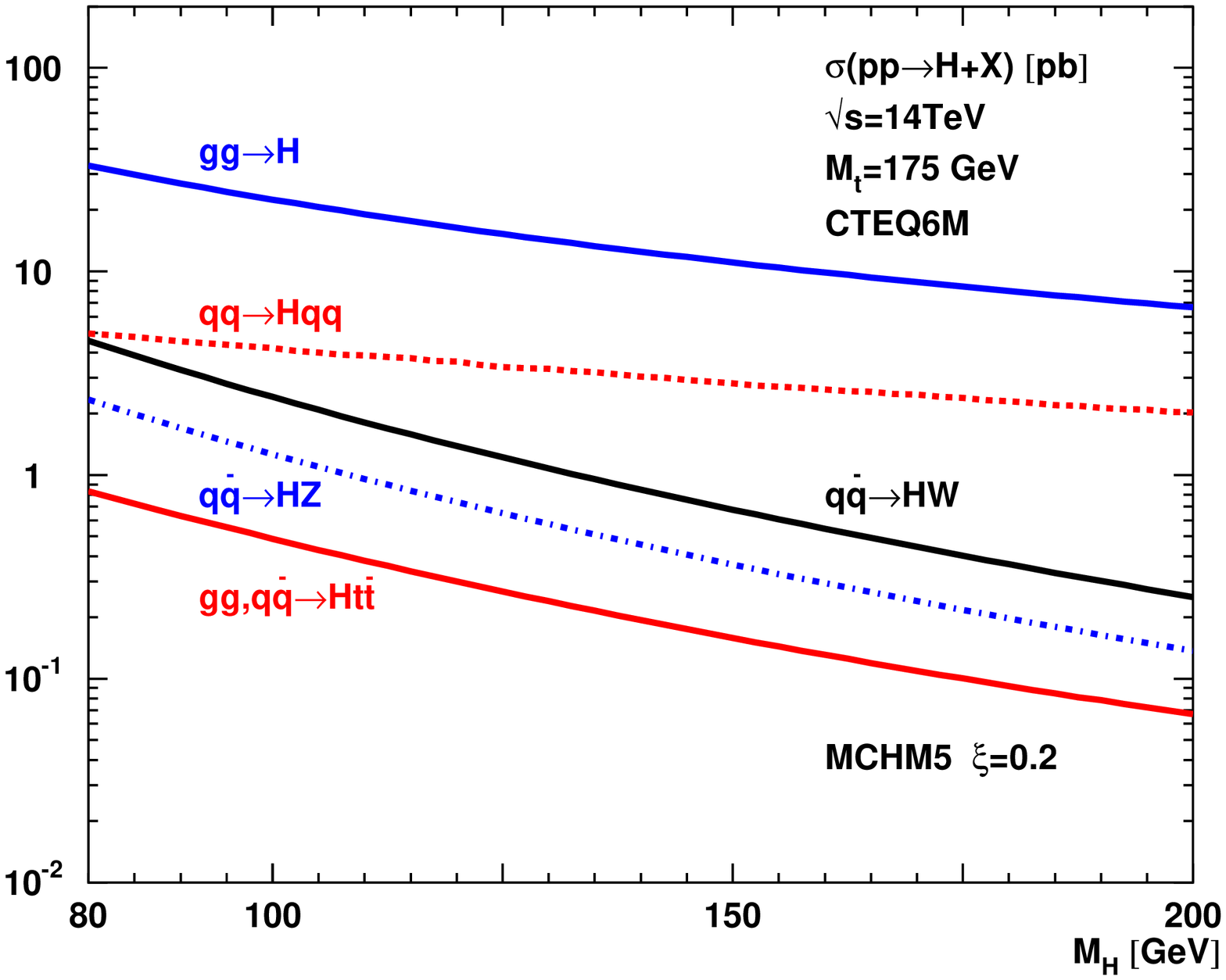,width=7cm}
\vskip 0.5 cm
\epsfig{figure=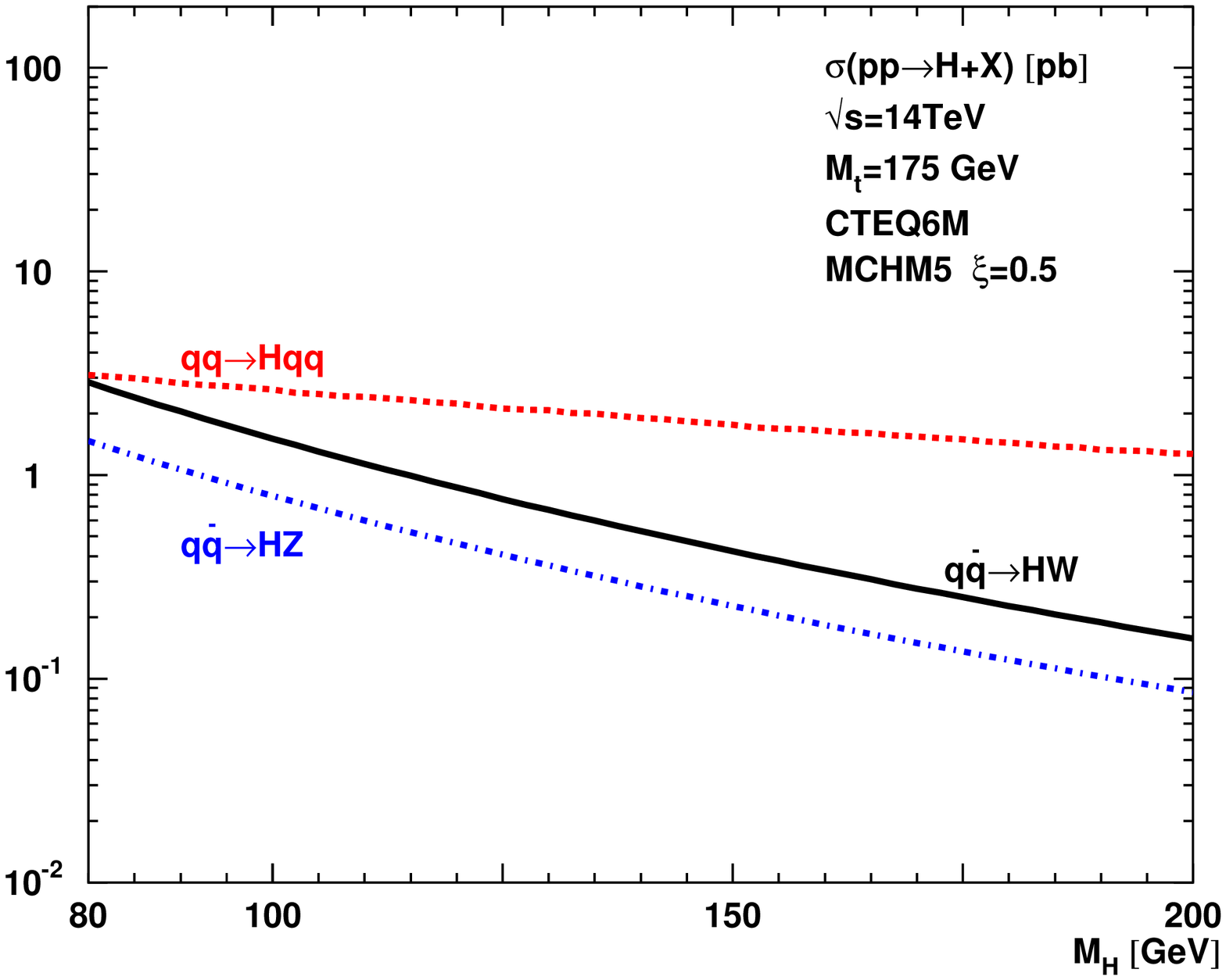,width=7cm}
\hspace*{0.5cm}
\epsfig{figure=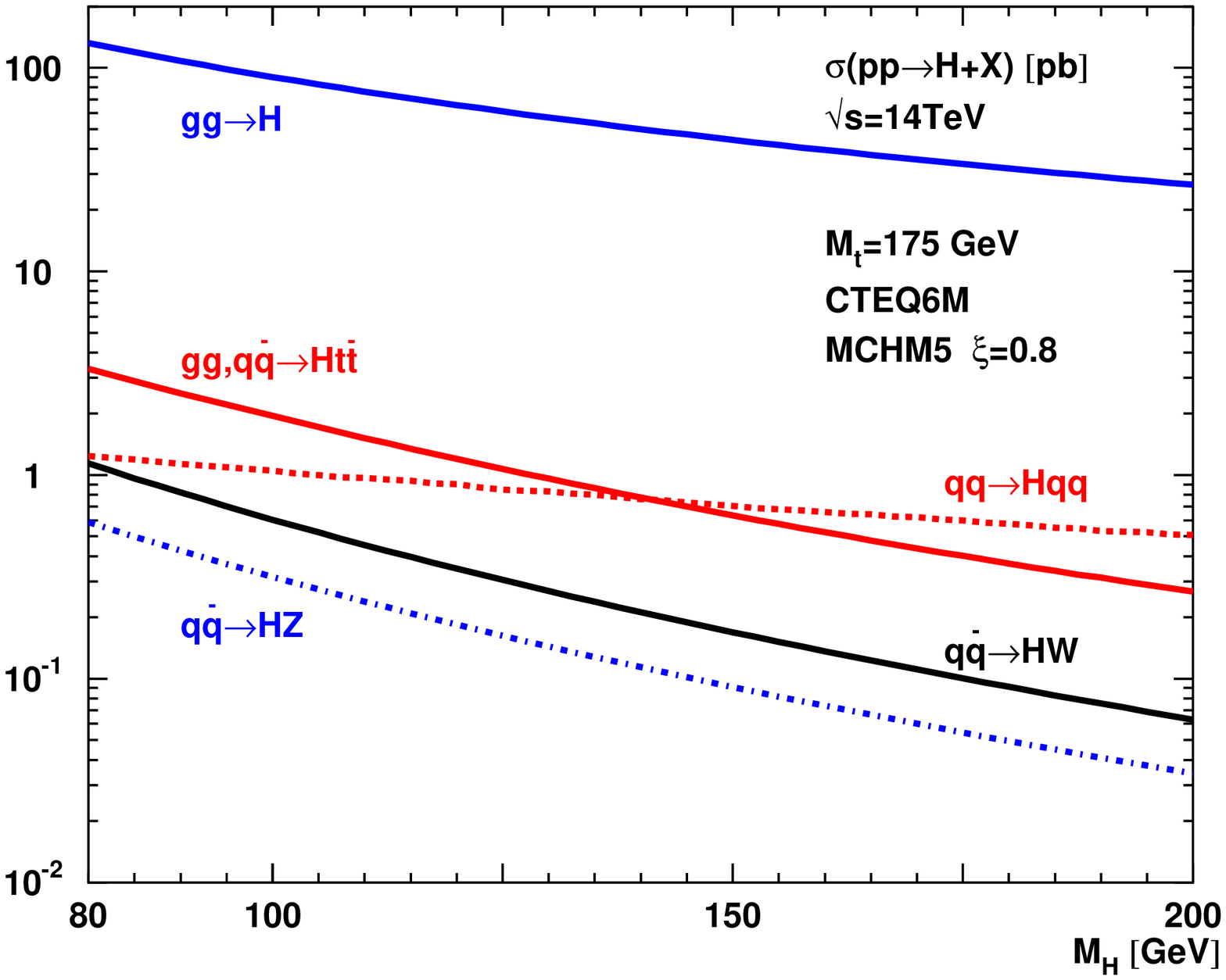,width=7cm}
\caption{\label{fig:prodXSC2}
The LHC Higgs boson production cross-sections as a function
of the Higgs boson mass in the SM ($\xi=0$, upper left) and for MCHM5 
with $\xi=0.2$ (upper right), 0.5 (bottom left) and 0.8 (bottom right).}
\end{center}
\end{figure}
\end{description}

In MCHM4, all Higgs production cross-sections are reduced by the overall
factor $(1-\xi)$. We do not show these cross-sections separately since
they can easily be obtained from the SM results, which are shown in
Fig.~\ref{fig:prodXSC2} upper left. In order to make contact with the
existing ATLAS and CMS experimental analyses of SM Higgs searches we use
$\sqrt{s}=14$ TeV, even if this is beyond the value the LHC will be able
to reach in its first years of running. The production cross-sections in
MCHM5 are also shown in Fig.~\ref{fig:prodXSC2}, as a function of the Higgs boson mass in the
interesting mass range $M_H=80\div200$~GeV for $\xi=0.2, 0.5$ and 0.8. In
the SM, the main production is given by gluon fusion, followed by gauge
boson fusion. The Higgs-strahlung processes $HW,HZ$ and the production in
association with top quarks are less important. For $\xi=0.2$ the
processes involving quarks, i.e., gluon fusion and $t\bar{t}H$ production,
are reduced by a factor 0.45 and the gauge boson processes, $WW,ZZ$ fusion
and Higgs-strahlung $HW,HZ$, are multiplied by a factor 0.8, according to
Eqs. (\ref{eq:ggfus}) to (\ref{eq:tth}). The inclusive Higgs production
will hence shrink considerably and might render the Higgs searches
difficult. The situation gets worse for $\xi=0.5$, where the gluon fusion
and $t\bar{t}H$ processes are completely absent and the gauge production
processes are diminished by a factor 2. For $\xi=0.8$, on the other hand,
the situation is reversed: while the gauge boson fusion and Higgs-strahlung
processes are only 20\% of the corresponding SM production processes and
might eventually not be exploitable for Higgs boson searches, the gluon
fusion and $t\bar{t}H$ production are enhanced by a factor~1.8.


\subsection{Statistical significances for different search channels}
 
In order to obtain the significances for the most important Higgs boson
search channels at the LHC, we refer to the analyses presented in the
CMS TDR~\cite{cmstdr}. Referring to the ATLAS TDR analyses~\cite{atlastdr}
would not lead to very different results.  The derivation of the
significances in the composite Higgs models is drastically simplified by
the fact that in our models we assumed only the couplings of the Higgs
bosons to deviate from the SM. Therefore only the numbers of the signal
events are modified while the numbers of the background events do not
change. More precisely, we proceed as follows. The experimental analyses
obtain the signal and background numbers in the investigated Higgs boson
search channels after application of cuts.  We
take the signal numbers and rescale them according to our model. The
rescaling factor $\varkappa$ is dictated by the change in the production
cross-sections and branching ratios compared to the Standard Model. For
the composite Higgs production in the process $p$ with subsequent decay
into a final state $X$, this factor is then given by
\beq
\varkappa = \frac{\sigma_{p} \, BR(H\to X)}{\sigma_{p}^{SM}
  BR(H^{SM} \to X)} \;.
\eeq
The number of signal events $s$ is obtained from
\beq
s = \varkappa\, \cdot \, s^{SM} \; ,
\eeq
where we take the number of SM model signal events after application of all
cuts, $s^{SM}$, from the experimental analyses. The signal events $s$ and the 
background events after cuts, i.e., $b \equiv b^{SM}$ are used to calculate 
the corresponding significances in the composite Higgs model.

\begin{figure}[ht]
\begin{center}
\epsfig{figure=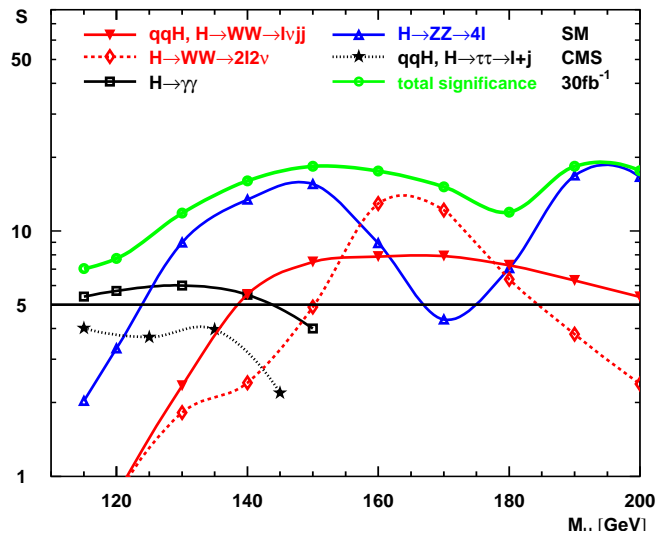,width=9cm}
\caption{\label{fig:sig0}
The significances in different Higgs search channels as a 
function of the Higgs boson mass in the SM with 
$\int{\cal L}=30\ \mathrm{fb}^{-1}$ and a CMS-type of analysis. The significances are computed using the data given in Tables~\ref{tab:Hgg}--\ref{tab:HtautauVBF}, they will slightly differ from the {\it official} CMS numbers which rely on a more sophisticated statistical analysis.} 
\end{center}
\end{figure}
Figure~\ref{fig:sig0} shows for reference the significances of different
channels searching for a light SM Higgs boson at CMS, as a function of
$M_H=115\div200$ GeV with $\int \, {\cal L } = 30$~fb$^{-1}$. The
gold-plated channel with 4 leptons in the final state from the Higgs decay
into $Z$ bosons reaches significances above 5 standard deviations for
Higgs masses $\gsim 125$~GeV. For Higgs masses around the $W$ boson
threshold inclusive production with subsequent decay into $W$ bosons takes
over. Higgs production in vector boson fusion, with decay into $WW$,
provides an efficient search channel in the intermediate region $\sim
140\div180$~GeV. The light Higgs mass region below $\sim 120$~GeV turns
out to be more difficult. Sophisticated cuts and combination of several
search channels are needed to achieve significances above 5$\sigma$.
In this way, the LHC will be able to cover the whole canonical Higgs mass
range up to $\sim 600$~GeV (not shown in the plot). The ATLAS experiment
provides a similar coverage.  We take these SM results as a benchmark and
study how the modified couplings of a composite Higgs will change this
picture.  The channels we investigate are 
\beq 
\begin{array}{lll}
\mbox{Inclusive production with subsequent decay}: & H\to \gamma\gamma \\
& H\to ZZ\to 2e 2\mu, 4e, 4\mu \\ 
& H\to WW \to 2l 2\nu \\ 
\mbox{Vector boson fusion with subsequent decay} : & H\to WW \to l \nu jj \\ 
& H\to \tau \tau \to l+j + E_T^{miss}\;.
\end{array} \eeq

We do not consider other Higgs channels which are of subleading
importance. For example, we do not discuss $t\bar{t}H$ production with
subsequent decay into $b\bar{b}$. This channel has been removed recently
from the list of possible search modes, since controlling the background 
appears to be too difficult to make reliable predictions. We have also
checked that we do not gain much significance by the inclusion of the
gluon-fusion Higgs production followed by $H\to\tau\tau$ decay, with an
additional resolved jet ($pp\to H+j\to \tau\tau+j$)~\cite{ellis}, which
has been recently revived in~\cite{moretti} as a promising channel for
light Higgs searches in models with an enhanced Higgs BR into $\tau\tau$.
In our models, whenever this channel has a sizeable significance, other 
channels already provide large significance.

In the following, we will discuss each channel in turn, giving the
expected significance as a function of the Higgs mass $M_H$ and the $\xi$
parameter. For concreteness, we fix the integrated luminosity to $\int
{\cal L}=30\ \mathrm{fb}^{-1}$. We base ourselves on the CMS TDR
\cite{cmstdr} and the relevant CMS Notes.  Similar results would be
expected for ATLAS.

Since the CMS analyses calculate the significances with different
definitions for the various channels, we take the pragmatic approach to
get our $\xi=0$ significances as close as possible to the SM result given
by CMS, so that any deviation at $\xi\neq 0$ can be attributed
to the composite character of the Higgs. We are forced then
to choose different significance definitions (as listed in the appendix).
Our combined significances have been obtained by adding in quadrature the
individual significances without caring about their heterogeneous nature
and, therefore, have to be taken as merely indicative.

\subsection{${\bma{H\to \gamma\gamma}}$}

This channel is of crucial importance for the Higgs search at low masses
(below $\sim 150$~GeV, see Fig.~\ref{fig:sig0}) where the decays into real
gauge bosons are closed. Furthermore, since the decay into photons is
loop-mediated, it is sensitive to new physics effects due to new
particles in the loop (see Ref.~\cite{hgg} for a recent study). 
The signature is characterized by two isolated high $E_T$
photons. While the photons can easily be identified, this channel is
very challenging due to the small signal rate compared to the large
background. The reason is that the Higgs boson dominantly decays into
$b\bar{b}$ in this mass region, which cannot be exploited though due
to the high QCD background. The $\gamma\gamma$ 
signal will appear as a narrow mass peak
above the large background. The latter can be measured from the
sidebands outside the peak and extrapolated into the signal region.

For the production cross-sections, we use the same as in the CMS
analyses\footnote{See section 2.1 of the CMS TDR \cite{cmstdr} and the CMS
Note 2006/112~\cite{CMS2006112}.} which are based on the Higgs production
in gluon fusion, vector boson fusion, associated production with $W,Z$
bosons and $Ht\bar{t}$ production. As SM benchmark data for the expected
significance we use the CMS standard cut-based analysis, see
Table~\ref{tab:Hgg}, which subdivides the total sample of events in a
number of different categories especially designed to improve the combined
significance. A more sophisticated analysis~\cite{cmstdr} leads to even
higher significances and our results for this channel are therefore
conservative.

\begin{table}[ht]
\begin{center}
  \begin{tabular}{ |c || c | c | c | c | c | }
    \hline
    $M_H$ (GeV) & 115 &  120 & 130 & 140 & 150 \\ \hline\hline
    $S_{CMS}$ & 5.4 & 5.7 & 6.0 & 5.5 & 4.0 \\
    \hline
  \end{tabular}
\end{center}
\vspace*{-0.5cm}
\caption{Expected significances for the SM Higgs search in the
  $H\to\gamma\gamma$ channel, with $\int {\cal L}=30\ \mathrm{fb}^{-1}$,
as given by the standard CMS cut-based analysis presented in  
Ref.~\cite{CMS2006112}, Fig.~5.}
\label{tab:Hgg}
\end{table}

At the different steps of this analysis the partial significances of the
different categories are well described by the simple formula
$s_i/\sqrt{b_i}$ so that the analysis for the composite Higgs can be
performed exactly in the same way, except for an overall universal factor
that takes into account the change in the signal yields, so that the
combined total significance is rescaled by that same universal factor.
While a dedicated CMS-type cut-based analysis for the composite Higgs case
could improve over the rescaled significance, this simple recipe allows us
to make smooth contact with the CMS results and to improve over simpler
significance estimates which use the total number of signal and background
events.
\begin{figure}[ht]
\begin{center}
\epsfig{figure=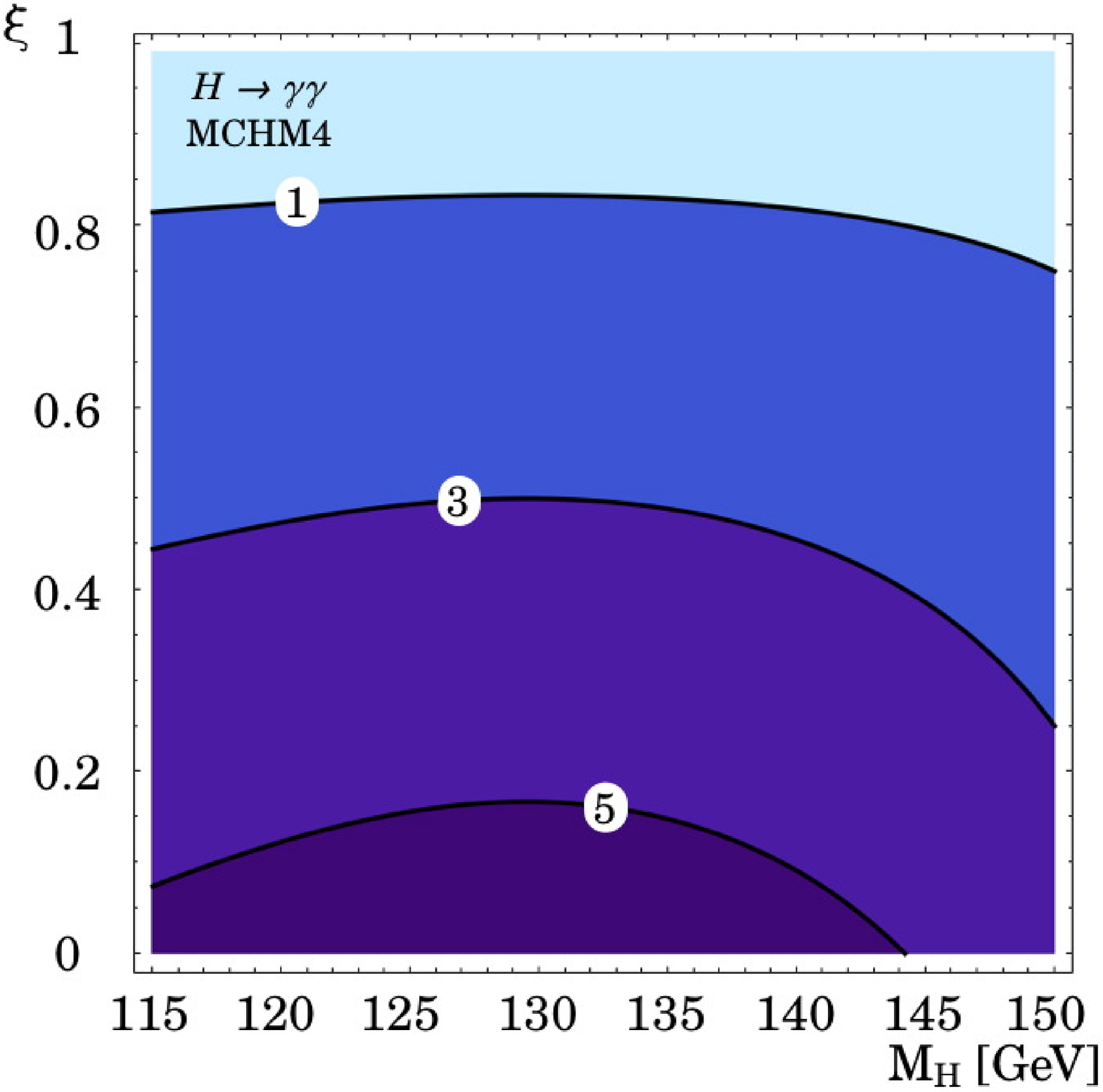,width=7cm}
\hspace*{0.5cm}
\epsfig{figure=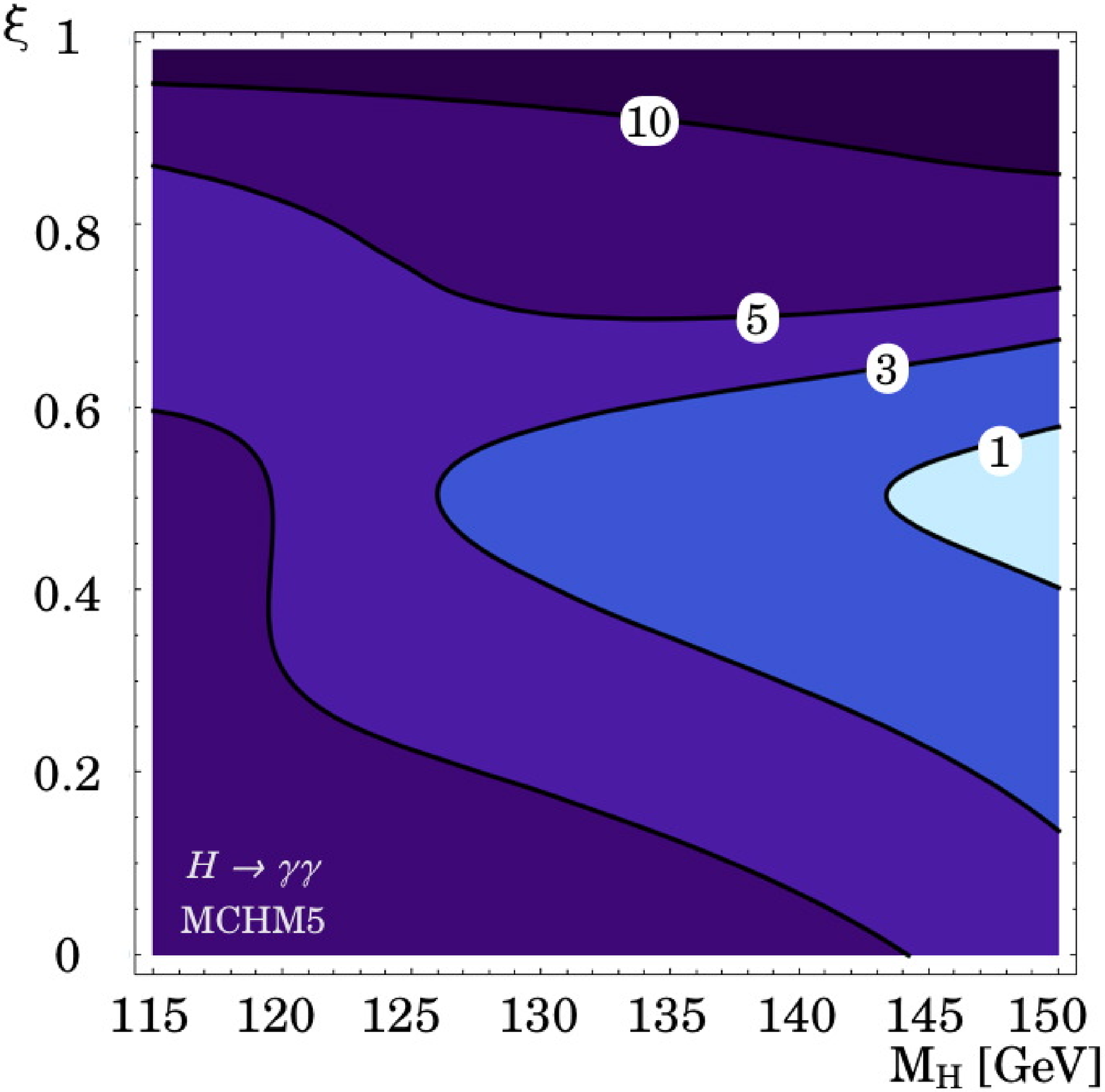,width=7cm}
\caption{\label{fig:signgaga}
The signal significance in the channel $H\to \gamma\gamma$ in the
$(M_H,\xi)$ plane with an integrated luminosity of $30\ \mathrm{fb}^{-1}$
for MCHM4~(left) and MCHM5~(right). The darker the color, the higher the 
significance. The contours delineate the regions corresponding to a 
significance of 1, 3, 5 and 10$\sigma$. }
\end{center}
\end{figure} 

The results for the significances in both MCHM4 and MCHM5 are presented in
Fig.~\ref{fig:signgaga} as contour lines in the plane $(M_H,\xi)$. The
values along the $\xi=0$ axis coincide with the SM numbers as given in
Table~\ref{tab:Hgg}. In MCHM4 (left plot), the significance degrades
quickly as $\xi$ gets larger as a result of the production cross-section
getting smaller with the decreasing rescaling factor $(1-\xi)$ (we remind
the reader, that in MCHM4 the branching ratios do not change compared to
the SM). This trend will recur in all channels.  In MCHM5 (right
plot), the significance is worst along intermediate values of $\xi$ where
the total production cross-section has a minimum, although this effect is
partially compensated at low Higgs masses by the increase in the
$H\to\gamma\gamma$ branching ratio in this $\xi$ region, see
Figs.~\ref{fig:BRs} and \ref{fig:prodXSC2}. Both behaviours are due to the
vanishing Yukawa couplings at $\xi=0.5$. The expected significance is
larger than $5\sigma$ in a large region of parameter space, especially for
$\xi$ near 1, where the production cross-sections mediated via Yukawa
couplings are largely enhanced. Here the significances can be larger than
those of the SM.

\subsection{$\bma{H\to ZZ\to 4l}$}

This clean channel, with the Higgs decaying through $ZZ^{(*)}$ into $4e$,
$2e2\mu$ and $4\mu$, is one of the most promising Higgs discovery channels
for Higgs masses above $\sim 130$ GeV, although, as shown in
Fig.~\ref{fig:sig0}, the expected significance drops in the neighbourhood
of $M_H\sim 160$ GeV where $H\to WW$ peaks (see Fig.~\ref{fig:BRs}).  The
production cross-section, dominated by gluon fusion in this mass range, is
large and so is the branching ratio into $ZZ^{(*)}$ which is sizeable for
$M_H \gsim 130$~GeV. The channel yields a significant, very clean and
simple multi-lepton final state signature. Furthermore, it provides a
precise determination of the Higgs boson mass and, to a lesser extent, cross-section and also
allows, via angular and mass distributions, the determination of the spin
and CP quantum numbers of the Higgs
boson~\cite{Accomando:2006ga,cppapers}.

The CMS analyses\footnote{We used Section~2.2 of the CMS TDR~\cite{cmstdr}
and CMS Note 2006/115~\cite{CMS2006115} (for $4e$); Section~3.1 of the CMS
TDR and CMS Note 2006/122 \cite{CMS2006122} (for $4\mu$); Section~10.2.1
of the CMS TDR and CMS Note 2006/136~\cite{CMS2006136} (for $2e2\mu$); and
the more recent CMS PAS HIG-08-003~\cite{CMSPASHIG08003}.} are based on
the production through gluon fusion and vector boson fusion.  The Higgs
boson signal is characterized by two pairs of
  isolated primary electrons and muons. One pair in general results
  from a $Z$ boson decay on its mass shell. In the analyses the main
  background processes considered are $t\bar{t}, Zb\bar{b}\to
  2lb\bar{b}$ and $ZZ\to 4l$. In order to extract the expected experimental
  sensitivity a sequential cut based approach is used and the search
  is performed with a window in the hypothetical mass $M_H.$ The SM signal
and background rates as well as the significances are given in
Table~\ref{tab:HZZll}. The resulting significances are very similar in the
three different subchannels (both in the SM and in the composite Higgs
models) so that we only discuss the significance for the combined
channels, which is shown in Fig.~\ref{fig:sig0} for the SM. For the
calculation of the composite Higgs significances with $30$~fb$^{-1}$
integrated luminosity, we use the Poisson significance $S_P$ as defined in
the appendix, neglecting the systematic uncertainty of the background,
which has only a small effect.

\begin{table}[ht]
\begin{center}
  \begin{tabular}{ |c || c | c | c | c | c | c | c | c | c | c |}
    \hline
$M_H$ (GeV) 
          & 115  & 120  & 130  & 140  & 150  & 160  & 170  & 180  & 190  & 200\\ \hline\hline
      $s$ & 0.18 & 0.33 & 1.27 & 2.43 & 3.05 & 1.51 & 0.73 & 1.78 & 6.5  & 7.08 \\ \hline
      $b$ & 0.16 & 0.19 & 0.29 & 0.42 & 0.47 & 0.47 & 0.61 & 1.38 & 2.74 & 3.52 \\ \hline
$S_{CMS}$ & ---  & 0.13 & 1.32 & 2.22 & 2.64 & 1.36 & 0.50 & 1.09 & 2.92 & 2.87 \\ \hline
    $S_P$  &  ---   & 0.13 & 1.35 & 2.24 & 2.64 & 1.38 & 0.50 & 1.09 & 2.96 & 2.92 \\ 
    \hline
  \end{tabular}
\end{center}
\vspace*{-0.5cm}
\caption{Number of signal $s$ and background events $b$ and resulting
  significance $S_{CMS}$ expected for the SM Higgs search in the
  channel $H\to ZZ\to 2l2l'$, with $\int{\cal L}=1\ \mathrm{fb}^{-1}$, as 
given in Ref.~\cite{CMSPASHIG08003}, Table~3. 
To extend the Higgs mass range, the point $M_H=115$ GeV 
has been added using results from
refs.~\cite{CMS2006115,CMS2006122,CMS2006136}. For comparison, the
last row gives the expected Poisson significance $S_P$ for $\int{\cal 
L}=1\ \mathrm{fb}^{-1}$,
with  systematic background uncertainties $\Delta b$ included (as
defined in the appendix),  and with $\Delta b/b=0.21\ (0.08)$, 
for low (high) Higgs masses.}
\label{tab:HZZll}
\end{table}

\begin{figure}[ht]
\begin{center}
\epsfig{figure=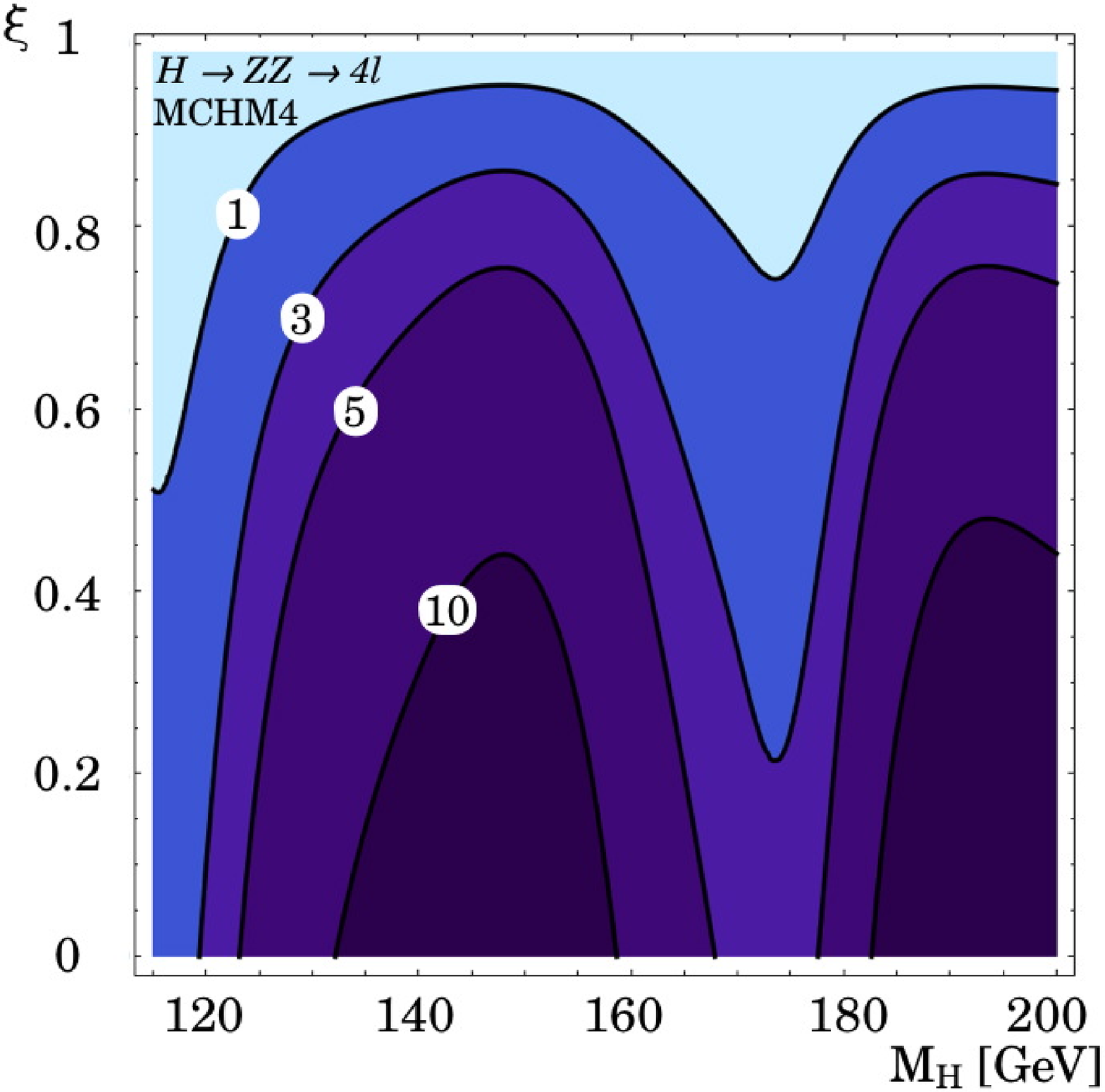,width=7cm}
\hspace*{0.5cm}
\epsfig{figure=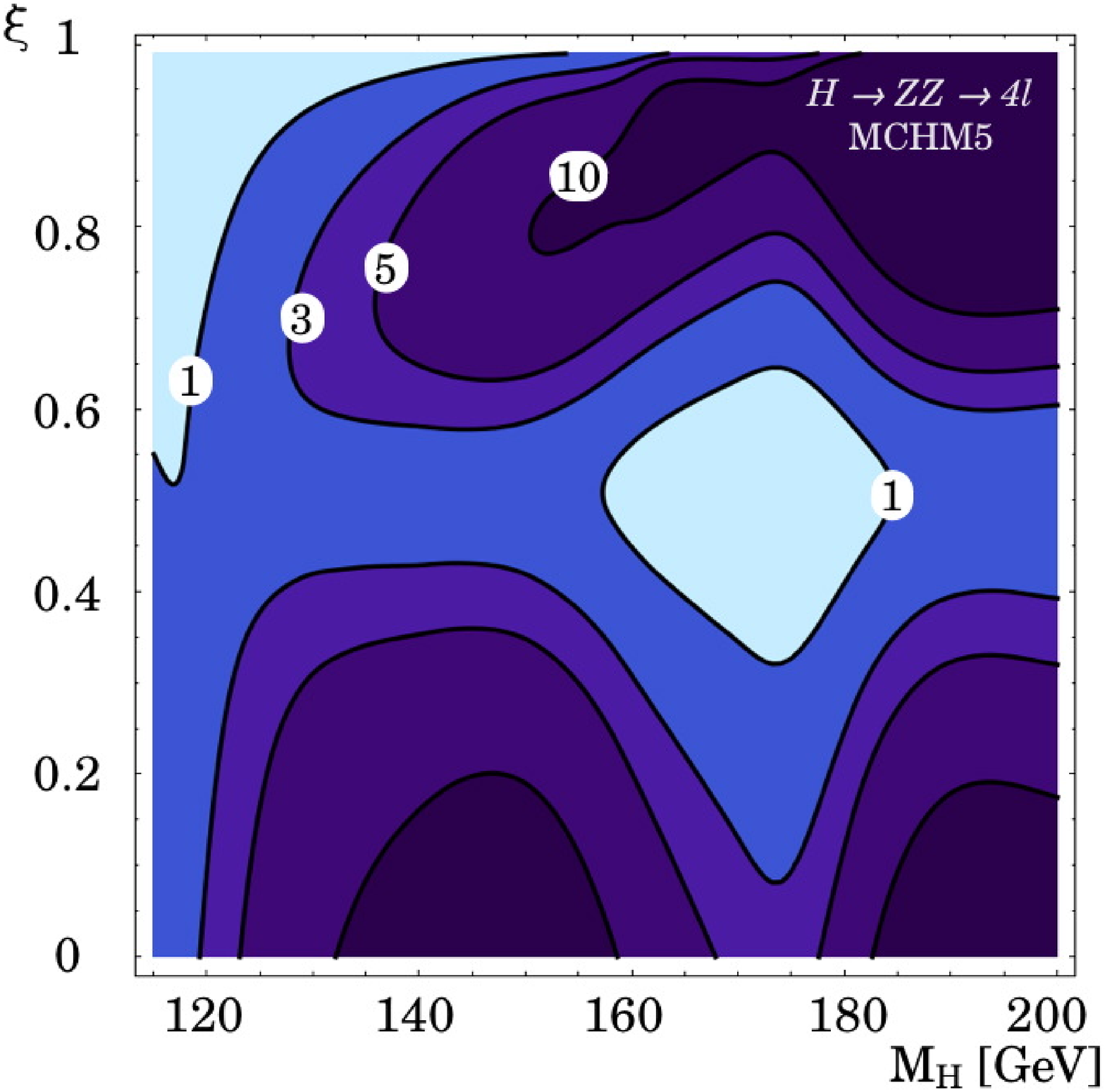,width=7cm}
\caption{\label{fig:signZZll}
The combined signal significance for the channels $H\to ZZ \to 4l$ in 
the $(M_H,\xi)$ plane with an integrated luminosity of $30\ 
\mathrm{fb}^{-1}$ for MCHM4~(left) and MCHM5~(right).
The darker the color, the higher the significance. The contours 
correspond to a significance of 1, 3, 5 and 10$\sigma$.}
\end{center}
\end{figure} 

The significances for both MCHM4 and MCHM5 are presented in
Fig.~\ref{fig:signZZll} as contour lines in the plane $(M_H,\xi)$. The
values along the $\xi=0$ axis coincide with the SM numbers scaled up to
30~fb$^{-1}$. In MCHM4~(left plot), the significance degrades as usual
with increasing $\xi$ due to the reduction in the production cross-section
but remains sizeable up to large values of $\xi$ due to its initially
large value at $\xi=0$.  In MCHM5~(right plot), the significance follows
in its horizontal behaviour the SM change in the significance with the
Higgs mass. The vertical behaviour as function of $\xi$ results mostly
from the variation of the cross-sections with $\xi$ which drops
considerably at $\xi=0.5$ where the Yukawa couplings are zero. The drop is
partially compensated by the enhancement in the $ZZ$ branching ratio in
this region (see Figs.~\ref{fig:BRs} and \ref{fig:prodXSC2}). Thus the
significance is worst along intermediate values of $\xi$ and higher for
large values of $\xi$ where the gluon-fusion cross-section is enhanced.
Here it can even exceed the SM significance for Higgs mass values above
$\sim 180$ GeV.

\subsection{$\bma{H\to WW\to 2l2\nu}$}

The Higgs decay into $WW$ which subsequently decay into leptons is the
main discovery channel in the intermediate region $2 M_W \lsim M_H \lsim 2
M_Z$ where the Higgs branching ratio into $WW$ is close to one. This
channel has seen its revival after it was realized that the spin
correlation in the $W^+W^-$ system can be exploited to extract the signal
from the background~\cite{Dittmar:1996ss}. The signature is characterized
by two leptons and missing high energy. Since no narrow mass peak can be
reconstructed, a good background control and a high signal to background
ratio are needed.  The considered production mechanisms used in the CMS
analyses\footnote{The relevant CMS documents are Section~10.2.2 of the CMS
TDR~\cite{cmstdr}, the CMS Note 2006/047~\cite{CMS2006047} and the most
recent CMS PAS HIG--08--006~\cite{CMSPASHIG08006}.} are both gluon fusion
and vector boson fusion. The SM data for this channel are collected in
Table~\ref{tab:HWW2l2nu}.  We use the $ScP2$ significance (see the
appendix) including a background systematic uncertainty estimated to be
$14.6\%$ at 1~fb$^{-1}$.  The SM result for 30~fb$^{-1}$, with background systematic uncertainty scaled down to 10\%, is shown in Fig.~\ref{fig:sig0} (we calculate the significance simply from the total numbers of signal and background events. The CMS analysis is performed first with the $ee, e\mu$ and $\mu\mu$ subchannels separately which are then combined).

\begin{table}[ht]
\begin{center}
  \begin{tabular}{ |c || c | c | c | c | c | c | c | c | c |}
    \hline
     $M_H$ (GeV)
         & 120  & 130  & 140   & 150  & 160  & 170  & 180  & 190  & 200 \\ \hline\hline
     $s$ &  7.5 & 17.3 &  31.4 & 24.4 & 67.5 & 66.8 & 50.9 & 31.2 &  29.6 \\ \hline
     $b$ & 87.3 & 89.4 & 121.4 & 42.5 & 37.4 & 40   & 67.3 & 73.3 & 115.8 \\ \hline
$S_{CMS}$& 0.55 & 1.0  & 1.55  &  2.4 & 5.93 & 6.1  & 3.35 & 1.95 &  1.45 \\ \hline
   $ScP2$ & 0.46 & 1.03 & 1.42  &  2.4 & 6.16 & 5.89 & 3.42 & 2.08 &  1.39 \\ 
    \hline
  \end{tabular}
\end{center}
\vspace*{-0.5cm}
\caption{Number of signal and background events and resulting 
significance expected
for the SM Higgs search in the channel $H\to WW\to 2l2\nu$, with 
$\int{\cal L}=1\ \mathrm{fb}^{-1}$,
as given in Ref.~\cite{CMSPASHIG08006}, Table~9 and Fig.~6. The last row 
gives the expected significance 
$ScP2(s,b,\Delta b)$ (as defined in the appendix), 
 with $\Delta b/b=0.146$ and for $\int{\cal L}=1\ \mathrm{fb}^{-1}$.}
\label{tab:HWW2l2nu}
\end{table}

\begin{figure}[ht]
\begin{center}
\epsfig{figure=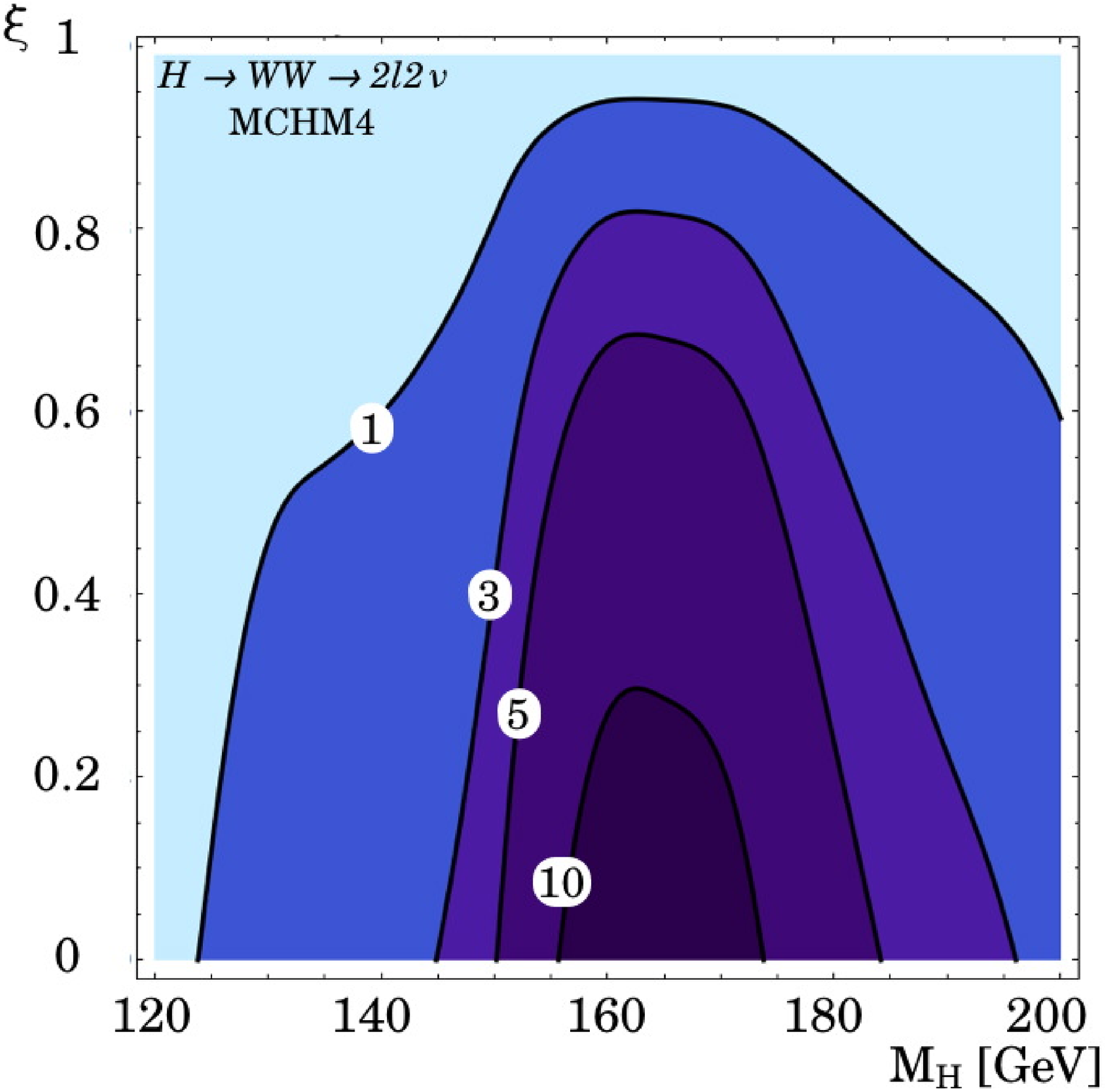,width=7cm}
\hspace*{0.5cm}
\epsfig{figure= 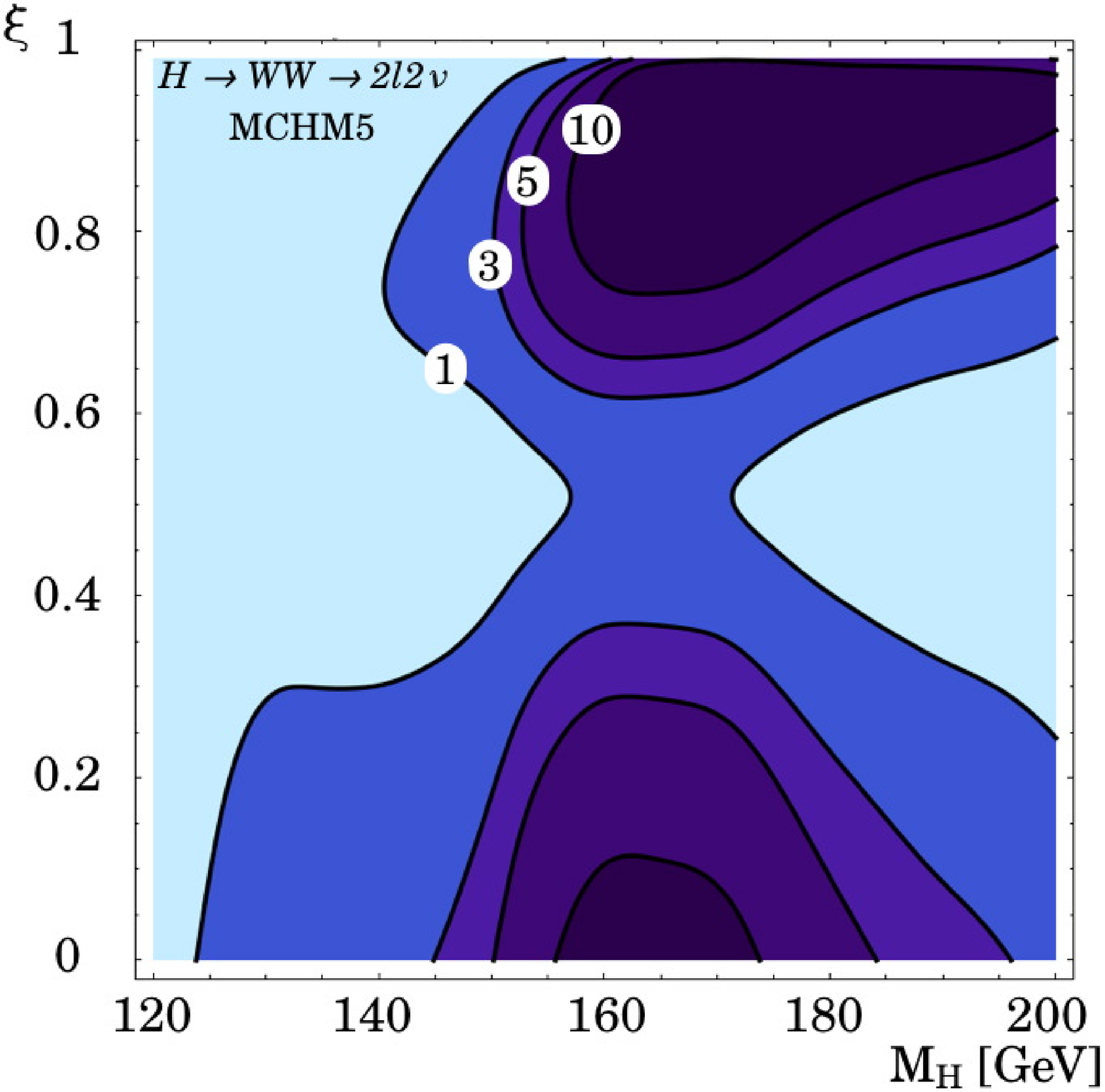,width=7cm}
\caption{\label{fig:signWWlnu}
The signal significance in the channel $H\to WW\to 2l 2\nu$ in 
the $(M_H,\xi)$ plane with an integrated luminosity of $30\ 
\mathrm{fb}^{-1}$ and 10\% background systematic uncertainty for 
MCHM4~(left) and MCHM5~(right). The contours correspond to a significance 
of 1, 3, 5 and 10$\sigma$.}
\end{center}
\end{figure} 

The results for the expected significances in MCHM4 and MCHM5
 are presented in Fig.~\ref{fig:signWWlnu} as contour lines in the plane
$(M_H,\xi)$. As usual, the values along $\xi=0$ agree well with the CMS SM
results. The significance in MCHM4 degrades with increasing $\xi$, but
remains sizeable up to large values of $\xi$ due to its initially large
value at $\xi=0$. The value of $\xi$ at which the significance really
deteriorates compared to the SM one depends on the Higgs mass. For MCHM5,
we find the usual behaviour, with the smallest significances at $\xi\sim
0.5$, the value that determines an approximate axis of symmetry for the
resulting significances.  As for the previous channels, the expected
significance is larger than $5\sigma$ in a sizeable portion of parameter
space and exceeds the SM significance at large values of $\xi$ where the
Yukawa couplings, and hence the gluon-fusion process, are significantly
enhanced.  These regions with largest significances, however, are already
being probed and a priori excluded by the Tevatron (see
Section~\ref{sec:constraints}), which exploits the same decay channel.

\subsection{$\bma{H\to WW\to l\nu jj}$}

The Higgs search in vector boson fusion with subsequent decay $H\to
W^+W^-\to l^\pm \nu jj$ is very important to cover the Higgs mass region
$160$~GeV$\lsim M_H \lsim 180$~GeV where the $H\to ZZ^*$ branching ratio
is largely suppressed because of the opening of $H\to W^+W^-$. Due to the
possibility of direct Higgs mass reconstruction, it complements the
previous search channel, which has two unobservable neutrinos in the final
state.  The event topology is characterized by two forward jets, two
central jets from the $W$ hadronic decay, and one high $p_T$ lepton and
missing transverse energy from the $W$ leptonic decay. Furthermore, an 
extra jet veto can be applied to efficiently reduce the background. 
The large background necessitates robust reconstruction and
selection strategies to extract the signal from the background and
minimize the systematic uncertainties. The SM data for this channel are
collected in Table~\ref{tab:HWWlnujj}\footnote{The relevant CMS documents
are Section~10.2.4 of the CMS TDR~\cite{cmstdr} and the CMS Note
2006/092~\cite{CMS2006092}.}.  For the calculation of the significances in
the composite models, we use the $ScL'$ significance (see the appendix)
including a background systematic uncertainty of $16\%$. The SM result at
30 fb$^{-1}$ is shown in Fig.~\ref{fig:sig0}. The significance is larger
than 5$\sigma$ for $M_H\gsim 135$~GeV.

\begin{table}[ht]
\begin{center}
  \begin{tabular}{ |c || c | c | c | c | c | c | c | c | c |}
    \hline
    $M_H$ (GeV) & 120 & 130 & 140 & 150 & 160 & 170 & 180 & 190 & 200 \\ \hline\hline
    $s$ & 6.93 & 19.92 & 49.68 & 69.51 & 89.67 & 90.18 & 82.14 & 70.2  & 59.49\\ \hline
    $b$ &34.92 & 34.92 & 34.92 & 34.92 & 46.95 & 46.95 & 46.95 & 46.95 & 46.95 \\ \hline
    $S_{CMS}$ &  0.8 &  2.25 & 5.3   & 7.3   &  8.1  & 8.15  & 7.3   & 6.25  & 5.3   \\  \hline
 $ScL'$ & 0.84 &  2.34 & 5.52  & 7.47  & 7.86  & 7.90  & 7.26  & 6.29  & 5.4   \\  
  \hline
  \end{tabular}
\end{center}
\vspace*{-0.5cm}
\caption{Number of signal and background events and resulting 
significance expected
for the SM Higgs search in the channel $H\to WW\to l\nu jj$, at 
$\int{\cal L}=30\ 
\mathrm{fb}^{-1}$ with full extra jet veto, as given in Section~10.2.4 of the 
CMS TDR~\cite{cmstdr}, Tables~10.12, 10.13 (rescaling the integrated 
luminosity) and 
Fig.~10.19. The last row gives the expected significance
$ScL'(s,b,\Delta b)$ (as defined in the appendix) with $\Delta b/b=0.16$ 
and for $\int{\cal L}=30\ \mathrm{fb}^{-1}$.}
\label{tab:HWWlnujj}
\end{table}

\begin{figure}[ht]
\begin{center}
\epsfig{figure= 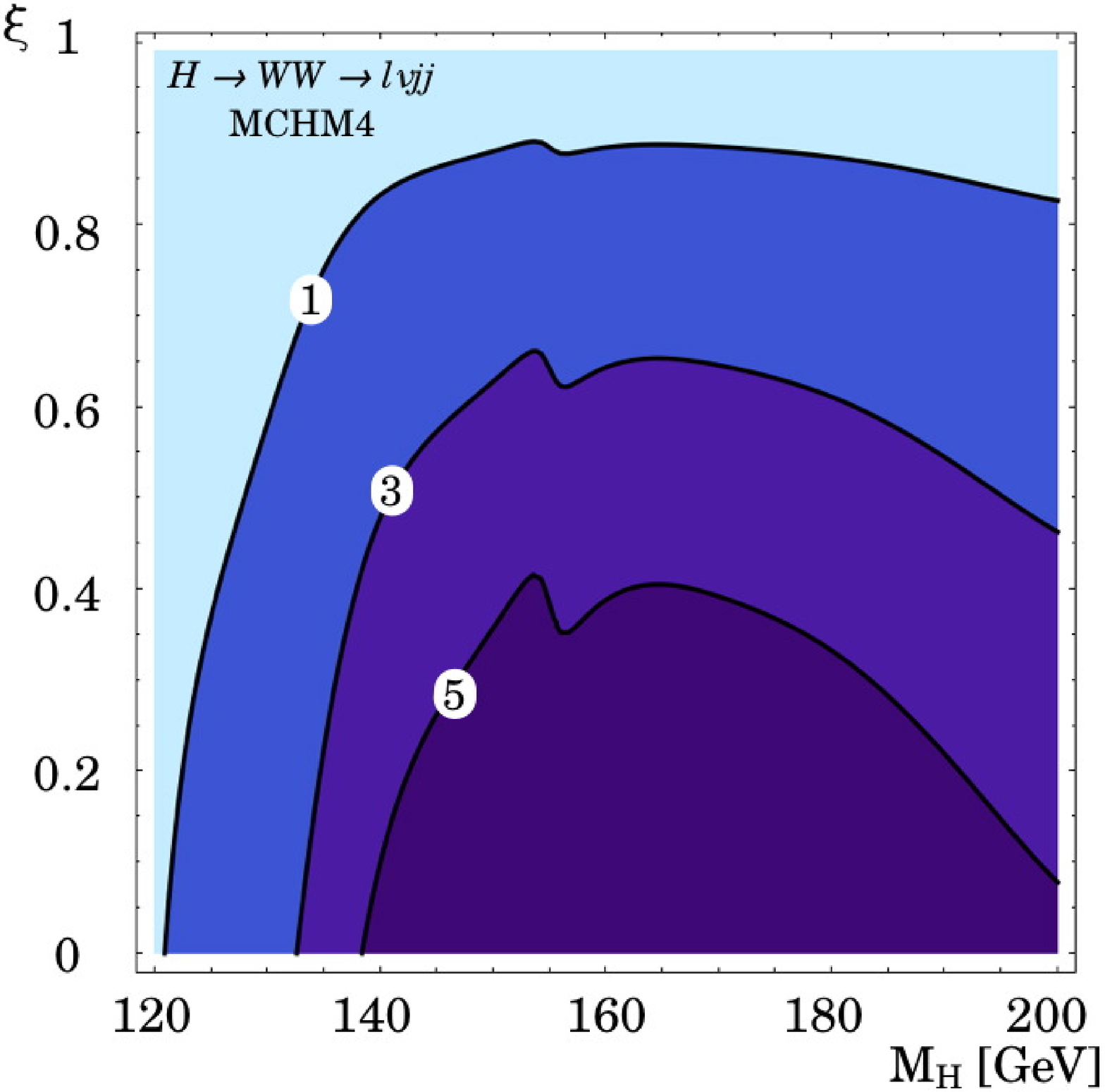,width=7cm}
\hspace*{0.5cm}
\epsfig{figure= 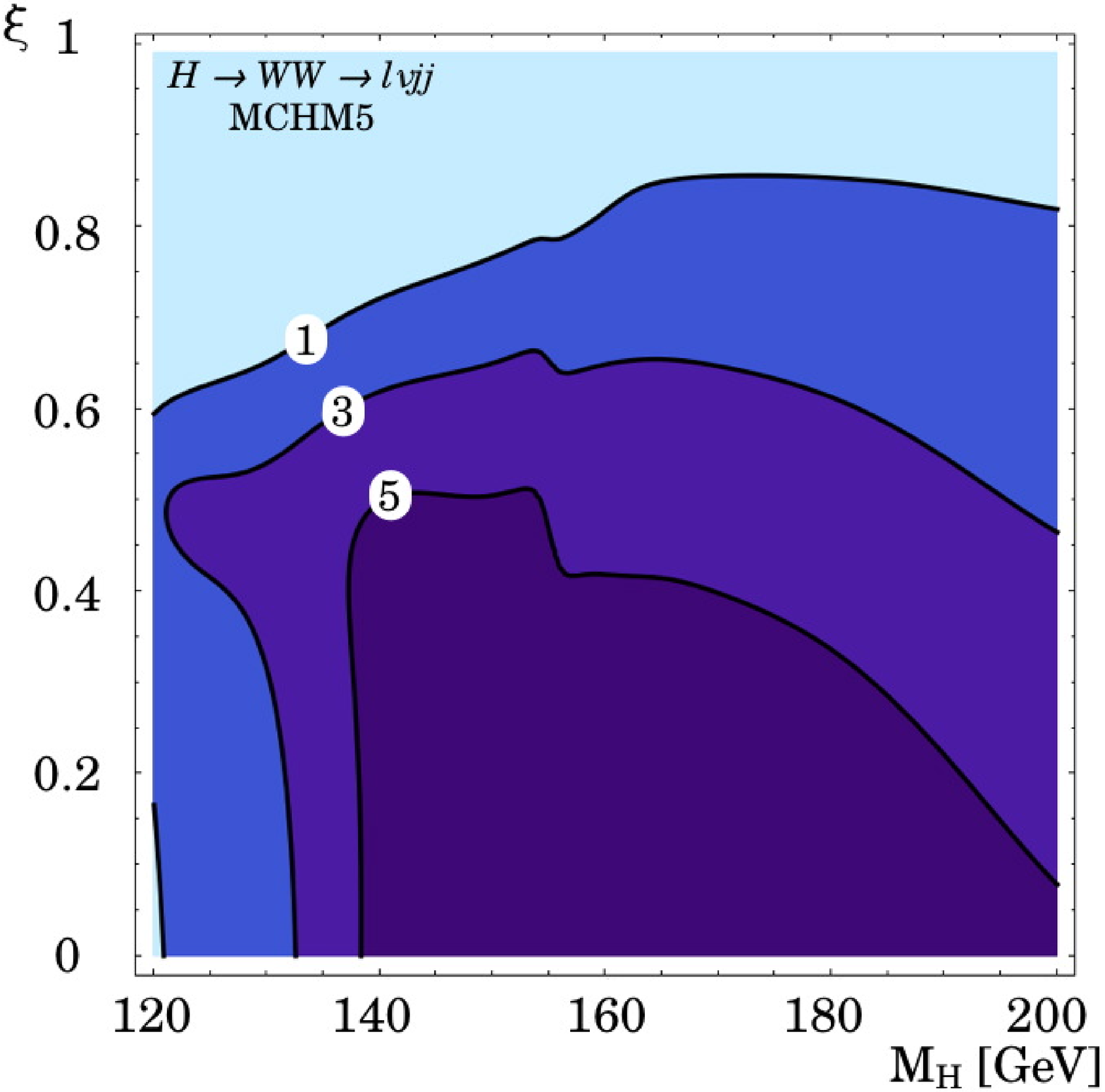,width=7cm}
\caption{\label{fig:signWWlnujj}
The signal significance in the channel $H\to WW\to l\nu jj$ in the
$(M_H,\xi)$ plane with an integrated luminosity of $30\ \mathrm{fb}^{-1}$ and 16\% background systematic uncertainty
for MCHM4~(left) and MCHM5~(right). The contours correspond to a 
significance of 1, 3, 5 and 10$\sigma$.}
\end{center}
\end{figure} 

The results for the expected significances in MCHM4 and MCHM5 are
presented in Fig.~\ref{fig:signWWlnujj} as contour lines in the plane
$(M_H,\xi)$. As usual, the values along $\xi=0$ agree well with the CMS SM
results. The significance in MCHM4 degrades with increasing $\xi$ but
remains sizeable up to large values of $\xi$ due to its initially large
value at $\xi=0$. The significance never exceeds the SM significance. Due
to the fact that only vector boson fusion (which is always suppressed
compared to the SM cross-section) production is considered, MCHM5 does not
exhibit the usual symmetric behaviour around the axis $\xi= 0.5$. The
behaviour is quite similar to that in MCHM4 with the differences that, in
MCHM5, the regions with higher significance are larger for low $\xi$
values, but smaller for values of $\xi \gsim 0.6$. The former is due to
the enhanced branching ratio into $WW$ in MCHM5, reaching its maximum at
$\xi=0.5$ where the Yukawa decay channel into $b\bar{b}$ is closed. The
latter is due to the suppressed branching ratio into $WW$ for values of
$\xi$ beyond 0.5, where in turn the decay into $b\bar{b}$ is enhanced.

Two regions are worthwhile discussing in more detail. Both MCHM4 and MCHM5
exhibit an edge in the significance around $M_H \approx 150$~GeV. This is
due to the larger background values used in the CMS analyses for $M_H \ge
160$~GeV (due to a different hadronic $W$ mass selection window for $M_H<
160$~GeV). Finally, MCHM5 shows a bulge with higher luminosity around $\xi
\sim 0.5$ extending to lower Higgs mass values. This is due to the
enhanced branching ratio into $WW$ for $\xi=0.5$. Altogether the expected
significance is at most as good as in the SM and larger than $5\sigma$
only for $\xi<0.5$.

\subsection{$\bma{H\to \tau\tau \to l+j+E_T^{miss}}$}

In parton-level analyses~\cite{Rainwater:1998kj}, as well as in studies
with detector simulation~\cite{Cavalli:2002vs}, it was shown that Higgs
production in vector boson fusion with subsequent decay into $\tau$
leptons is an important search channel at low Higgs masses, $M_H\simlt 140$
GeV. In this mass region the $H\to \tau\tau$ decay is second in importance
after the $b\bar{b}$ decay (which cannot be exploited because of the large
QCD background). Although this is not the main channel in that region, it
can contribute to improve the total significance when combined with other
channels.  Furthermore this channel adds to the determination of the Higgs
couplings~\cite{Rainwater:1998kj}.

The signature of the signal process are a high $p_T$ lepton and a
$\tau$-jet, two energetic forward jets and the total missing $E_T$ of the
system. The backgrounds considered in the analysis\footnote{The relevant
CMS documents are Section~10.2.3 of the CMS TDR~\cite{cmstdr}, the CMS
Note 2006/088~\cite{CMS2006088}.} are the irreducible ones from QCD and
electroweak $Z/\gamma^\star$ boson production with 2 or 3 associated jets and the
reducible background processes from $W+$ multi-jet and $t\bar{t}$ events.
The background can efficiently be reduced by using the characteristics of
the weak boson fusion process, which are the wide rapidity separation of
the two leading quark jets and the suppressed hadronic activity in the
central region due to the absence of colour exchange between the forward
quark jets. The SM data for this channel are collected in
Table~\ref{tab:HtautauVBF}. We use the Poisson significance (see the
appendix) including a background systematic uncertainty estimated to be
$7.8\%$. The SM result is shown in Fig.~\ref{fig:sig0}.

\begin{table}[ht]
\begin{center}
  \begin{tabular}{ |c || c | c | c | c | }
    \hline
    $M_H$ (GeV) & 115 & 125  & 135 & 145  \\ \hline\hline
    $s$ & 10.5 & 7.8  & 7.9  & 3.6 \\ \hline
    $b$ & 3.7  & 2.2  & 1.8  & 1.4 \\ \hline
    $S_{CMS}$ & 3.97 & 3.67 & 3.94 & 2.18  \\ \hline
    $S_P$ & 4.01 & 3.70 & 3.97 & 2.19  \\ 
    \hline
  \end{tabular}
\end{center}
\vspace*{-0.5cm}
\caption{Number of signal and background events and resulting 
significance expected
for the SM Higgs search in the channel $H\to \tau\tau\to l + j +
E_T^{miss}$, with $\int{\cal L}=30\ \mathrm{fb}^{-1}$ as given in
Section~10.2.3 of the CMS TDR~\cite{cmstdr}, Table~10.10. The last
row gives the expected Poisson significance $S_P$ (as defined 
in the appendix) with $\Delta b/b=0.078$ and for $\int{\cal L}=30\ 
\mathrm{fb}^{-1}$.}
\label{tab:HtautauVBF}
\end{table}

\begin{figure}[ht]
\begin{center}
\epsfig{figure=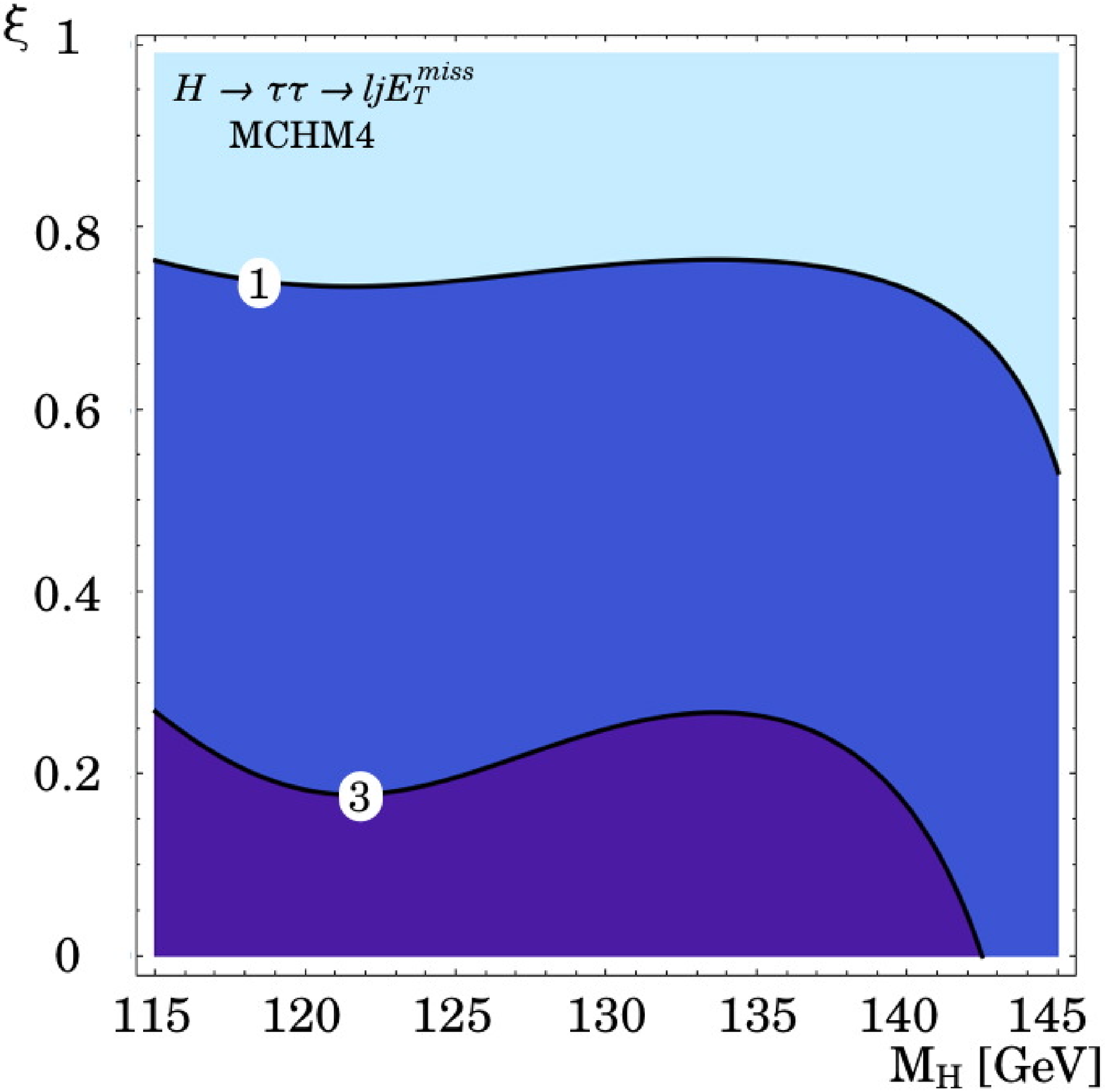,width=7cm}
\hspace*{0.5cm}
\epsfig{figure= 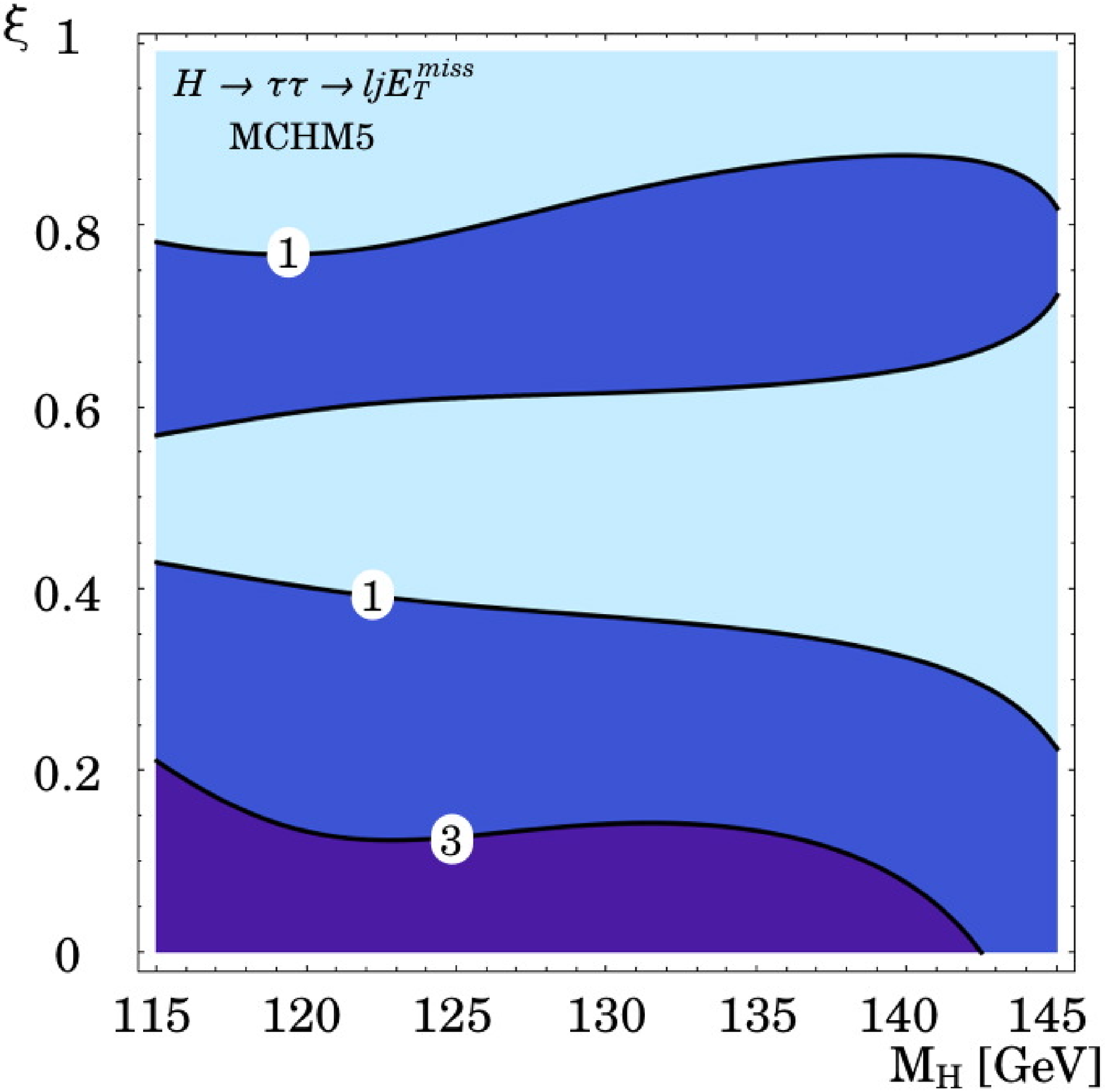,width=7cm}
\caption{\label{fig:signtautauvbf}
The signal significance in the channel $H\to \tau\tau\to l + j + 
E_T^{miss}$ in the $(M_H,\xi)$ plane with an integrated luminosity of 
$30\ \mathrm{fb}^{-1}$ and 7.8\% background systematic uncertainty for MCHM4~(left) and MCHM5~(right). The contours 
correspond to a significance of 1, 3 and 5$\sigma$.}
\end{center}
\end{figure} 

The results for the expected significances in MCHM4 and MCHM5
 are presented in Fig.~\ref{fig:signtautauvbf} as contour lines in the
plane $(M_H,\xi)$. As usual, the values along $\xi=0$ agree well with the
CMS SM results. The significance in MCHM4 degrades with increasing $\xi$.
MCHM5 produces a similar decrease in significance, since only vector boson
fusion production is considered\footnote{In this region with $\xi\sim 1$
the gluon-fusion process $pp\to H+j\to \tau\tau+j$ of
Refs.~\cite{ellis,moretti} would be important. Nevertheless, this region
is already covered by the channel $H\to ZZ\to 2l 2l'$.}. For $\xi\lsim
0.6$ the significance is lower than in MCHM4 because of the suppressed
branching ratio into $\tau\tau$ which vanishes finally at $\xi = 0.5$.
Beyond this value it increases with rising $\xi$ so that at high $\xi$
values, MCHM4 and MCHM5 show a similar behaviour. Unlike in the previous
channels, the expected significances in both models are always less than
$5\sigma$, since already in the SM the significances range below this
value. MCHM4 and MCHM5 cannot compensate for that as with the vector boson
fusion process the production cross-section is always smaller than in the
SM and the increase in the branching ratio into $\tau\tau$ in MCHM5 cannot
keep up with that.

\begin{figure}[ht]
\begin{center}
\epsfig{figure=egm0Hcomb.eps,width=7cm}
\hspace*{0.5cm}
\epsfig{figure=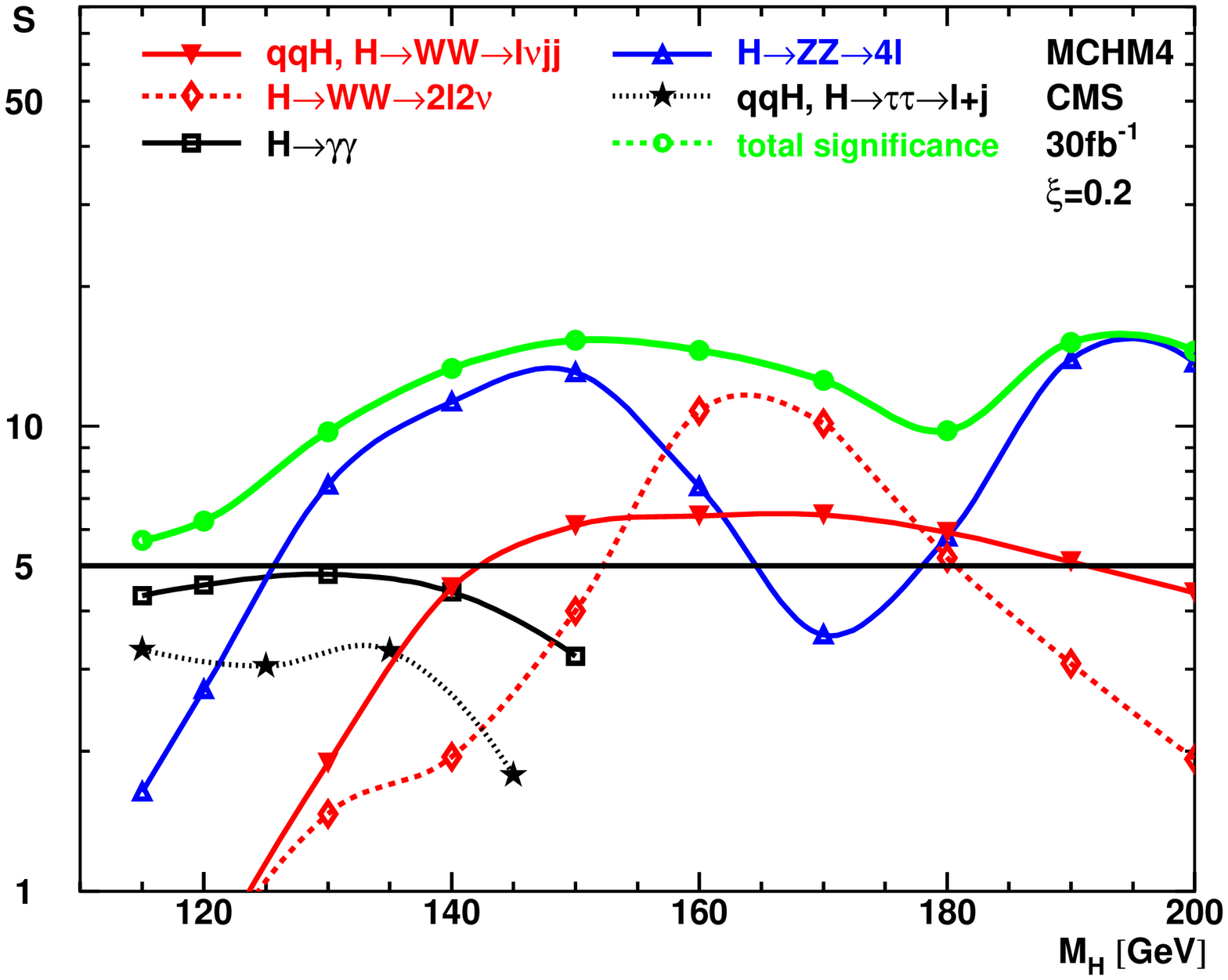,width=7cm}
\vskip 0.5 cm
\epsfig{figure=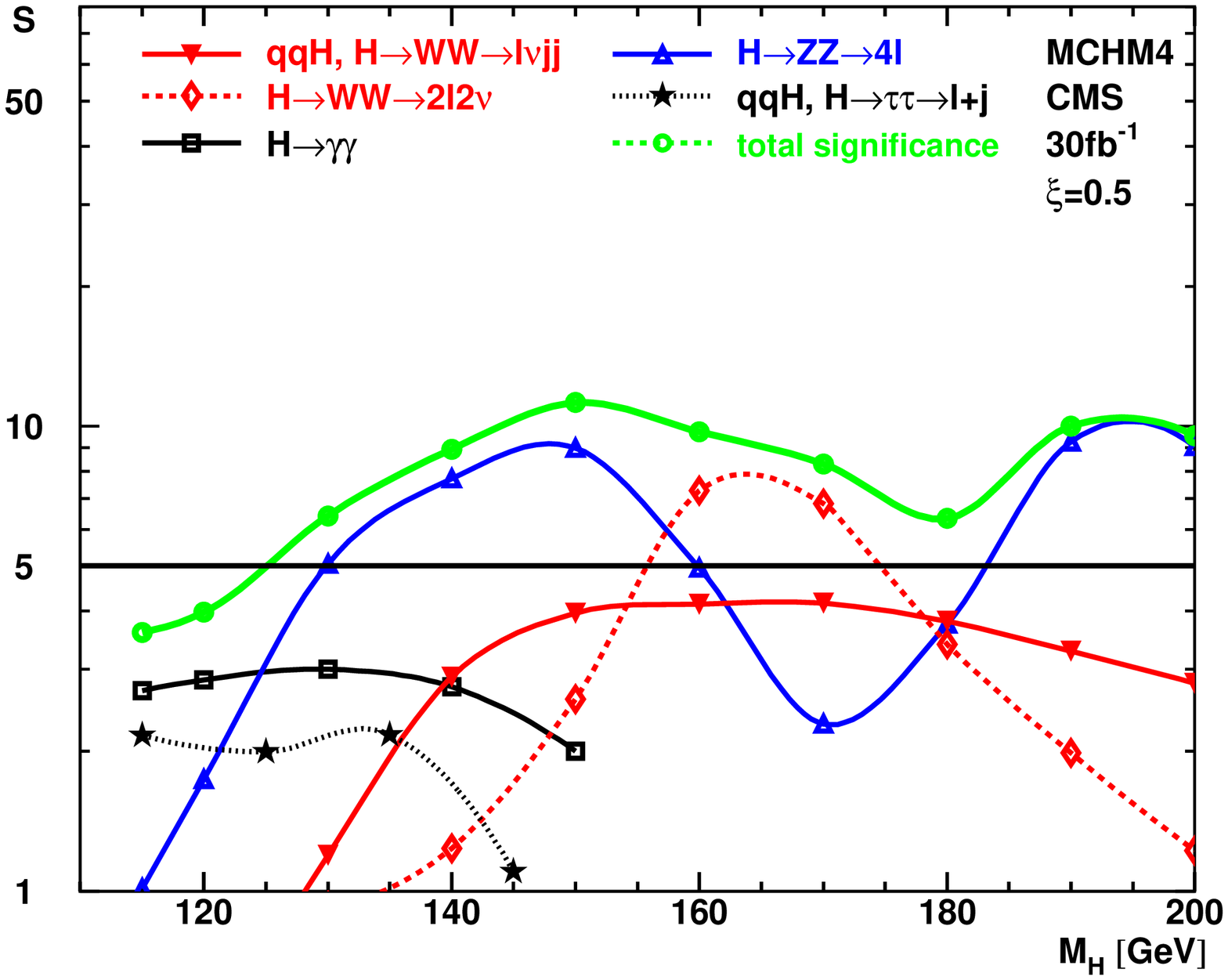,width=7cm}
\hspace*{0.5cm}
\epsfig{figure=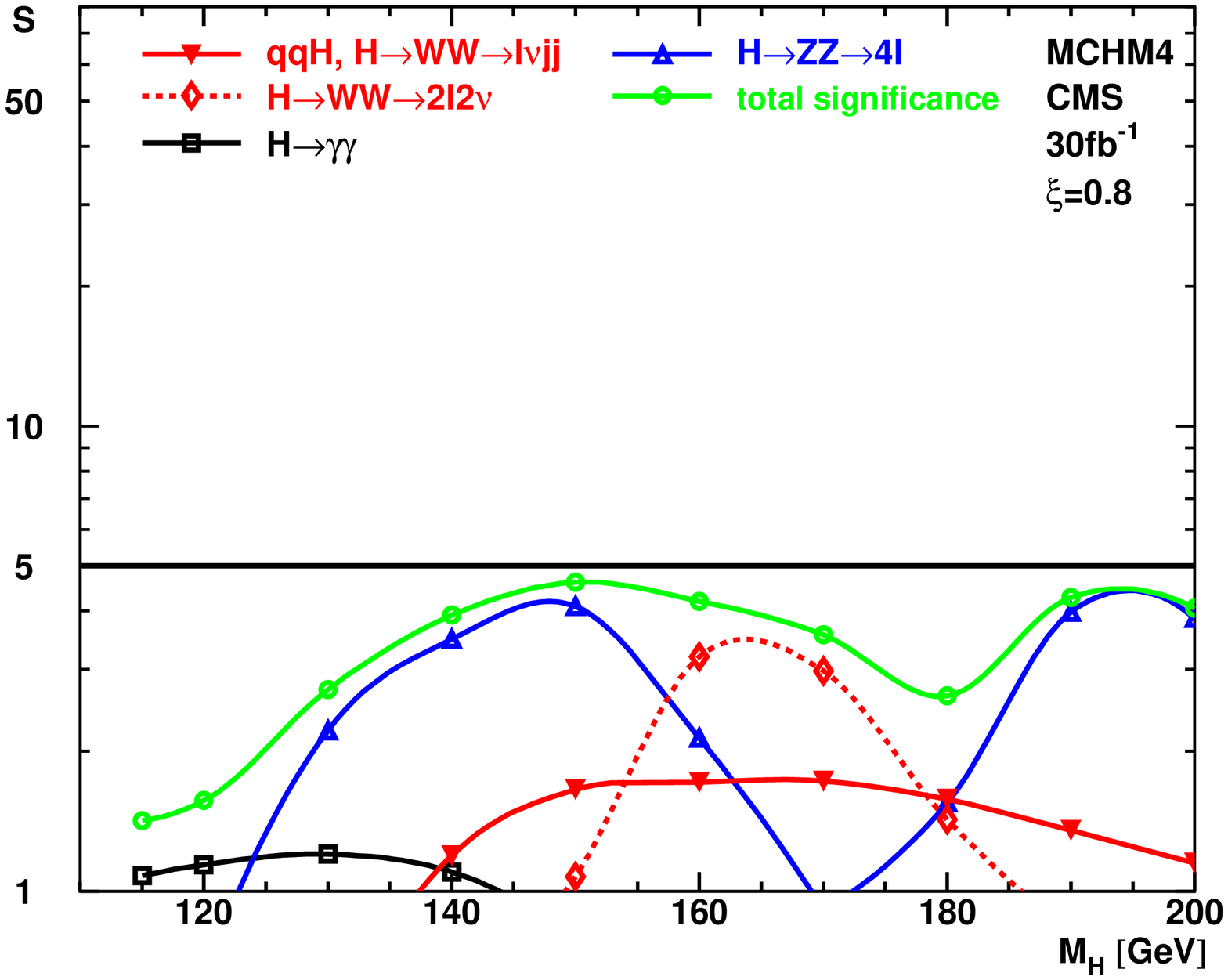,width=7cm}
\caption{\label{fig:sig1}
The significances in different channels as a function of the Higgs boson 
mass in the SM ($\xi=0$, upper left) and for 
MCHM4 with $\xi=0.2$ (upper right), 0.5 (bottom left) and 0.8 (bottom 
right).} \end{center}
\end{figure}

\section{Summary of results and conclusions}
\label{sec:Conclusions}

\begin{figure}[ht]
\begin{center}
\epsfig{figure=egm0Hcomb.eps,width=7cm}
\hspace*{0.5cm}
\epsfig{figure=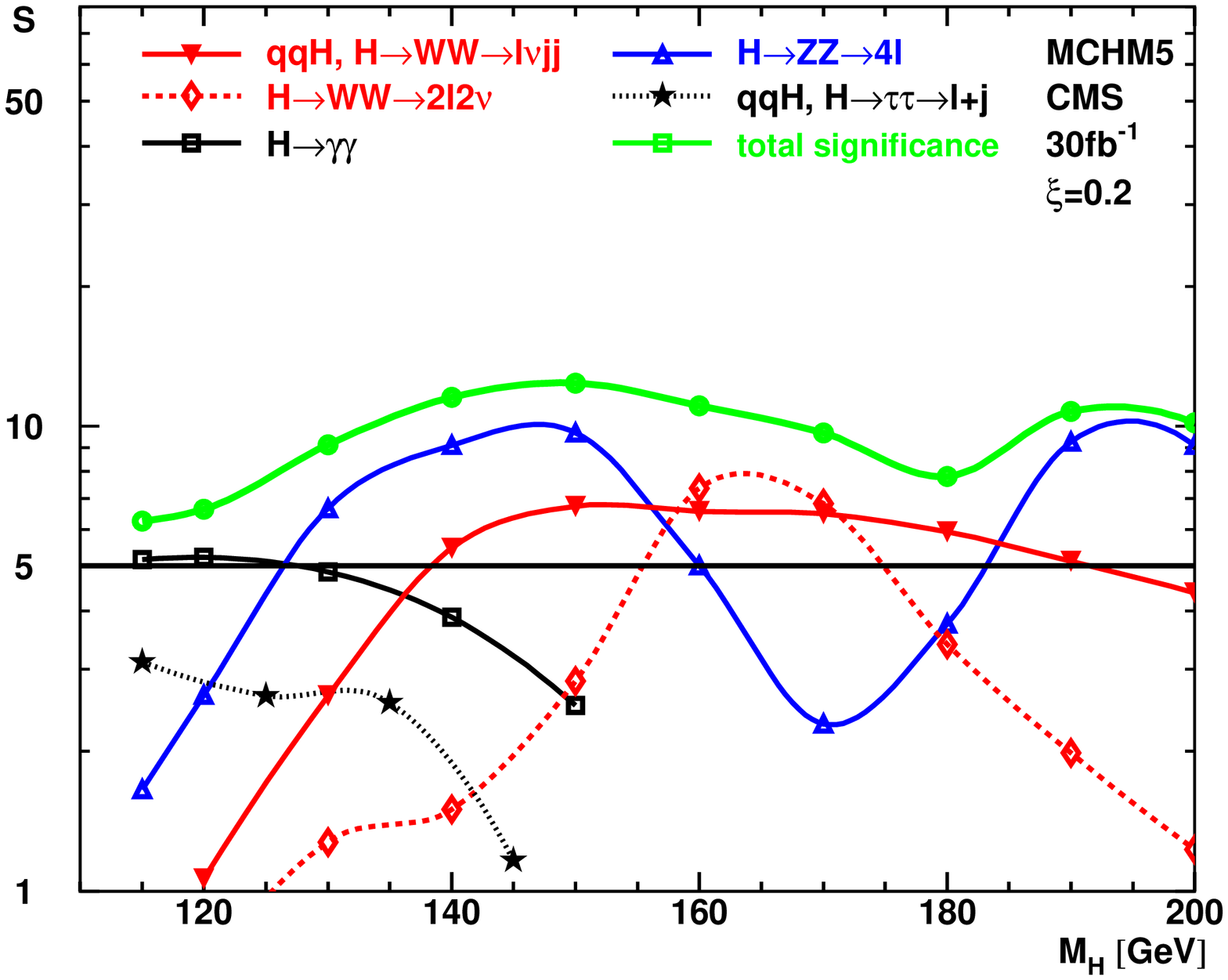,width=7cm}
\vskip 0.5 cm
\epsfig{figure=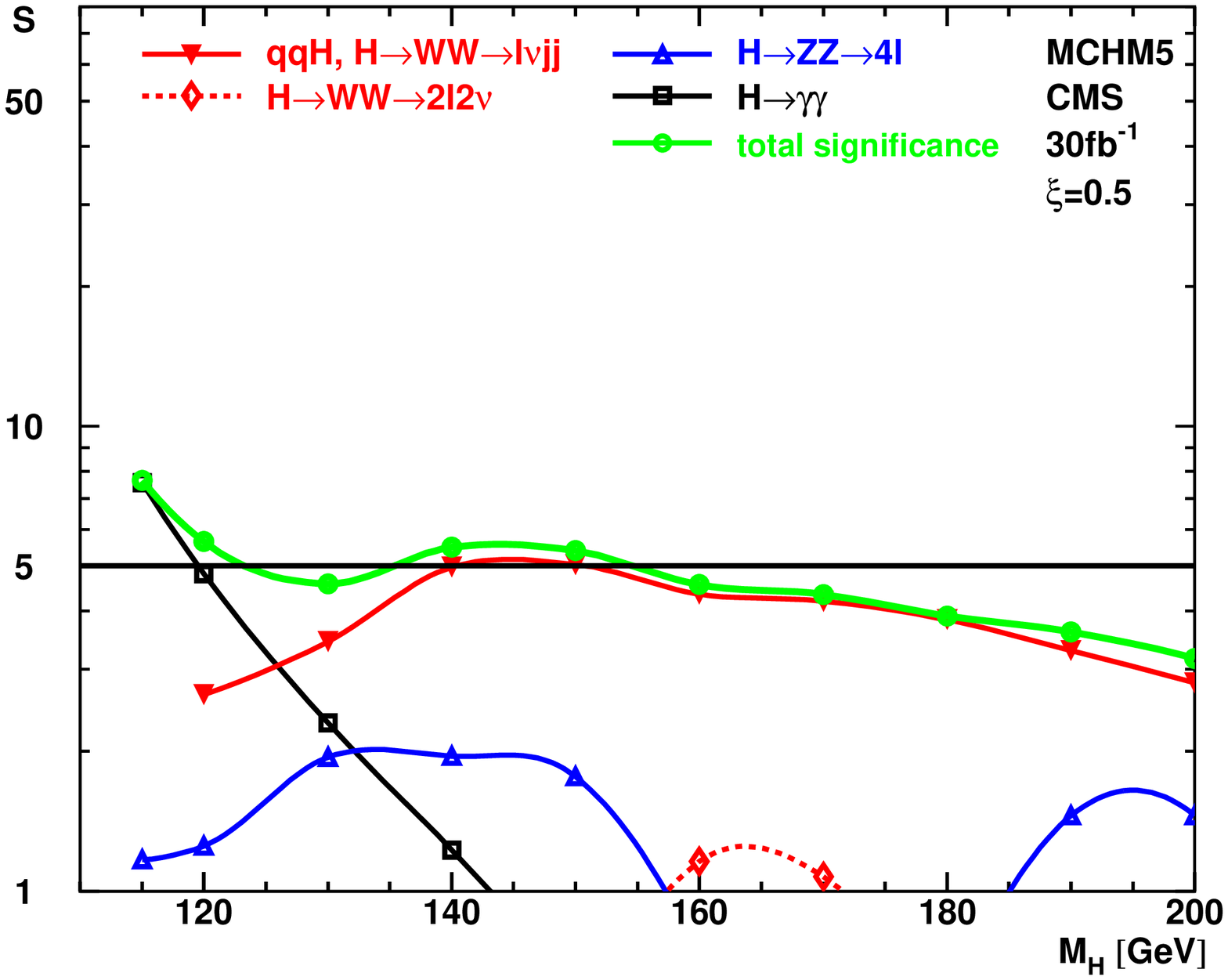,width=7cm}
\hspace*{0.5cm}
\epsfig{figure=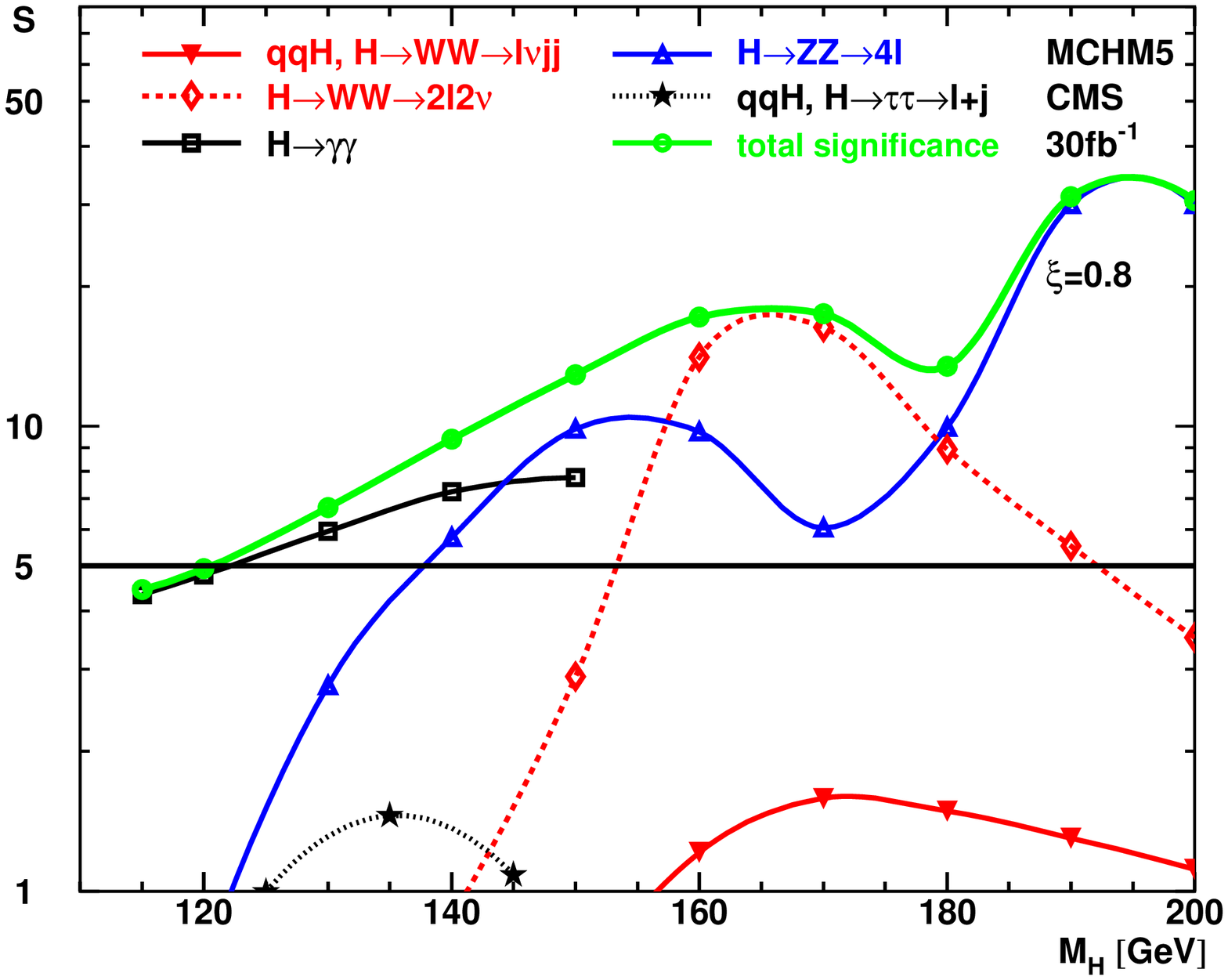,width=7cm}
\caption{\label{fig:sig2}
The significances in different channels as a function
  of the Higgs boson mass in the SM ($\xi=0$, upper left) and for 
MCHM5 with $\xi=0.2$ (upper right), 0.5 (bottom left) and 0.8 (bottom 
right).} \end{center}
\end{figure}

Combining the various channels discussed in the previous section 
gives an overall view of the expected significances 
and interplay of the different search channels we have discussed. 
Figures~\ref{fig:sig1} and \ref{fig:sig2} summarize the situation in 
MCHM4 and MCHM5, respectively, presenting as a function of $M_H$ the 
different expected significances and the total combined one. We choose the 
three representative values $\xi=0.2, 0.5$ and 0.8 and also show the SM case
($\xi=0$) for comparison.

In both models, for $\xi = 0.2$ the expectations are less promising than
in the SM. Both the gluon fusion and the gauge boson fusion production
cross-sections are reduced, so that all the significances move downwards.
Also, in MCHM5, this cannot be compensated by the enhancement of the
branching ratios into $\gamma\gamma$ and massive gauge bosons. The overall
significance in MCHM5 is worse than in MCHM4 because the gluon-fusion
process, which contributes to the main channels $H\to ZZ \to 4l$ and
$H\to WW\to 2l2\nu$, is more strongly suppressed than in MCHM4. Nevertheless,
by combining several search modes, discovery with an integrated luminosity
of 30 fb$^{-1}$ will still be possible.

The situation looks worse for $\xi=0.5$. In MCHM4, the combined
significance drops below $5\sigma$ for the interesting range $M_H \lsim
125$~GeV. A more sophisticated treatment of the $H\to\gamma\gamma$ channel
(like the one performed by CMS~\cite{cmstdr}) would be required to improve
the significance in that region, or other alternative search channels
should be exploited (see below). In MCHM5, the combined significance is
much worse (as the inclusive production cannot be exploited here anymore)
and barely reaches $5\sigma$ in some ranges, although these include the
interesting low mass range, thanks to the enhanced Higgs branching ratio
into photons.  In fact, this channel seems to be good for searches for
Higgs masses below 120~GeV, as the tendency of the curve implies. However,
in order to confirm this, experimental analyses for masses below 115~GeV,
which are not yet excluded in the composite model, would be needed. For
higher masses only the weak boson fusion with subsequent decay into $WW$
can be exploited.

For $\xi=0.8$, the situation is totally different in the two models. In
MCHM4, the progressive deterioration of the significance continues and the
combined significance is always below $5\sigma$. Instead, for MCHM5,
things look much better for masses above $\sim 120$~GeV.  The production
is completely taken over by the gluon-fusion process and leads to large
significances in the massive gauge boson final states. Also the
$\gamma\gamma$ final state contributes significantly above $\sim 120$~GeV.
The tendency of the curve shows that, for masses above 150~GeV, this
channel will still have large significance.  However, also here
experimental analyses are needed to confirm this. At low masses, the
situation does not look as good. Since the vector boson fusion and
Higgs-strahlung processes are largely suppressed they cannot contribute to
the search channels in this difficult region. One has to rely on inclusive
production with subsequent decay into photons. Besides an improved
analysis of the $H\to\gamma\gamma$ mode, perhaps $t\bar{t}H$ production
with $H\to b\bar{b}$ might help. Although, as we said, this channel is no
longer considered to be very useful in the SM, the enhancement of the
gluon-fusion cross-section (by a factor 1.8 for $\xi=0.8$) might reopen
this option.

\begin{figure}[!t]
\begin{center}
\epsfig{figure=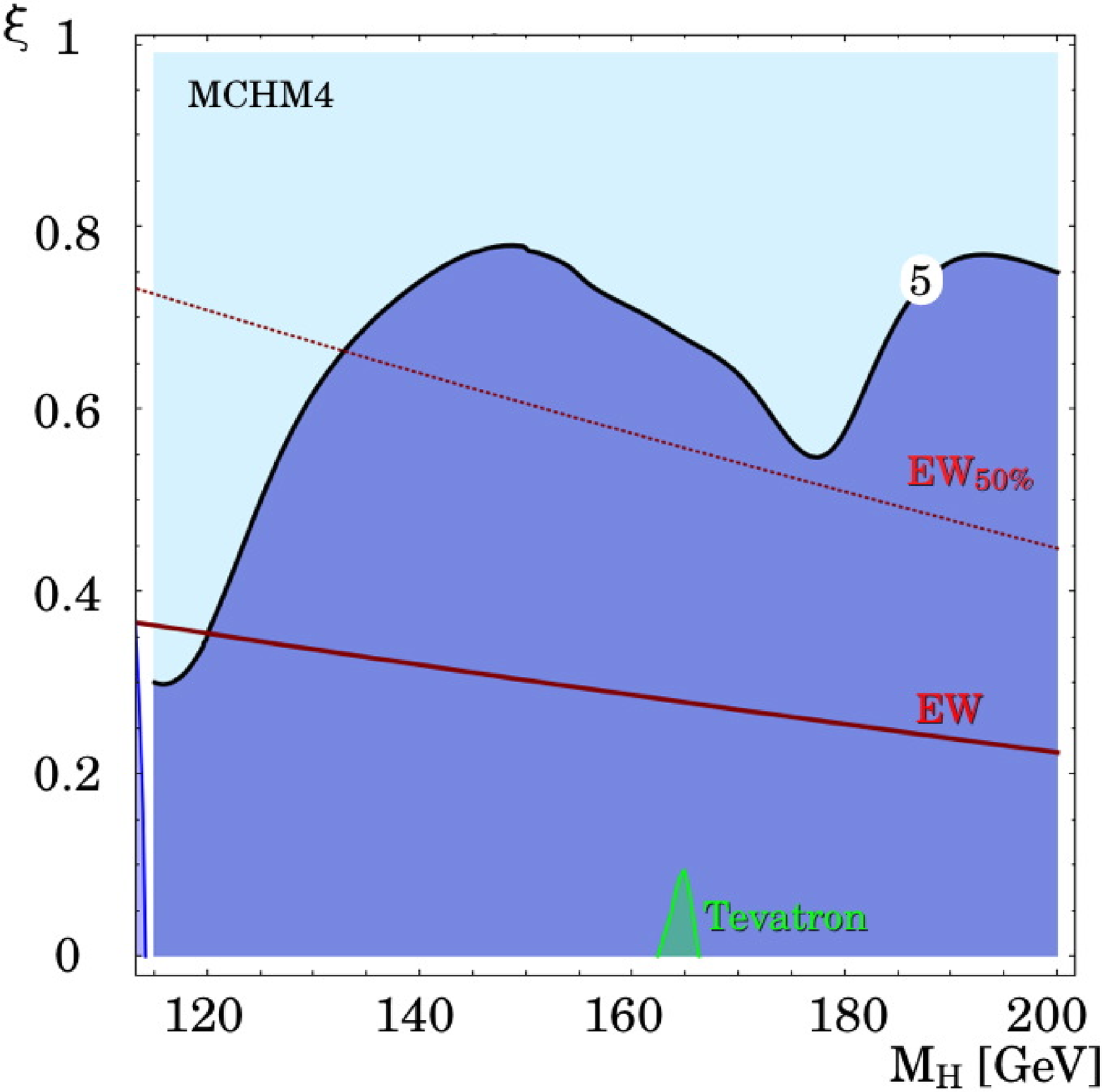,width=7cm}
\hspace*{0.5cm}
\epsfig{figure= 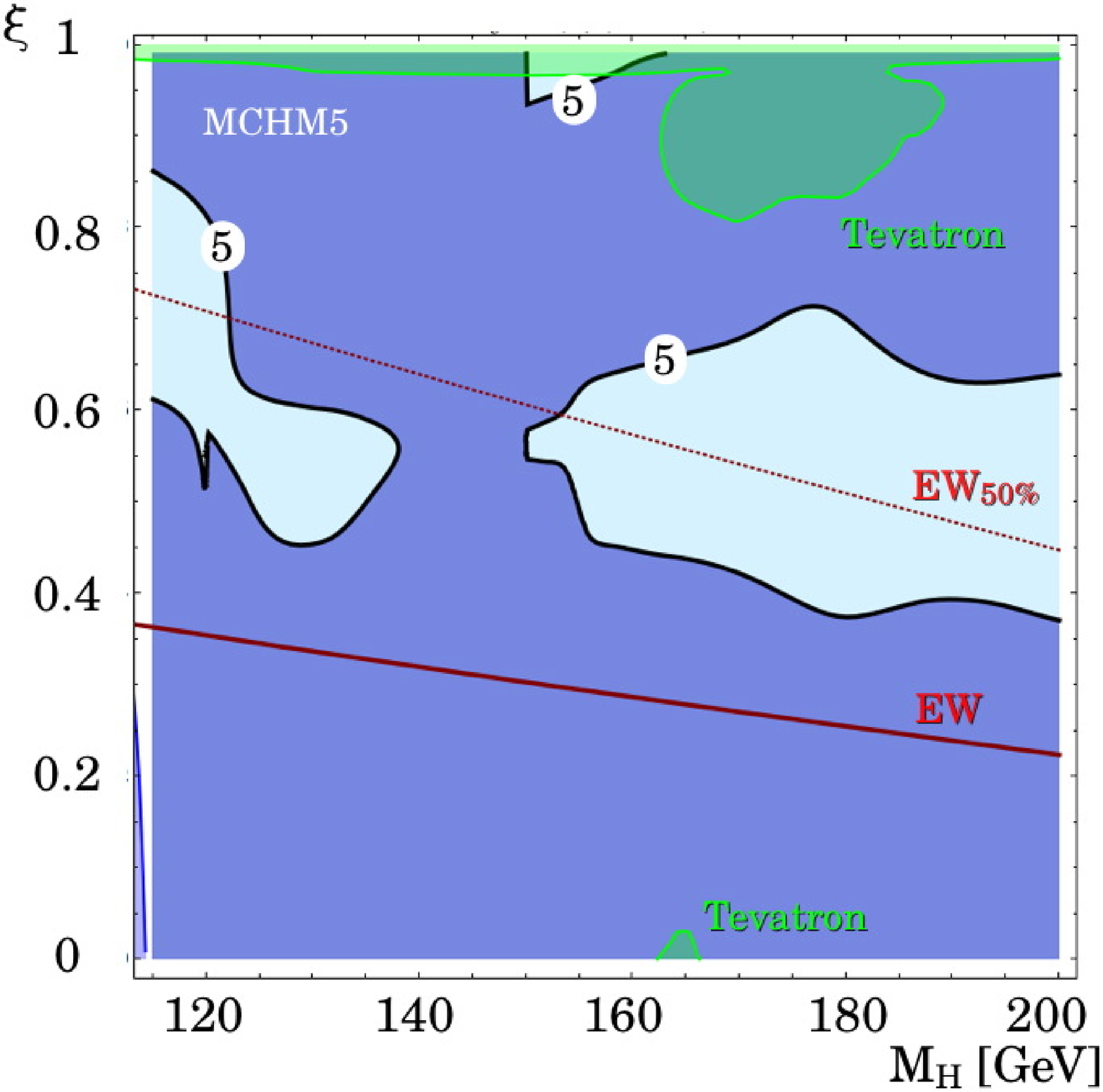,width=7cm}
\caption{\label{fig:signcomb}
The total 5$\sigma$ significance for the combined channels in 
the $(M_H,\xi)$ plane with an integrated luminosity of $30\ 
\mathrm{fb}^{-1}$ for MCHM4~(left) and MCHM5~(right), as in the previous figures the darker region corresponds to the higher significance.
The Tevatron exclusion bounds (green) from Section~\ref{sec:constraints} are also reported.
The red continuous line delineates the region favored at 99\% CL (with a cutoff scale fixed at 
2.5~TeV) while the region below the red dashed line 
survives if there is an additional 50\% cancellation of the oblique 
parameters. The LEP exclusion bounds are only marginally visible (tiny blue region in the bottom left corner) since the search analyses carried out by CMS deal exclusively with Higgs masses above the SM LEP exclusion bound.}
\end{center}
\end{figure} 

Figure~\ref{fig:signcomb} gives the $5\sigma$-significance contour line in the plane $(M_H,\xi)$ for MCHM4 and MCHM5.
Most of the CMS analyses available did not consider Higgs mass below the SM LEP exclusion bound. However, by comparing Fig.~\ref{fig:bounds} and Fig.~\ref{fig:signcomb}, we can infer that it might be worth extending these analyses for lower values of the Higgs mass, in particular in the region $\xi\sim 0.5$, where the LEP exclusion limit really deteriorates and the conflict with EW precision measurements is still not too severe. 


In summary, the search modes and corresponding significances can
substantially depart from the SM case, even at moderately low value of
$\xi$, i.e., for large compositeness scale of the Higgs boson. We did not
perform a full exploration of the 2D parameter space that controls the
deviations of the Higgs couplings, but, focusing on two particular
directions in this parameter space, we have identified interesting and
distinctive behaviors. In the first explicit model we consider, all the
Higgs couplings are reduced compared to the SM ones and, as a result, the
Higgs searches deteriorate. On the contrary, in the second model, for low
enough composite scales, the Higgs production by gluon fusion is enhanced
and results in searches with higher statistical significances.

After more than 40 years of theoretical existence, the Higgs boson has a
chance to show its face soon in the LHC detectors. Its discovery will
certainly also provide us with useful information about the nature of the
Higgs sector since the relative importance of the various production and
decay channels measures, to a certain extent, the dynamics of this Higgs
sector and will tell whether the force behind the phenomenon of
electroweak symmetry breaking is weak or strong.



\section*{Appendix: Significance estimators} 

For a given number of expected signal and background events ($s$ and $b$, 
respectively) there are many alternative ways in the literature to compute 
the corresponding expected significance, taking also into account the 
possible presence of a systematic uncertainty $\Delta b$ on $b$ (see 
\cite{Cousins} for a comparison of different possibilities). Besides the 
simple estimate $s/\sqrt{b}$, we make use in this paper of the following 
definitions of significance: 
\bea 
ScL[s,b] &\equiv &\sqrt{2[(s+b)\log(1+s/b)-s]}\ ,\nn\\
ScL'[s,b,\Delta b] &\equiv & ScL[s,b+\Delta b^2]\ ,
\eea
and
\be
ScP2[s,b,\Delta b] \equiv  
2\left(\sqrt{s+b}-\sqrt{b}\right)\sqrt{\frac{b}{b+\Delta b^2}}\ .
\ee
Note that this last definition, advocated in Eq.~(A.5) of Appendix~A of 
the CMS TDR~\cite{cmstdr} is incorrectly written there~\cite{Tom}.

Finally, the Poisson significance is defined as the number of standard
deviations that a Gaussian variable would fluctuate in one direction to
give the same \texttt{p}-value computed using the Poisson distribution
given the numbers of signal and background events, i.e., the Poisson
significance, $S_P$, is the solution of the equation
\begin{equation}
\sum_{i=0}^{s+b-1}\frac{e^{-b} b^i}{i!} = 
\int_{-\infty}^{S_P}dx\frac{e^{-x^2/2}}{\sqrt{2\pi}}.
\end{equation}

\section*{Acknowledgments}

\noindent
We thank Alexander Belyaev, Roberto Contino, Javier Cuevas, Michael Dittmar, Abdelhak Djouadi, Tommaso
Dorigo, Gian Giudice, Chiara Mariotti, Riccardo Rattazzi and Michael Spira
for useful discussions and for providing relevant information.  J.R.E. and M.M.
thank CERN PH-TH for partial financial support.  This work has been
partly supported by the European Commission under the contract ERC
advanced grant 226371 `MassTeV', the contract PITN-GA-2009-237920
`UNILHC', and the contract MRTN-CT-2006-035863 `ForcesUniverse', as well
as by the Spanish Consolider--Ingenio 2010 Programme CPAN (CSD2007-00042)
and the Spanish Ministry MICNN under contract FPA 2007-60252.


\begin{thebibliography}{10}
%
\bibitem{Giudice:2007fh}
  G.~F.~Giudice, C.~Grojean, A.~Pomarol and R.~Rattazzi,
  JHEP {\bf 0706} (2007) 045
  [hep-ph/0703164].
%
\bibitem{Contino:2010mh}
  R.~Contino, C.~Grojean, M.~Moretti, F.~Piccinini and R.~Rattazzi,
  [hep-ph/1002.1011].
%
\bibitem{dilaton}
  W.~D.~Goldberger, B.~Grinstein and W.~Skiba,
  Phys.\ Rev.\ Lett.\  {\bf 100} (2008) 111802
  [hep-ph/0708.1463];
  J.~Fan, W.~D.~Goldberger, A.~Ross and W.~Skiba,
  Phys.\ Rev.\  D {\bf 79} (2009) 035017
  [hep-ph/0803.2040];
  L.~Vecchi,
  [hep-ph/1002.1721].
%
\bibitem{Kaplan:1983fs}
  D.~B.~Kaplan and H.~Georgi,
  Phys.\ Lett.\  B {\bf 136} (1984) 183.
%
\bibitem{othercompositeHiggs}
S.~Dimopoulos and J.~Preskill,
  Nucl.\ Phys.\  B {\bf 199}, 206 (1982);
T.~Banks,
  Nucl.\ Phys.\  B {\bf 243}, 125 (1984);
D.~B.~Kaplan, H.~Georgi and S.~Dimopoulos,
  Phys.\ Lett.\  B {\bf 136}, 187 (1984);
H.~Georgi, D.~B.~Kaplan and P.~Galison,
  Phys.\ Lett.\  B {\bf 143}, 152 (1984);
H.~Georgi and D.~B.~Kaplan,
  Phys.\ Lett.\  B {\bf 145}, 216 (1984);
M.~J.~Dugan, H.~Georgi and D.~B.~Kaplan,
  Nucl.\ Phys.\  B {\bf 254}, 299 (1985).
%
\bibitem{Falkowski:2007hz}
  A.~Falkowski,
  Phys.\ Rev.\  D {\bf 77} (2008) 055018
  [hep-ph/0711.0828].
%
\bibitem{otherhgg}
  A.~Djouadi and G.~Moreau,
  Phys.\ Lett.\  B {\bf 660} (2008) 67
  [hep-ph/0707.3800];
  N.~Maru,
  Mod.\ Phys.\ Lett.\  A {\bf 23} (2008) 2737
  [hep-ph/0803.0380];
  G.~Bhattacharyya and T.~S.~Ray,
  Phys.\ Lett.\  B {\bf 675} (2009) 222
  [hep-ph/0902.1893];
  N.~Maru, T.~Nomura, J.~Sato and M.~Yamanaka,
  [hep-ph/0905.4554].
%
\bibitem{reviewtechnicolor}
For recent reviews on technicolor models, see
  C.~T.~Hill and E.~H.~Simmons,
  Phys.\ Rept.\  {\bf 381} (2003) 235
  [hep-ph/0203079]; 
  F.~Sannino,
  hep-ph:0911.0931.
  
%
\bibitem{Giudice:2000av}
  G.~F.~Giudice, R.~Rattazzi and J.~D.~Wells,
  Nucl.\ Phys.\  B {\bf 595} (2001) 250
  [hep-ph/0002178].
%
\bibitem{higgsless}
  C.~Csaki, C.~Grojean, H.~Murayama, L.~Pilo and J.~Terning,
  Phys.\ Rev.\  D {\bf 69} (2004) 055006
  [hep-ph/0305237];
  C.~Csaki, C.~Grojean, L.~Pilo and J.~Terning,
  Phys.\ Rev.\ Lett.\  {\bf 92} (2004) 101802
  [hep-ph/0308038].
%
\bibitem{Cacciapaglia:2006mz}
  G.~Cacciapaglia, C.~Csaki, G.~Marandella and J.~Terning,
  JHEP {\bf 0702} (2007) 036
  [hep-ph/0611358].
%
\bibitem{Stancato:2008mp}
  D.~Stancato and J.~Terning,
  JHEP {\bf 0911} (2009) 101
  [hep-ph/0807.3961].
%
\bibitem{Luty:2004ye}
  M.~A.~Luty and T.~Okui,
  JHEP {\bf 0609}, 070 (2006)
  [hep-ph/0409274].
%
\bibitem{reviewsNewPhysics}
  C.~Grojean,
  [hep-ph/0910.4976];
  D.~E.~Morrissey, T.~Plehn and T.~M.~P.~Tait,
  [hep-ph/0912.3259].
%
\bibitem{generalhgg}
  A.~V.~Manohar and M.~B.~Wise,
  Phys.\ Lett.\  B {\bf 636} (2006) 107
  [hep-ph/0601212];
  A.~Pierce, J.~Thaler and L.~T.~Wang,
  JHEP {\bf 0705} (2007) 070
  [hep-ph/0609049].
%
\bibitem{hgg}
  G.~Cacciapaglia, A.~Deandrea and J.~Llodra-Perez,
  JHEP {\bf 0906} (2009) 054
  [hep-ph/0901.0927].
%
\bibitem{Contino:2003ve}
  R.~Contino, Y.~Nomura and A.~Pomarol,
  Nucl.\ Phys.\  B {\bf 671} (2003) 148
  [hep-ph/0306259].
%
\bibitem{Agashe:2004rs}
  K.~Agashe, R.~Contino and A.~Pomarol,
  Nucl.\ Phys.\  B {\bf 719} (2005) 165
  [hep-ph/0412089].
%
\bibitem{Contino:2006qr}
  R.~Contino, L.~Da Rold and A.~Pomarol,
  Phys.\ Rev.\  D {\bf 75} (2007) 055014
  [hep-ph/0612048].
%
\bibitem{Low:2009di}
  I.~Low, R.~Rattazzi and A.~Vichi,
  [hep-ph/0907.5413].
%
\bibitem{Spira:1997ce}
  M.~Spira and J.~D.~Wells,
  Nucl.\ Phys.\  B {\bf 523} (1998) 3
  [hep-ph/9711410].
%
\bibitem{hdecay}
A.~Djouadi, J.~Kalinowski and M.~Spira,
Comput.\ Phys.\ Commun.\  {\bf 108} (1998) 56
[hep-ph/9704448].
For an update, see 
 A. Djouadi, J. Kalinowski, M. Muhlleitner and M. Spira
 in J.~M.~Butterworth {\it et al.},
  [hep-ph/1003.1643].
%
\bibitem{higgsbounds}
P.~Bechtle, O.~Brein, S.~Heinemeyer, G.~Weiglein and K.~E.~Williams,
Comput.\ Phys.\ Commun.\  {\bf 181} (2010) 138
[hep-ph:0811.4169].
(See also http://www.ippp.dur.ac.uk/HiggsBounds)
%
\bibitem{LEPHbb}
R.~Barate {\it et al.}  [LEP Working Group for Higgs boson searches],
Phys.\ Lett.\  B {\bf 565} (2003) 61
[hep-ex/0306033];
S.~Schael {\it et al.}  [ALEPH and DELPHI and L3 and OPAL Collaborations],
Eur.\ Phys.\ J.\  C {\bf 47} (2006) 547
[hep-ex/0602042].
%
\bibitem{LEPHgg} [LEP Higgs Working Group], LHWG Note 2002-02.
%
\bibitem{TEVH}
  T.~Aaltonen {\it et al.}  [CDF and D0 Collaborations],
  Phys.\ Rev.\ Lett.\  {\bf 104} (2010) 061802
  [hep-ex/1001.4162];
%
\bibitem{TEVtautau}
The TEVNPH Working Group for the CDF and D0 Collaborations,
FERMILAB-PUB-09-394-E, CDF Note 9888, D0 Note 5980-CONF.
%
\bibitem{Peskin:1991sw}
  M.~E.~Peskin and T.~Takeuchi,
  Phys.\ Rev.\  D {\bf 46} (1992) 381.
%
\bibitem{Barbieri:2007bh}
  R.~Barbieri, B.~Bellazzini, V.~S.~Rychkov and A.~Varagnolo,
  Phys.\ Rev.\  D {\bf 76} (2007) 115008
  [hep-ph/0706.0432].
%
\bibitem{Altarelli:1990zd}
  G.~Altarelli and R.~Barbieri,
  Phys.\ Lett.\  B {\bf 253} (1991) 161.
%
\bibitem{fortsch} 
  M.~Spira,
  Fortsch.\ Phys.\  {\bf 46}, 203 (1998)
  [hep-ph/9705337].
%
\bibitem{Djouadi:2005gi}
  A.~Djouadi,
  Phys.\ Rept.\  {\bf 457} (2008) 1
  [hep-ph/0503172].
  %
\bibitem{georgi}
H.~M.~Georgi, S.~L.~Glashow, M.~E.~Machacek and D.~V.~Nanopoulos,
  Phys.\ Rev.\ Lett.\  {\bf 40}, 692 (1978).
%
\bibitem{graudenz}
  M.~Spira, A.~Djouadi, D.~Graudenz and P.~M.~Zerwas,
  Phys.\ Lett.\  B {\bf 318} (1993) 347;
  Nucl.\ Phys.\  B {\bf 453} (1995) 17
  [hep-ph/9504378].
%
\bibitem{laenen}
  D.~Graudenz, M.~Spira and P.~M.~Zerwas,
  Phys.\ Rev.\ Lett.\  {\bf 70} (1993) 1372.
%
  S.~Dawson,
  Nucl.\ Phys.\  B {\bf 359}, 283 (1991);
  R.~P.~Kauffman and W.~Schaffer,
  Phys.\ Rev.\  D {\bf 49}, 551 (1994)
  [hep-ph/9305279];
  S.~Dawson and R.~Kauffman,
  Phys.\ Rev.\  D {\bf 49}, 2298 (1994)
  [hep-ph/9310281];
  M.~Kramer, E.~Laenen and M.~Spira,
  Nucl.\ Phys.\  B {\bf 511}, 523 (1998)
  [hep-ph/9611272].
%
\bibitem{harlander}
  R.~V.~Harlander and W.~B.~Kilgore,
  Phys.\ Rev.\ Lett.\  {\bf 88} (2002) 201801
  [hep-ph/0201206];
  C.~Anastasiou and K.~Melnikov,
  Nucl.\ Phys.\  B {\bf 646}, 220 (2002)
  [hep-ph/0207004];
  V.~Ravindran, J.~Smith and W.~L.~van Neerven,
  Nucl.\ Phys.\  B {\bf 665} (2003) 325
  [hep-ph/0302135].
%
\bibitem{resum}
  S.~Catani, D.~de Florian, M.~Grazzini and P.~Nason,
  JHEP {\bf 0307}, 028 (2003)
  [hep-ph/0306211].
%
\bibitem{ozeren} 
  R.~V.~Harlander and K.~J.~Ozeren,
  Phys.\ Lett.\  B {\bf 679}, 467 (2009)
  [hep-ph/0907.2997];
  R.~V.~Harlander and K.~J.~Ozeren,
  JHEP {\bf 0911}, 088 (2009)
  [hep-ph/0909.3420];
  A.~Pak, M.~Rogal and M.~Steinhauser,
  Phys.\ Lett.\  B {\bf 679}, 473 (2009)
  [hep-ph/0907.2998];
  A.~Pak, M.~Rogal and M.~Steinhauser,
  [hep-ph/0911.4662].
%
\bibitem{ewcorr} 
  A.~Djouadi and P.~Gambino,
  Phys.\ Rev.\ Lett.\  {\bf 73}, 2528 (1994)
  [hep-ph/9406432];
  A.~Ghinculov and J.~J.~van der Bij,
  Nucl.\ Phys.\  B {\bf 482}, 59 (1996)
  [hep-ph/9511414];
  A.~Djouadi, P.~Gambino and B.~A.~Kniehl,
  Nucl.\ Phys.\  B {\bf 523}, 17 (1998)
  [hep-ph/9712330];
  G.~Degrassi and F.~Maltoni,
  Phys.\ Lett.\  B {\bf 600} (2004) 255;
  [hep-ph/0407249].
  U.~Aglietti, R.~Bonciani, G.~Degrassi and A.~Vicini,
  [hep-ph/0610033];
  S.~Actis, G.~Passarino, C.~Sturm and S.~Uccirati,
  Phys.\ Lett.\  B {\bf 670}, 12 (2008)
  [hep-ph/0809.1301];
  C.~Anastasiou, R.~Boughezal and F.~Petriello,
  JHEP {\bf 0904}, 003 (2009)
  [hep-ph/0811.3458].
%
\bibitem{higlu}
  M.~Spira,
  ``HIGLU: A Program for the Calculation of the Total Higgs Production 
Cross Section at Hadron Colliders via Gluon Fusion including QCD 
Corrections,''
  [hep-ph/9510347].
%
\bibitem{wzfusion} 
  R.~N.~Cahn and S.~Dawson,
  Phys.\ Lett.\  B {\bf 136}, 196 (1984)
  [Erratum-ibid.\  B {\bf 138}, 464 (1984)];
  K.~I.~Hikasa,
  Phys.\ Lett.\  B {\bf 164}, 385 (1985)
  [Erratum-ibid.\  {\bf 195B}, 623 (1987)];
  G.~Altarelli, B.~Mele and F.~Pitolli,
  Nucl.\ Phys.\  B {\bf 287}, 205 (1987).
%
\bibitem{wzfusqcd}
  T.~Han, G.~Valencia and S.~Willenbrock,
  Phys.\ Rev.\ Lett.\  {\bf 69}, 3274 (1992)
  [hep-ph/9206246].
%
\bibitem{wzfusew} 
  T.~Figy, C.~Oleari and D.~Zeppenfeld,
  Phys.\ Rev.\  D {\bf 68}, 073005 (2003)
  [hep-ph/0306109];
  E.~L.~Berger and J.~M.~Campbell,
  Phys.\ Rev.\  D {\bf 70}, 073011 (2004)
  [hep-ph/0403194];
  M.~Ciccolini, A.~Denner and S.~Dittmaier,
  Phys.\ Rev.\  D {\bf 77}, 013002 (2008)
  [hep-ph/0710.4749].
%
\bibitem{programs} URL: {\tt http://people.web.psi.ch/spira/proglist.html}
%
\bibitem{higgsrad} 
  S.~L.~Glashow, D.~V.~Nanopoulos and A.~Yildiz,
  Phys.\ Rev.\  D {\bf 18}, 1724 (1978);
  Z.~Kunszt, Z.~Trocsanyi and W.~J.~Stirling,
  Phys.\ Lett.\  B {\bf 271}, 247 (1991).
%
\bibitem{qcdhrad} 
  T.~Han and S.~Willenbrock,
  Phys.\ Lett.\  B {\bf 273}, 167 (1991).
%
\bibitem{nnlohrad} 
  O.~Brein, A.~Djouadi and R.~Harlander,
  Phys.\ Lett.\  B {\bf 579}, 149 (2004)
  [hep-ph/0307206].
%
\bibitem{ewhrad} 
  M.~L.~Ciccolini, S.~Dittmaier and M.~Kramer,
  Phys.\ Rev.\  D {\bf 68}, 073003 (2003)
  [hep-ph/0306234].
%
\bibitem{lotth} 
  R.~Raitio and W.~W.~Wada,
  Phys.\ Rev.\  D {\bf 19}, 941 (1979);
  J.~N.~Ng and P.~Zakarauskas,
  Phys.\ Rev.\  D {\bf 29}, 876 (1984);
  Z.~Kunszt,
  Nucl.\ Phys.\  B {\bf 247}, 339 (1984);
  W.~J.~Marciano and F.~E.~Paige,
  Phys.\ Rev.\ Lett.\  {\bf 66}, 2433 (1991).
%
\bibitem{nlotth} 
  W.~Beenakker, S.~Dittmaier, M.~Kramer, B.~Plumper, M.~Spira and 
P.~M.~Zerwas,
  Phys.\ Rev.\ Lett.\  {\bf 87} (2001) 201805
  [hep-ph/0107081];
  Nucl.\ Phys.\  B {\bf 653} (2003) 151
  [hep-ph/0211352];
  S.~Dawson, L.~H.~Orr, L.~Reina and D.~Wackeroth,
  Phys.\ Rev.\  D {\bf 67}, 071503 (2003)
  [hep-ph/0211438].
%
\bibitem{cmstdr}
  G.~L.~Bayatian {\it et al.}  [CMS Collaboration],
  J.\ Phys.\ G {\bf 34} (2007) 995.
%
\bibitem{atlastdr}
  G.~Aad {\it et al.}  [The ATLAS Collaboration],
  [hep-ex/0901.0512].
%
\bibitem{ellis}
R.~K.~Ellis, I.~Hinchliffe, M.~Soldate and J.~J.~van der Bij,
Nucl.\ Phys.\  B {\bf 297}~(1988)~221.
%
\bibitem{moretti}
  A.~Belyaev, R.~Guedes, S.~Moretti and R.~Santos,
  [hep-ph/0912.2620].
%
\bibitem{CMS2006112}
M.~Pieri et al., [CMS Collaboration],
CERN-CMS-NOTE-2006/112.
%
\bibitem{Accomando:2006ga}
  E.~Accomando {\it et al.},
  [hep-ph/0608079], and references therein.
%
\bibitem{cppapers}
  V.~D.~Barger, K.~M.~Cheung, A.~Djouadi, B.~A.~Kniehl and P.~M.~Zerwas,
  Phys.\ Rev.\  D {\bf 49} (1994) 79
  [hep-ph/9306270].
 S.~Y.~Choi, D.~J.~.~Miller, M.~M.~Muhlleitner and P.~M.~Zerwas,
 Phys.\ Lett.\  B {\bf 553} (2003) 61;
 [hep-ph/0210077];
 C.~P.~Buszello, I.~Fleck, P.~Marquard and J.~J.~van der Bij,
  Eur.\ Phys.\ J.\  C {\bf 32} (2004) 209;
 [hep-ph/0212396];
  R.~M.~Godbole, D.~J.~Miller and M.~M.~Muhlleitner,
  JHEP {\bf 0712} (2007) 031.
  [0708.0458 [hep-ph]].
%
\bibitem{CMS2006115}
S.~Baffioni {\it et al.}, [CMS Collaboration], 
J.\ Phys.\ G {\bf 34} (2007) N23; CMS Note 2006/115.
%
\bibitem{CMS2006122}
S.~Abdullin {\it et al.}, [CMS Collaboration],
Acta Phys.\ Polon.\  B {\bf 38} (2007) 731, CMS Note 2006/122.
%
\bibitem{CMS2006136}
D.~Futyan, D.~Fortin and D.~Giordano, [CMS Collaboration],
J.\ Phys.\ G {\bf 34} (2007) N315, CMS Note 2006/136.
%
\bibitem{CMSPASHIG08003}
CMS Collaboration, 
CMS PAS HIG-08-003.
%
\bibitem{Dittmar:1996ss}
  M.~Dittmar and H.~K.~Dreiner,
  Phys.\ Rev.\  D {\bf 55} (1997) 167
  [hep-ph/9608317].
\bibitem{CMS2006047}
  G.~Davatz, M.~Dittmar and A.~S.~Giolo-Nicollerat, [CMS Collaboration],
  J.\ Phys.\ G {\bf 33} (2007) N85, CMS Note 2006/047.
%
\bibitem{CMSPASHIG08006}
CMS Collaboration, 
CMS PAS HIG--08--006.
%
\bibitem{CMS2006092}
  H.~F.~Pi, P.~Avery, J.~Rohlf, C. Tully and S.~Kunori, [CMS Collaboration],
CMS Note 2006/092.  
%
\bibitem{Rainwater:1998kj}
  D.~L.~Rainwater, D.~Zeppenfeld and K.~Hagiwara,
  Phys.\ Rev.\  D {\bf 59} (1999) 014037
  [hep-ph/9808468];
  T.~Plehn, D.~L.~Rainwater and D.~Zeppenfeld,
  Phys.\ Rev.\  D {\bf 61} (2000) 093005
  [hep-ph/9911385].
%
\bibitem{Cavalli:2002vs}
  D.~Cavalli {\it et al.},
  [hep-ph/0203056].
%
\bibitem{CMS2006088}
 C.~Foudas, A.~Nikitenko and M.~Takahashi, [CMS Collaboration],
CMS Note 2006/088.  
%
\bibitem{Cousins}
R.~D.~Cousins, J.~T.~Linnemann and J.~Tucker,
Nucl.\ Instrum.\ Meth.\  A {\bf 595} (2008) 480.
%
\bibitem{Tom} Tommaso Dorigo, private communication.
%
\end{thebibliography}
\end{document}